\documentclass[fleqn,usenatbib]{mnras}

\usepackage{newtxtext,newtxmath}

\usepackage[T1]{fontenc}

\DeclareRobustCommand{\VAN}[3]{#2}
\let\VANthebibliography\thebibliography
\def\thebibliography{\DeclareRobustCommand{\VAN}[3]{##3}\VANthebibliography}

\usepackage{graphicx}	
\usepackage{amsmath}	
\usepackage{subcaption}
\usepackage{MnSymbol}
\usepackage{xcolor}

\title[WISDOM BH mass in NGC\,0612, NGC\,1574, and NGC\,4261]{WISDOM project - XIV. SMBH mass in the early-type galaxies NGC\,0612, NGC\,1574, and NGC\,4261 from CO dynamical modelling}

\author[I. Ruffa et al.]{
Ilaria Ruffa,$^{1,2}$\thanks{E-mail: RuffaI@cardiff.ac.uk}
Timothy A. Davis,$^{1}$
Michele Cappellari,$^{3}$
Martin Bureau,$^{3}$
Jacob Elford,$^{1}$
Satoru Iguchi,$^{4,5}$
\newauthor
Federico Lelli,$^{6}$
Fu-Heng Liang,$^{3}$
Lijie Liu,$^{3}$
Anan Lu,$^{7}$
Marc Sarzi,$^{8}$
Thomas G. Williams$^{9,3}$
\\
$^{1}$Cardiff Hub for Astrophysics Research \&\ Technology, School of Physics \&\ Astronomy, Cardiff University, Queens Buildings, The Parade, Cardiff, CF24 3AA, UK\\
$^{2}$INAF - Istituto di Radioastronomia, via P.\ Gobetti 101, 40129 Bologna, Italy\\
$^{3}$Sub-department of Astrophysics, Department of Physics, University of Oxford, Keble Road, Oxford, OX1 3RH, UK\\
$^{4}$Department of Astronomical Science, SOKENDAI (The Graduate University of Advanced Studies), Mitaka, Tokyo 181-8588, Japan\\
$^{5}$National Astronomical Observatory of Japan, National Institutes of Natural Sciences, Mitaka, Tokyo 181-8588, Japan\\
$^{6}$INAF, Arcetri Astrophysical Observatory, Largo Enrico Fermi 5, I-50125, Florence, Italy\\
$^{7}$Trottier Space Institute and Department of Physics, McGill University, 3600 rue University, Montreal, QC H3A 2T8, Canada\\
$^{8}$Armagh Observatory and Planetarium, College Hill, Armagh BT61 9DG, UK\\
$^{9}$Max Planck Institut f\"{u}r Astronomie, K\"{o}nigstuhl 17, 69117 Heidelberg, Germany
}

\date{Accepted XXX. Received YYY; in original form ZZZ}

\pubyear{2015}

\begin{document}
\label{firstpage}
\pagerange{\pageref{firstpage}--\pageref{lastpage}}
\maketitle

\begin{abstract}
We present a CO dynamical estimate of the mass of the super-massive black hole (SMBH) in three nearby early-type galaxies: NGC\,0612, NGC\,1574 and NGC\,4261. Our analysis is based on Atacama Large Millimeter/submillimeter Array (ALMA) Cycle 3-6 observations of the $^{12}$CO(2-1) emission line with spatial resolutions of $14-58$~pc ($0.01\arcsec-0.26\arcsec$). We detect disc-like CO distributions on scales from $\lesssim200$~pc (NGC\,1574 and NGC\,4261) to $\approx10$~kpc (NGC\,0612). In NGC\,0612 and NGC\,1574 the bulk of the gas is regularly rotating. The data also provide evidence for the presence of a massive dark object at the centre of NGC\,1574, allowing us to obtain the first measure of its mass,  $M_{\rm BH}=(1.0\pm0.2)\times10^{8}$~M$_{\odot}$ (1$\sigma$ uncertainty). In NGC\,4261, the CO kinematics is clearly dominated by the SMBH gravitational influence, allowing us to determine an accurate black hole mass of $(1.62{\pm 0.04})\times10^{9}$~M$_{\odot}$ ($1\sigma$ uncertainty). This is fully consistent with a previous CO dynamical estimate obtained using a different modelling technique. Signs of non-circular gas motions (likely outflow) are also identified in the inner regions of NGC\,4261. In NGC\,0612, we are only able to obtain a (conservative) upper limit of $M_{\rm BH}\lesssim3.2\times10^{9}$~M$_{\odot}$. This has likely to be ascribed to the presence of a central CO hole (with a radius much larger than that of the SMBH sphere of influence), combined with the inability of obtaining a robust prediction for the CO velocity curve. The three SMBH mass estimates are overall in agreement with predictions from the $M_{\rm BH}-\sigma_{\rm \filledstar}$ relation.
\end{abstract}




\section{Introduction}\label{sec:intro}
One of the most demanding requirements of modern galaxy formation theories is to shed light on the physical processes driving and regulating the connection between the growth of super-massive black holes (SMBHs, found at the centre of almost every galaxy with stellar mass $M_{\filledstar} \gtrsim 10^{9}~M_{\odot}$) and the evolution of their host galaxies \citep[the so-called "co-evolution"; see e.g.][]{Kormendy13}.
The observational efforts carried out over the last three decades have indeed demonstrated that the mass of such SMBHs correlates with a number of host galaxy properties, suggesting that the two may co-evolve in a self-regulating manner. The correlation between the SMBH mass and stellar velocity dispersion (commonly indicated as $M_{\rm BH}-\sigma_{\rm \filledstar}$, where $\sigma_{\rm \filledstar}$ is the stellar velocity dispersion within one effective radius) has been initially found to be the tightest \citep[e.g.][]{Ferrarese00,Gebhardt00,Gultekin09}, but evidence have been accumulated over the years of divergence between galaxies of different morphological types or masses \citep[e.g.][]{McConnell13,Vandenbosch16,Kraj18}. In the current scenario, active galactic nuclei (AGN) and associated energetic output (i.e.\,feedback) are believed to play a crucial role in setting up self-regulating co-evolutionary processes, being responsible for changing the physical conditions of the surrounding interstellar medium (ISM) or expelling it from the nuclear regions \citep[e.g.][]{Morganti17,Harrison18}. The many details of these processes, however, are still poorly understood, and thus the mechanisms regulating the SMBH-host galaxy connection are still hotly debated \citep[see e.g.][for a recent review]{Donofrio21}.

A key issue is the fact that SMBH-host galaxy correlations strongly rely on the accuracy of the SMBH mass measurements. The most reliable estimates can be obtained through dynamical studies of matter orbiting within the gravitational sphere of influence (SOI) of the SMBH\footnote{The SOI is the region inside which the gravitational potential of the SMBH dominates over that of the host galaxy, and is typically defined as $R_{\rm SOI} = \dfrac{GM_{\rm BH}}{\sigma_{\rm \filledstar}^{2}}$, where $G$ is the gravitational constant, $M_{\rm BH}$ is the mass of the SMBH, and $\sigma_{\rm \filledstar}$ is the stellar velocity dispersion within one effective radius.}. Method used so far includes modelling the stellar \citep[e.g.][]{Cappellari02b,Kraj09}, ionised gas \citep[e.g][]{Ferrarese96,Sarzi01,Dallabonta09,Walsh13}, and maser kinematics \citep[e.g.][]{Miyoshi95,Greene10,Kuo11}. Each of these methods, however, have unavoidable limitations. For instance, studies of stellar kinematics can be strongly affected by dust extinction and are often restricted to axisymmetric objects, resulting in samples heavily biased towards early-type galaxies (ETGs). They also require very high resolution to reliably probe the line-of-sight stellar velocity distribution. Dynamical measurements from ionised gas kinematics would be technically simpler, but they are often (if not always) challenged by the presence of non-gravitational forces (e.g.\,shocks) and/or turbulent gas motions superimposed on (quasi-)circular motion. High-precision SMBH mass estimates can be obtained by modelling maser motions in the accretion disc, as these probe material very close to the SMBH. However, maser emission is usually observed only at the centre of certain types of AGN (mostly Seyfert 2, typically in late-type hosts) and detectable only if the maser disc is observed edge-on. 

Over the past decade, thanks to the unprecedented resolution and sensitivity provided by the latest generation of (sub-)millimeter interferometers, a new method for SMBH mass measurements has been developed and successfully applied: probing the kinematics of the molecular gas down to the SOI using CO emission lines \citep[e.g.][]{Davis13a,Barth16,Onishi17,Davis17,Boizelle17,North19,Smith19,Smith21b,Boizelle21}. CO is a promising kinematic tracer: the bulk of the gas is usually found to be dynamically cold (i.e.\,large v/$\sigma$ ratios), and unaffected by dust attenuation. Non-circular motions (such as outflows) can challenge the $M_{\rm BH}$ determination, but few galaxies are expected (and observed) to have strong outflows on the typical scales probed in these studies (i.e.\,10-100~pc; \citealp[e.g.][]{Stuber21}). The presence of CO morphological asymmetries (e.g.\,warps) and/or nuclear gas deficiencies can be equally problematic, but it has been already demonstrated that reliable mass estimates can be obtained also in cases like these \citep[e.g.][]{Davis13a,Onishi17,Davis18,Smith19}. This technique also has the great advantage of allowing SMBH mass estimations in the same manner in all types of galaxies (late- and early-types, active and inactive), thus offering a promising alternative method that can be consistently applied across the Hubble sequence. 

The mm-Wave Interferometric Survey of Dark Object Masses (WISDOM) project has the primary aim of measuring SMBH masses using high-resolution CO observations from the Atacama Large Millimeter/submillimeter Array (ALMA) and the Combined Array for Research in Millimeter-wave Astronomy (CARMA). The WISDOM sample probes a morphologically diverse sample of galaxy hosts, including both early-type (ellipticals and lenticulars) and late-type (LTGs; spirals and irregulars), active and inactive galaxies (see Elford et al.\,, subm.\,). To date, WISDOM has provided accurate SMBH masses in six typical ETGs \citep[][]{Onishi17,Davis17,Davis18,Smith19,North19,Smith21b}, a peculiar luminous infrared galaxy (LIRG) with a central spiral disc \citep[][]{Lelli22}, and even a dwarf ETG \citep[][]{Davis20b}. SMBH mass measurements based on the same technique have been also presented by other groups for six additional ETGs \citep[][]{Barth16,Boizelle19,Nagai19,Ruffa19b,Boizelle21}, seven LTGs with AGN \citep[][]{Combes19}, and a barred spiral galaxy \citep[][]{Nguyen20}.

This paper presents the WISDOM 3D kinematic modelling of the CO(2-1) line in three ETGs observed at high-resolution with ALMA: NGC\,0612, NGC\,1574, and NGC\,4261. The main aim is to obtain molecular gas dynamical measurements of the SMBH mass in these objects. We note that a CO dynamical SMBH mass estimate for NGC\,4261 has been already presented in \citet[][]{Boizelle21} using ALMA data at lower resolution, and its inclusion here allows us to perform a crucial cross check between the different methodologies developing in this field.

The paper is structured as follows. In Section~\ref{sec:targets} we summarise the main properties of the targets. The ALMA data used in this work and their reduction is described in Section~\ref{sec:obs}. We describe the basic CO data cube analysis in Section~\ref{sec:cube_analysis}. The molecular gas dynamical modelling is described in Section~\ref{sec:method}. We discuss the results in Section~\ref{sec:discussion}, before summarising and concluding in Section~\ref{sec:conclusions}. In Appendices~\ref{sec:mom_maps_appendix} and \ref{sec:corner_plots_appendix} we provide a comparison between the full set of data and model CO products and the corner plots of the best-fitting models of all the three targets, respectively. In Appendix \ref{sec:MGE_solutions_appendix} we report the solutions adopted for the parametrisation of the stellar light profiles in NGC\,1574 and NGC\,4261.

Throughout this work we assume a $\Lambda$CDM cosmology with H$_{\rm 0}=70$\,km\,s$^{-1}$\,Mpc$^{\rm -1}$, $\Omega_{\rm \Lambda}=0.7$ and $\Omega_{\rm M}=0.3$. All of the velocities in this paper are given in the optical convention and Kinematic Local Standard of Rest (LSRK) frame.

\begin{figure*}
\centering
\includegraphics[scale=0.55]{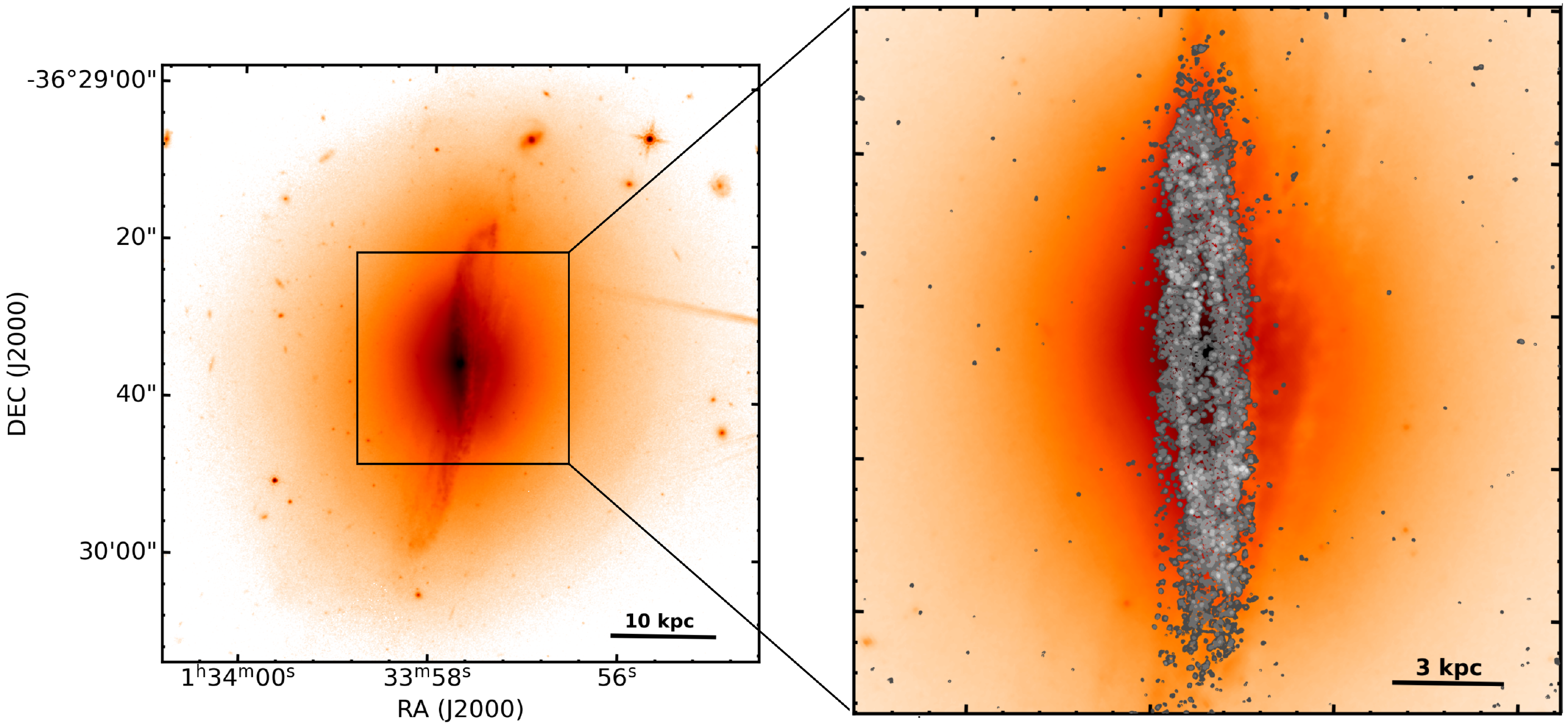}
\caption[]{{\bf Left panel:} {\it Hubble Space Telescope} (HST) image of NGC\,0612 retrieved from the Hubble Legacy Archive (HLA) at the Canadian Astronomy Data Centre (CADC). The image is taken with the Wide-Field Camera 3 (WFC3) in the F160W filter. {\bf Right panel:} Zoom of the map on the left in the central $29\arcsec \times 29\arcsec$ ($\approx 17 \times 17$~kpc$^{2}$) made using a different colour stretching. CO(2-1) integrated intensity (moment 0) contours from the ALMA observations analysed in this paper (see Section~\ref{sec:obs}) are overlaid in black-to-grey colours. The contours are linearly spaced from 0.15 to 0.31~Jy~beam$^{-1}$~km~s$^{-1}$. A scale bar is shown in the bottom-right corner of each panel. East is to the left and North to the top. \label{fig:NGC612_optical}}
\end{figure*}

\section{Targets}\label{sec:targets}
\subsection{NGC 0612}
NGC\,0612 is optically classified as an S0a galaxy viewed close to edge-on, characterised by an extensive and strongly warped dust distribution in the equatorial plane and by a massive, regularly-rotating stellar disc (\citealp{West66,Fasano96,Holt07}; Figure~\ref{fig:NGC612_optical}, left panel). Such disc has a star formation rate of $\approx7$~M$_{\odot}$~yr$^{-1}$ \citep{Asabere16}, suggesting that the galaxy may be a misclassified late-type galaxy, rather than an ETG. 
NGC\,0612 resides in a poor group of seven galaxies \citep[][]{Ramella02}, and forms a pair with the barred spiral galaxy NGC\,0619 \citep[][]{Emonts08}. A redshift-independent distance for this galaxy is currently not available. Assuming a redshift $z = 0.0298$ (as in \citealp[][]{Ruffa19a}) and the cosmology declared in Section~\ref{sec:intro}, we derive an Hubble flow luminosity distance of 130.4~Mpc and an angular size distance of 123~Mpc (being this latter the distance parameter used to compute the scale for the dynamical models described in Section~\ref{sec:method}). The angular scale for this target is thus 630~pc~arcsec$^{-1}$.
Some general galaxy properties are provided in Table~\ref{tab:general}.

NGC\,0612 hosts an enormous disc of atomic hydrogen (HI), extending $\approx$140~kpc along its major axis and with an estimated mass of $1.8\times10^{9}$~M$_{\odot}$ \citep{Emonts08}. A faint HI bridge connects NGC\,0612 with its companion, suggesting an interaction between the two galaxies.

NGC\,0612 is the host galaxy of the radio source PKS~0131-36, characterised by a peculiar core-double lobe morphology (oriented almost perpendicularly to the dust disc visible in Figure~\ref{fig:NGC612_optical}; see \citealp[][]{Ruffa19a}), and a total radio power of $P$\textsubscript{1.4~GHz}$=1.5\times10^{25}$~W~Hz$^{-1}$. The combination of these characteristics make PKS~0131-36 the prototypical hybrid Fanaroff-Riley type I/II (FR I/II) radio galaxy \citep{Fanaroff74,Kris00}.

The AGN in NGC\,0612 is found to be optically faint \citep[][]{Holt07}. This, along with the weak, LINER-like forbidden emission lines observed in its optical spectra \citep[][]{Tadhunter93,Holt07}, place NGC\,0612 into the typical regime of low-excitation radio galaxy (LERG; \citealp[e.g.][]{Best12}). However, the high dust extinction of its nucleus leads to a substantial obscuration even in the mid-infrared and X-ray regimes \citep[e.g.][]{Ueda15}, making its optical classification still controversial \citep[see also][]{Parisi09}. 

NGC\,0612 does not have a direct SMBH mass measurement yet. A rough estimate can be obtained assuming an effective stellar velocity dispersion of $\sigma_{\rm \filledstar} = 290 \pm 31$~km~s$^{-1}$, which has been recently measured through integral-field spectroscopy (Warren et al.\,in prep.\,). Adopting the $M_{\rm BH}-\sigma_{\rm \filledstar}$ relation of \citet[][]{Vandenbosch16}, a SMBH with a mass of $\approx1.5 \times 10^{9}$~M$_{\odot}$ is expected to reside at the centre of NGC\,0612. This corresponds to a $R_{\rm SOI}$ of $\approx$77.4~pc ($\approx$0.13$\arcsec$), which is about 1.3 times larger than the synthesized beam size of the observations analysed in this paper (see Section~\ref{sec:obs}).

We note that the intermediate-resolution CO(2-1) ALMA 12-m observations of NGC\,0612 used in this work (see Section~\ref{sec:obs}) were already published by \citet[][]{Ruffa19a}, and an accurate study of the CO kinematics based on such observations presented in \citet[][]{Ruffa19b}. We compare our results to these two previous works in Section~\ref{sec:mom_maps} and \ref{sec:mod_results}, respectively.

\begin{figure*}
\centering
\includegraphics[scale=0.55]{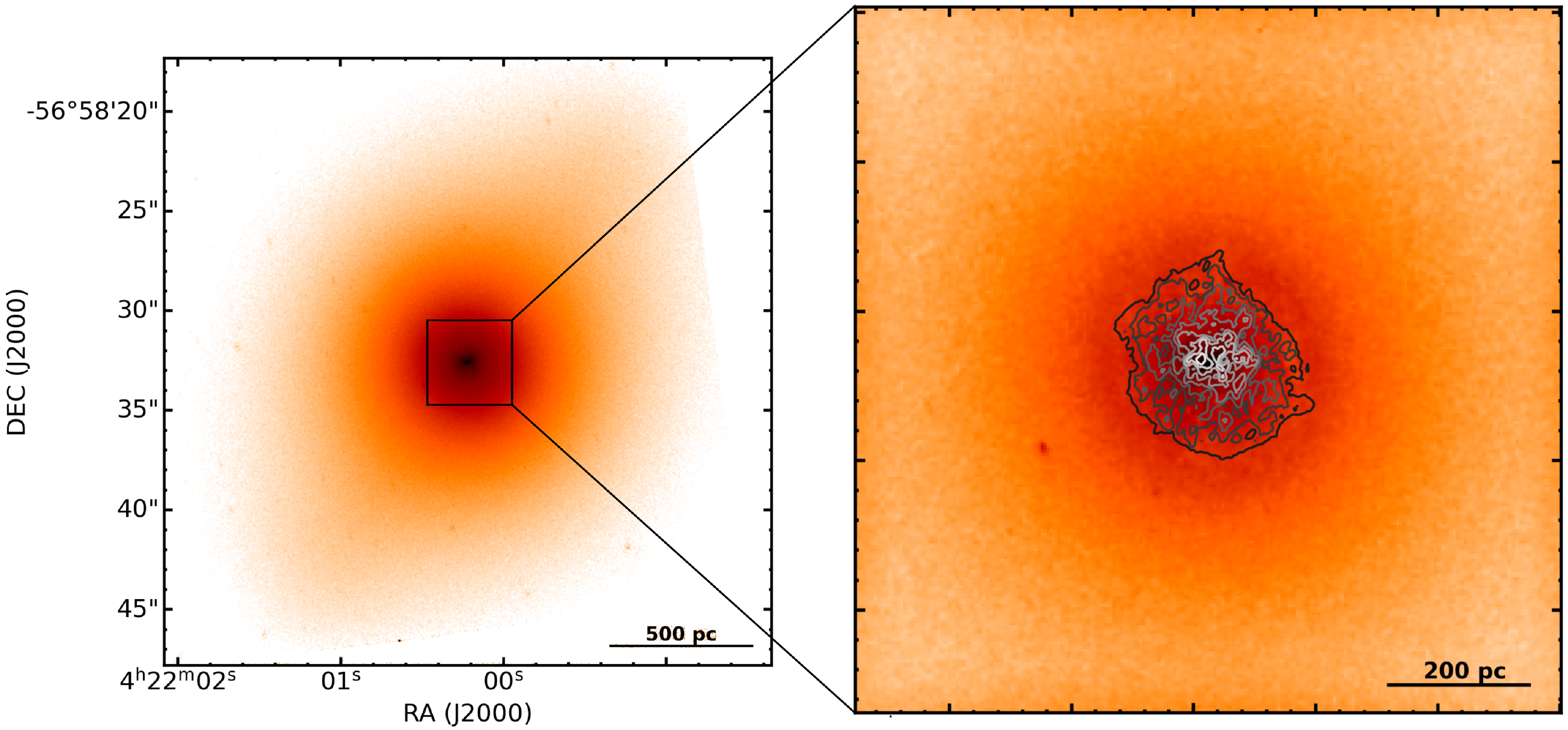}
\caption[]{As in Figure~\ref{fig:NGC612_optical}, but for NGC\,1574. The HST image is taken with the Wide-Field and Planetary Camera 2 (WFPC2) in the F606W filter. The zoom-in on the right shows the central $9\arcsec \times 9\arcsec$ ($\approx 600 \times 600$~pc$^{2}$). The CO(2-1) contours are linearly spaced from 0.17 to 3.10~Jy~beam$^{-1}$~km~s$^{-1}$. \label{fig:NGC1574_optical}}
\end{figure*}
 
\subsection{NGC 1574}
NGC\,1574 is a nearly face-on S0 galaxy with a small-scale bar in its nuclear regions (\citealp[][]{Phillips96,Lau06,Lau11,Gao19}; Figure~\ref{fig:NGC1574_optical}, left panel). The galaxy is one of the 46 members of the Dorado group \citep[][]{Maia89}. A surface brightness fluctuation (SBF) distance modulus for NGC\,1574 has been reported by \citet[][]{Tonry01}, and corresponds to a redshift-independent luminosity distance of 19.9~Mpc. Assuming a redshift $z=0.0035$, an angular size distance of 19.3~Mpc is derived. The angular scale adopted here is thus 93~pc~arcsec$^{-1}$. Some general galaxy properties are summarised in Table~\ref{tab:general}.

NGC\,1574 is one of the few Dorado group members not detected in HI \citep[][]{Kilborn05}. This galaxy also does not show any sign of nuclear activity that may point towards the presence of an AGN. The radio power at 5~GHz is estimated to be $< 5.6 \times 10^{20}$~W~Hz$^{-1}$ \citep[][]{Fabbiano89}.

NGC\,1574 does not have a direct SMBH mass measurement yet. Assuming an effective stellar velocity dispersion $\sigma_{\rm \filledstar} = 216 \pm 16$~km~s$^{-1}$ \citep[][]{Bernardi02} and adopting the $M_{\rm BH}-\sigma_{\rm \filledstar}$ relation of \citet[][]{Vandenbosch16}, a SMBH with a mass of $\approx2\times 10^{8}$~M$_{\odot}$ is expected to reside at the centre of NGC\,1574. This gives $R_{\rm SOI}\approx$12~pc (or $\approx$0.15$\arcsec$), roughly corresponding to the synthesized beam size of the observations analysed in this paper (see Section~\ref{sec:obs}).

\begin{figure*}
\centering
\includegraphics[scale=0.55]{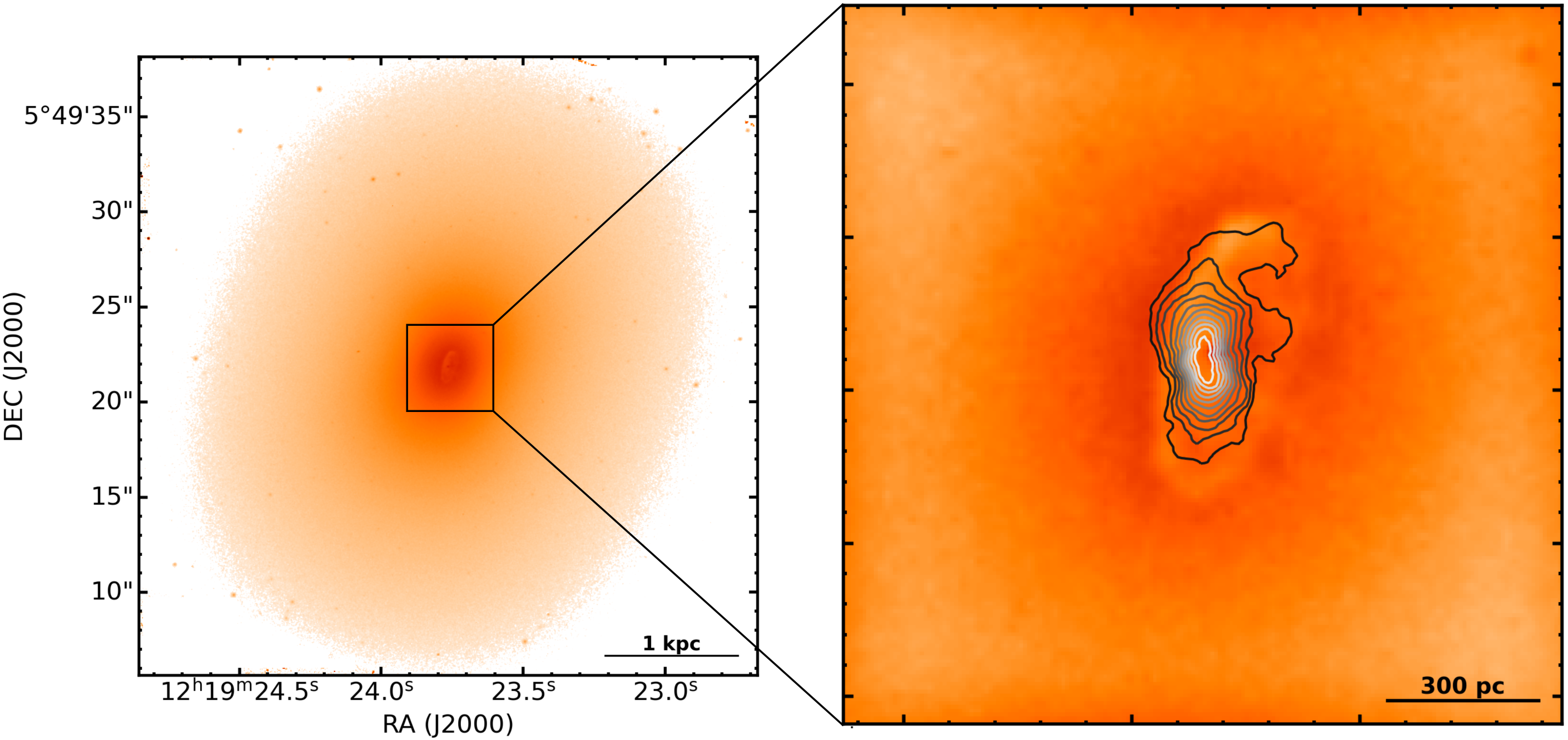}
\caption[]{As in Figure~\ref{fig:NGC612_optical}, but for NGC\,4261. The HST image is taken in the WFPC2 F791W filter. The zoom-in on the right shows the central $4\arcsec \times 4\arcsec$ ($\approx 600 \times 600$~pc$^{2}$). The CO(2-1) contours are linearly spaced from 0.15 to 15.2~Jy~beam$^{-1}$~km~s$^{-1}$. \label{fig:NGC4261_optical}}
\end{figure*}

\begin{table*}
\centering
\caption{General properties of the galaxies analysed in this paper.}\label{tab:general}
\begin{tabular}{l l c c c c c c c}
\hline
\multicolumn{1}{c}{ Galaxy } &
\multicolumn{1}{c}{ RA } &
\multicolumn{1}{c}{ DEC } &
\multicolumn{1}{c}{ T-type } & 
\multicolumn{1}{c}{ $z$ } &
\multicolumn{1}{c}{ D$_{\rm L}$ } &
\multicolumn{1}{c}{  Radio} &
\multicolumn{1}{c}{  $\log$P$_{\rm 1.4GHz}$} \\
\multicolumn{1}{c}{name} &
\multicolumn{1}{c}{ (J2000) }  &
\multicolumn{1}{c}{ (J2000) }  &
\multicolumn{1}{c}{ }  &
\multicolumn{1}{c}{ } &
\multicolumn{1}{c}{ }  &
\multicolumn{1}{c}{ source}   &
\multicolumn{1}{c}{ }   \\
\multicolumn{1}{c}{ } &
\multicolumn{1}{c}{ } &
\multicolumn{1}{c}{ } &
\multicolumn{1}{c}{  } &
\multicolumn{1}{c}{ } &
\multicolumn{1}{c}{ (Mpc)} &
\multicolumn{1}{c}{ }  &
\multicolumn{1}{c}{  (W~Hz$^{-1}$)}   \\
\multicolumn{1}{c}{(1)} &
\multicolumn{1}{c}{ (2)} &
\multicolumn{1}{c}{ (3)} &
\multicolumn{1}{c}{ (4)} &
\multicolumn{1}{c}{ (5) } &
\multicolumn{1}{c}{ (6) } &
\multicolumn{1}{c}{ (7)} &
\multicolumn{1}{c}{ (8)} \\
\hline
NGC\,0612 & 01$^{\rm h}$33$^{\rm m}$57$^{\rm s}$.74 & -36$^{\circ}$29$^{'}$35$\arcsec$.7 & $-1.2\pm0.6$ & 0.0298 & 130.4 &  PKS~0131-36 & 25.1 \\
NGC\,1574 & 04$^{\rm h}$21$^{\rm m}$58$^{\rm s}$.82 & -56$^{\circ}$58$^{'}$29$\arcsec$.1 & $-2.9 \pm 0.5$ & 0.0035 & 19.9 & -- & -- \\
NGC\,4261 & 12$^{\rm h}$19$^{\rm m}$23$^{\rm s}$.22 & +05$^{\circ}$49$^{'}$30$\arcsec$.77 & $-4.8 \pm 0.4$ & 0.0075 & 31.6 &  3C\,270 & 24.3 \\
\hline
\end{tabular}
\parbox[t]{1\textwidth}{ \textit{Notes.} Columns: (1) Primary galaxy identification name. (2)-(3) Galaxy Right Ascension and Declination in the J2000 reference frame. (4) Galaxy morphological classification as reported in the HyperLeda database (\url{http://leda.univ-lyon1.fr}), which defines as early-type galaxies (i.e.\,E and S0) those having $T<-0.5$. Specifically, according to this scheme, a galaxies is E if ${\rm T}\leq-3.5$ and S0 if $-3.5\leq{\rm T}\leq-0.5$. (5) Best estimate of the galaxy redshift taken from the NASA/IPAC extragalactic database (NED; \url{http://ned.ipac.caltech.edu}). For NGC\,0612 and NGC\,4261, these corresponds to the ones adopted by \citet{Ruffa19a} and \citet[][]{Boizelle21}, respectively. (6) Best estimate of the galaxy luminosity distance. From this and the $z$ value listed in Column (5), an estimate of the angular size distances (reported in Section~\ref{sec:targets}) is obtained using standard relations. (7) Name of the radio source. (8) Radio power at 1.4~GHz derived from the best flux density estimate including all the radio emission associated with the source.}
\end{table*}

\subsection{NGC 4261}
NGC\,4261 is a well-known elliptical galaxy hosting a famous nuclear dust disc (among the first ones detected in an ETG; \citealp[][]{Jaffe93}; see Figure~\ref{fig:NGC4261_optical}). The galaxy is classified as a slow rotator \citep[][]{Emsellem11}, with prolate stellar rotation \citep[][]{Krajnovic11}. 

It is the brightest galaxy in a group of 33 objects located in the direction of the Virgo West cloud \citep[][]{Nolthenius93,Davis95}, and forms a pair with the barred lenticular galaxy NGC\,4264 \citep[e.g.][]{Martel00}. Some general galaxy properties are provided in Table~\ref{tab:general}.

NGC\,4261 is the host of the FRI radio source 3C\,270, characterised by a prominent two-sided jet emanating from a bright core, and symmetric lobes on larger scales \citep[e.g.][]{Morganti93,Worrall10}. Such radio structure is oriented almost perpendicularly to the dust disc visible in Figure~\ref{fig:NGC4261_optical}. LINER-like forbidden emission lines are observed in high-quality optical spectra of NGC\,4261 \citep[e.g.][]{Ho97b}, such that it is classified as a LERG \citep[][]{Ogle10}.

Two early studies based on the kinematics of optical emission lines detected the presence of a central massive dark object in NGC\,4261 with an estimated mass of $\approx5 \times 10^{8}$~M$_{\odot}$  \citep[][]{Ferrarese96,Humphrey09}. More recently, \citet[][]{Boizelle21} reported $M_{\rm BH}=1.67 \times 10^{9}$~M$_{\odot}$ for NGC\,4261, based on a dynamical modelling using CO(2-1) ALMA 12-m observations at a resolution of $\approx0.31\arcsec$. CO(2-1) in NGC\,4261, however, has also been observed with ALMA as part of the WISDOM project, using both a more extended 12-m array configuration and the 7-m Atacama Compact Array (ACA). Our purpose here is thus to cross-check the CO dynamical $M_{\rm BH}$ measurements reported by \citet[][]{Boizelle21} using the combined higher- and lower-resolution 12-m plus the ACA CO(2-1) observations of NGC\,4261 and a different modelling technique (see Section~\ref{sec:method} for details). To make a coherent comparison with the Boizelle et al.\,work, we adopt the same luminosity distance of 31.6~Mpc (derived from a SBF distance modulus measurement reported by \citealp[][]{Tonry01}) and redshift of $z=0.00746$. The angular size distance for this target is thus 31.1~Mpc, and the derived angular scale is 151~pc~arcsec$^{-1}$. Assuming an effective stellar velocity dispersion $\sigma_{\rm \filledstar} = 263 \pm 12$~km~s$^{-1}$ \citep[][]{Vandenbosch16} and the black hole mass reported by \citet[][]{Boizelle21}, we estimate that for NGC\,4261 $R_{\rm SOI}\approx100$~pc (or $0.66\arcsec$), which is about 2.5 times larger than the minimum spatial scales that can be probed with the data used in this work (see Section~\ref{sec:obs}).

\begin{table*}
\centering
\caption{Main properties of the ALMA observations used in this work.}
\label{tab:ALMA observations summary}
\begin{tabular}{l c c c c c c c c}
\hline
\multicolumn{1}{c}{ Target } &
\multicolumn{1}{c}{ Project } & 
\multicolumn{1}{c}{ Observation } & 
\multicolumn{1}{c}{ Array } &
\multicolumn{1}{c}{ Baseline } &
\multicolumn{1}{c}{ Time } & 
\multicolumn{1}{c}{   MRS } & 
\multicolumn{1}{c}{   FOV} \\
\multicolumn{1}{c}{  } &
\multicolumn{1}{c}{ code } &
\multicolumn{1}{c}{  date} &
\multicolumn{1}{c}{  } &
\multicolumn{1}{c}{ range } &
\multicolumn{1}{c}{ on-source } &
\multicolumn{1}{c}{  } &
\multicolumn{1}{c}{  } \\
\multicolumn{1}{c}{  } & 
\multicolumn{1}{c}{  } &
\multicolumn{1}{c}{  } &  
\multicolumn{1}{c}{  } &
\multicolumn{1}{c}{  } &
\multicolumn{1}{c}{   (min)} &  
\multicolumn{1}{c}{  (arcsec) (kpc)} & 
\multicolumn{1}{c}{ (arcsec) (kpc) } \\
\multicolumn{1}{c}{   (1) } &   
\multicolumn{1}{c}{   (2) } &
\multicolumn{1}{c}{   (3) } &
\multicolumn{1}{c}{   (4) } &
\multicolumn{1}{c}{   (5) } &
\multicolumn{1}{c}{   (6) } &
\multicolumn{1}{c}{   (7) } & 
\multicolumn{1}{c}{   (8) } \\
\hline
NGC\,0612  &   \begin{tabular}[c]{@{}c@{}} 2015.1.01572.S \\ 2016.2.00046.S \\ 2017.1.00904.S \end{tabular} & \begin{tabular}[c]{@{}c@{}} 30-07-2016 \\ 30-07-2017 \\ 19-11-2017 \end{tabular}   & \begin{tabular}[c]{@{}c@{}} 12-m \\ 7-m \\ 12-m \end{tabular}   &   \begin{tabular}[c]{@{}c@{}} 15~m - 1.2~km \\ 9~m - 43~m \\ 92~m - 8.5~km \end{tabular} &    \begin{tabular}[c]{@{}c@{}} 3 \\ 17 \\ 7 \end{tabular}  &  \begin{tabular}[c]{@{}c@{}} 3.6 (2.1) \\ 29.9 (17.9) \\ 0.8 (0.5) \end{tabular} & \begin{tabular}[c]{@{}c@{}} 25.3 (15.1) \\ 42.9 (25.6) \\ 25.1 (15.0) \end{tabular} \\ 
\hline
NGC\,1574  &  \begin{tabular}[c]{@{}c@{}} 2015.1.00419.S \\ 2015.1.00419.S \\ 2016.2.00053.S \end{tabular} &  \begin{tabular}[c]{@{}c@{}}  12-06-2016 \\ 27-09-2016 \\ 24-07-2017 \end{tabular}    &   \begin{tabular}[c]{@{}c@{}} 12-m \\ 12-m \\ 7-m \end{tabular}   &  \begin{tabular}[c]{@{}c@{}} 15~m - 0.7~km \\ 80~m - 3.2~km \\ 9~m - 49~m \end{tabular}  &    \begin{tabular}[c]{@{}c@{}} 6 \\ 13 \\ 39 \end{tabular}  &     \begin{tabular}[c]{@{}c@{}} 5.5 (0.4) \\ 3.3 (0.2) \\ 30.2 (2.1) \end{tabular} & \begin{tabular}[c]{@{}c@{}} 24.4 (1.7) \\ 24.4 (1.7) \\ 41.8 (2.9) \end{tabular} \\ 
\hline
NGC\,4261  & \begin{tabular}[c]{@{}c@{}} 2016.2.00046.S \\ 2017.1.00301.S \\ 2018.1.00397.S \end{tabular} & \begin{tabular}[c]{@{}c@{}}    20-08-2017 \\ 19-01-2018 \\ 16-08-2019 \end{tabular}   &   \begin{tabular}[c]{@{}c@{}} 7-m \\ 12-m \\ 12-m \end{tabular} & \begin{tabular}[c]{@{}c@{}} 9~m - 49~m \\ 15~m - 1.4~km \\ 41~m - 3.6~km \end{tabular}  &  \begin{tabular}[c]{@{}c@{}} 19 \\ 31 \\ 5 \end{tabular}   &     \begin{tabular}[c]{@{}c@{}} 28.9 (4.3) \\ 4.1 (0.6) \\ 1.9 (0.3) \end{tabular} &  \begin{tabular}[c]{@{}c@{}} 41.9 (6.3) \\ 24.7 (3.7) \\ 24.5 (3.6) \end{tabular}   \\  
\hline
\end{tabular}
\parbox[t]{1\textwidth}{ \textit{Notes.} $-$ Columns: (1) Target name. (2) ALMA project code. (3) Observation dates. (4) Type of ALMA antennas used for the observation. (5) Minimum and maximum baseline lengths. (6) Total integration time on-source. (7) Maximum recoverable scale (MRS), i.e.\,largest angular scale structure that can be recovered with the array configuration in arcseconds, and corresponding scale in kiloparsec within parentheses. (7) Field of view (FOV), i.e.\,primary beam FWHM in arcseconds, and corresponding scale in kiloparsec within parentheses.}
\end{table*}

\section{ALMA observations and data reduction}\label{sec:obs}
The $^{12}$CO(2-1) ALMA observations analysed in this paper were obtained during Cycles 3–6 (2015–2018) as part of a variety of programmes, acquired in the context of both WISDOM and other projects. To ensure adequate uv-plane coverage, for each target we used ALMA data in two different configurations of the 12-m array plus a complementary ACA observation. Details on the main properties of each observing track, including number of antennas, maximum baseline lengths, maximum recoverable spatial scale (MRS) and field of view (FOV), are reported in Table~\ref{tab:ALMA observations summary}.
 
For all of the observations, ALMA Band 6 with the same spectral configuration was used: a total of four spectral windows (SPWs), one centred at the redshifted frequency ($\nu_{\rm sky}$) of the \textsuperscript{12}CO(2-1) line (rest frequency, $\nu_{\rm rest}=230.5380$~GHz) and divided into a large number (between 1920 and 4096) of small channels (between 0.976 and 7.812~MHz-wide); the other three used to map the continuum and divided into 128 31.25-MHz-wide channels. A standard calibration strategy was adopted: for each session, a single bright quasar was used as both flux and bandpass calibrator, a second one as phase calibrator.

We calibrated all data using the Common Astronomy Software Application \citep[{\sc casa};][]{McMullin07} package, version 5.4.1, reducing each dataset separately with standard data reduction scripts. Data from different array configurations were then combined into a single calibrated dataset, using the {\sc CASA} task {\tt concat}. The analysis described in the following has been carried out using the combined data only.

\subsection{Line imaging}
Point-like continuum emission was detected in all the sources. The analysis of such emission, however, is beyond the purpose of this work and will be not further discussed here (see Elford et al., subm.\,).

Line emission was isolated in the visibility plane using the {\sc CASA} task {\tt uvcontsub} to form a continuum model from linear fits in frequency to line-free channels and to subtract it from the visibilities. In this procedure, both the line-free channels of the line SPW and the closest (in frequency) continuum SPW were used to estimate the continuum level. We then produced the CO data cubes using the {\tt tclean} task and adopting Briggs weighting with a robust parameter of 0.5, which gave a good trade-off between angular resolution and signal-to-noise ratio (S/N). The channel velocities were computed in the source frame with zero-points corresponding to the redshifted frequency of the CO(2-1) line ($\nu$\textsubscript{sky}, Table~\ref{tab:line_images}). The continuum-subtracted dirty cubes were cleaned in regions of line emission (identified interactively) to a threshold equal to 1.5 times the rms noise level, determined in line-free channels. Several channel widths (i.e.\ spectral bins) were tested to find a good compromise between S/N and resolution of the line profiles; the final channel widths range from 10 to 20~km~s$^{-1}$. We note that for NGC\,4261 the final channel width of 15~km~s$^{-1}$ is about 2.6 times smaller than that adopted by \citet[][]{Boizelle21}, allowing us to study more in detail the kinematics of the molecular gas. Table~\ref{tab:line_images} summarises the main properties of the cleaned CO data cubes, which are characterised by rms noise levels (determined in line-free channels) between 0.5 and 0.7~mJy~beam$^{-1}$ for synthesized beams of 0.097$\arcsec$ - 0.265$\arcsec$ major axis full width at half-maximum (FWHM). 

\begin{table*}
\centering
\caption{Properties of the cleaned CO data cubes.}
\label{tab:line_images}
\begin{tabular}{l c c c c c c c c c c}
\hline
\multicolumn{1}{c}{ Target } &
\multicolumn{1}{c}{ $\nu$\textsubscript{sky}  (v\textsubscript{cen})} &
\multicolumn{1}{c}{   rms } & 
\multicolumn{1}{c}{   S/N } & 
\multicolumn{1}{c}{   $\Delta v_{\rm chan}$ } & 
\multicolumn{1}{c}{   $\theta$\textsubscript{maj} } & 
\multicolumn{1}{c}{   $\theta$\textsubscript{min}  } & 
\multicolumn{1}{c}{   PA} & 
\multicolumn{1}{c}{   Scale }\\        
\multicolumn{1}{c}{  } & 
\multicolumn{1}{c}{   (GHz) (km s$^{-1}$) } &          
\multicolumn{1}{c}{   (mJy beam$^{-1}$)} &
\multicolumn{1}{c}{  } &
\multicolumn{1}{c}{ (km s$^{-1}$) } &
\multicolumn{2}{c}{ (arcsec) } &   
\multicolumn{1}{c}{   (deg) } &
\multicolumn{1}{c}{   (pc)} \\       
\multicolumn{1}{c}{   (1) } &   
\multicolumn{1}{c}{   (2) } &
\multicolumn{1}{c}{   (3) } &
\multicolumn{1}{c}{   (4) } &
\multicolumn{1}{c}{   (5) } &
\multicolumn{1}{c}{   (6) } &
\multicolumn{1}{c}{   (7) } &
\multicolumn{1}{c}{   (8) } &
\multicolumn{1}{c}{   (9) } \\
\hline
NGC\,0612  &    223.8426  (8974)  & 0.7 &  10 & 20 & 0.097 & 0.087 &   69.964 &  58 \\ 
NGC\,1574  &  229.7339 (1050)   &  0.7  &  11  &  10 & 0.202    &    0.138    &    -17.010     &    14    \\ 
NGC\,4261  &  228.8218  (2250)  &   0.4  &  9 & 15  & 0.265    &     0.232    &    -70.267    &    40   \\  
\hline
\end{tabular}
\parbox[t]{1\textwidth}{ \textit{Notes.} $-$ Columns: (1) Target name. (2) CO(2-1) redshifted (sky) centre frequency estimated using the redshift listed in Table~\ref{tab:general}; the corresponding velocity (v\textsubscript{cen}; LSRK system, optical convention) is reported within parentheses. (3) 1$\sigma$ rms noise level measured in line-free channels with the widths listed in column (5). (4) Peak signal-to-noise ratio of the CO detection. (5) Final channel width of the data cube. (6)-(8) Major and minor axes FWHM, and position angle of the achieved synthesized beam. (9) Spatial scale corresponding to the major axis FWHM of the synthesized beam.}
\end{table*}

\section{Image cube analysis}\label{sec:cube_analysis}
\subsection{CO data products}\label{sec:mom_maps}
Integrated intensity (moment 0), mean line-of-sight velocity (moment 1), and observed velocity dispersion (moment 2) maps of the three CO detections were created from the cleaned continuum-subtracted data cubes using the masked moment technique as described by \citet[][see also \citealt{Bosma81a,Bosma81b,Kruit82,Rupen99}]{Dame11}. In this technique, a copy of the cleaned data cube is first Gaussian-smoothed spatially (with a FWHM equal to 1.5 times that of the synthesised beam) and then Hanning-smoothed in velocity. A three-dimensional mask is then defined by selecting all the pixels above a fixed flux-density threshold. This threshold is chosen so as to recover as much flux as possible while minimising the noise (higher thresholds usually need to be set for noisier maps; see also \citealt[][]{Ruffa19a}). Given the significance of our CO detections (see Table~\ref{tab:line_images}) and the need to recover even the faintest gas structures, a threshold of 0.9$\sigma$ (where $\sigma$ is the rms noise level measured in the un-smoothed data cube) has been used in all the cases. The moment maps were then produced from the original un-smoothed cubes using the masked regions only. Major-axis position-velocity diagrams (PVDs) were also constructed by summing pixels in the masked cube within a pseudo-slit whose long axis is orientated along the CO kinematic position angle (PA; see Section~\ref{sec:method}). The resulting CO moment maps and major axis PVDs are shown in Figures~\ref{fig:NGC612}-\ref{fig:NGC4261}. A brief description of the observed features for each target is provided in the following.

\subsubsection*{NGC\,0612}
\begin{figure*}
\centering
\begin{subfigure}[t]{0.3\textheight}
\centering
 \caption{Moment zero}\label{fig:ngc612_mom0}
\includegraphics[scale=0.5]{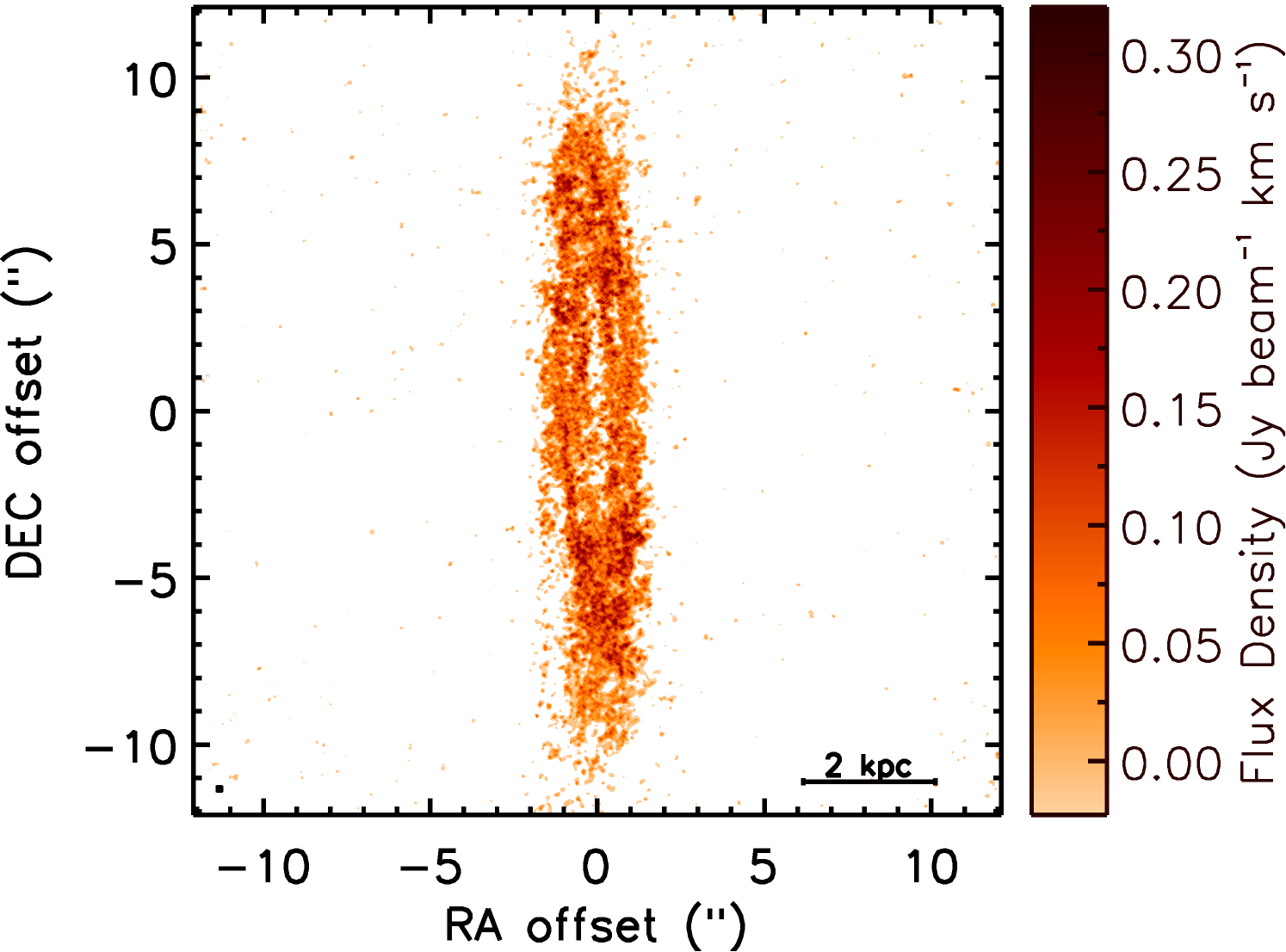}
\end{subfigure}
\hspace{2mm}
\begin{subfigure}[t]{0.3\textheight}
\centering
\caption{Moment one}\label{fig:ngc612_mom1}
\includegraphics[scale=0.5]{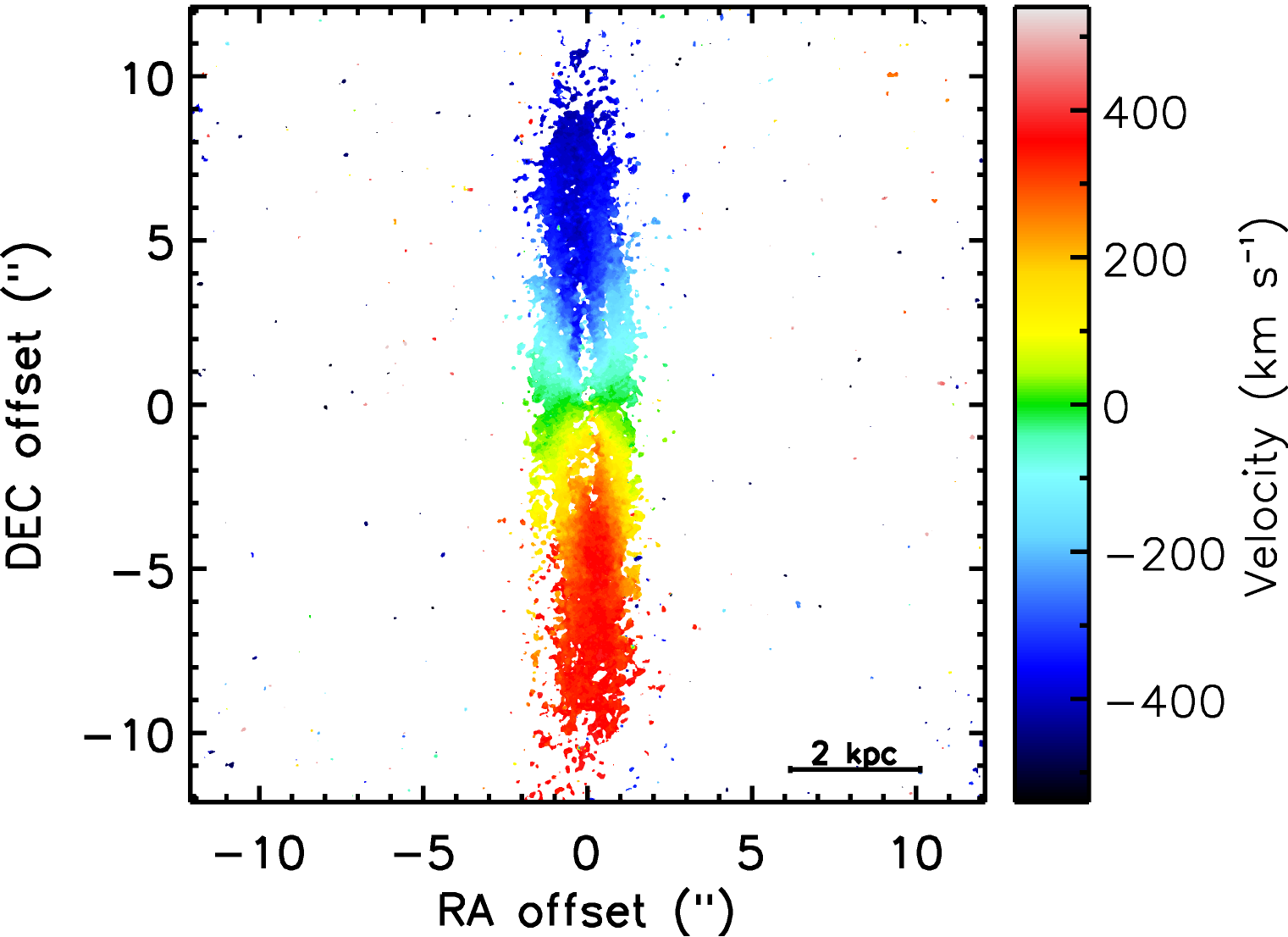}
\end{subfigure}

\medskip

\begin{subfigure}[t]{0.3\textheight}
\centering
\vspace{0pt}
\caption{Moment two}\label{fig:ngc612_mom2}
\includegraphics[scale=0.5]{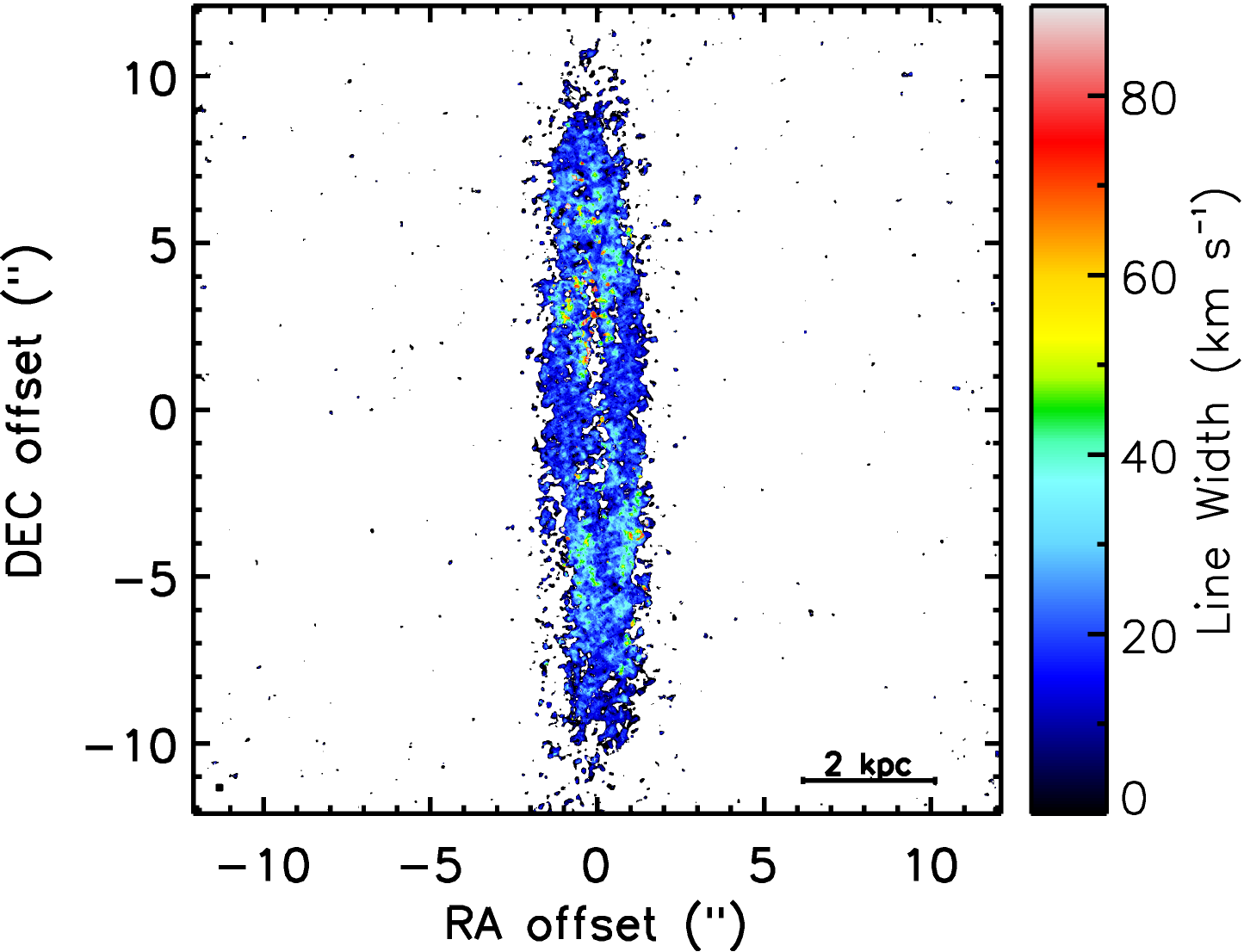}
\end{subfigure}
\begin{subfigure}[t]{0.3\textheight}
\centering
\vspace{0pt}
\caption{Major axis PVD}\label{fig:ngc612_PVD}
\includegraphics[scale=0.48]{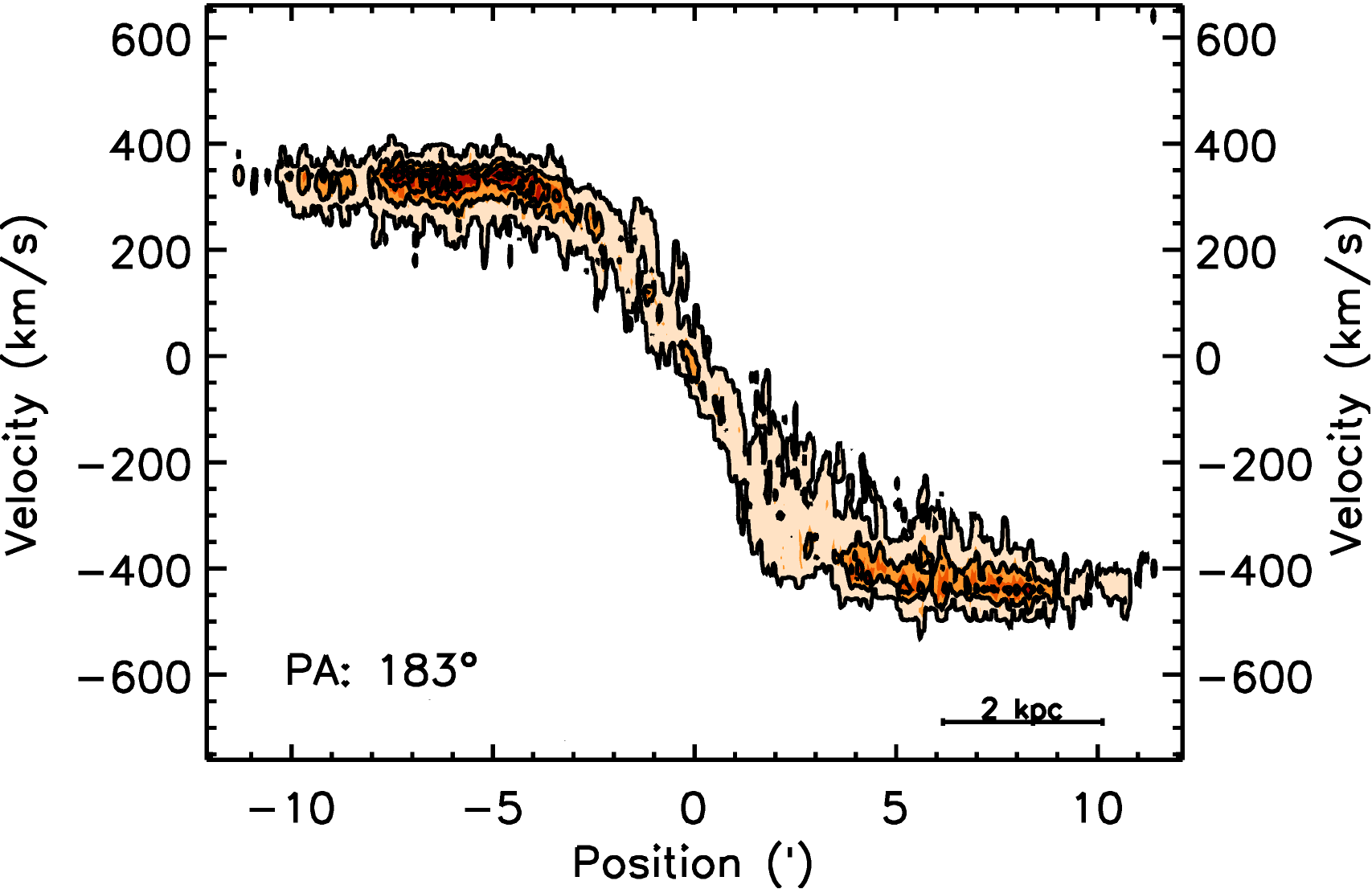}
\end{subfigure}
\caption{\textbf{NGC\,0612} moment 0 (integrated intensity; panel~\ref{fig:ngc612_mom0}), moment 1 (intensity-weighted mean line-of-sight velocity; panel~\ref{fig:ngc612_mom1}) and moment 2 (intensity weighted line-of-sight velocity dispersion; panel~\ref{fig:ngc612_mom2}) maps. These were created with the masked moment technique described in Section~\ref{sec:mom_maps} using a data cube with a channel width of 20~km~s$^{-1}$ (Table~\ref{tab:line_images}). The synthesised beam and a scale bar are shown in the bottom-left and bottom-right corners, respectively, of each map. The bar to the right of each panel shows the colour scales. Major axis PVD (panel~\ref{fig:ngc612_PVD}) extracted within a rectangular area whose long axis is orientated according to the position angle indicated in the bottom-left corner of the panel (i.e.\,along the CO major axis in this case). As usual, the PA is measured counterclockwise from North through East. A scale bar is shown in the bottom-right corner. Velocities are measured in the source frame. The left y-axis reports $v_{\rm obs} - v_{\rm sys}$ (where $v_{\rm obs}$ corresponds to the observed central frequency of the line), whereas the right y-axis shows $v_{\rm exp} - v_{\rm sys}$ (where $v_{\rm exp}$ corresponds to the expected, redshifted central frequency of the CO line reported in column (2) of Table~\ref{tab:line_images}). Coordinates are with respect to the image phase centre; East is to the left and North to the top. }\label{fig:NGC612}
\end{figure*}

Based only on the intermediate-resolution CO(2-1) ALMA 12-m observations (Table~\ref{tab:ALMA observations summary}), \citet[][]{Ruffa19a} reported the detection of a clumpy large-scale disc of molecular gas in NGC\,0612, extending up to $\approx10$~kpc along its major axis. The disc was found to be mostly regularly rotating, with signs of a mild warp on large scales. The line-of-sight gas velocity dispersion ($\sigma_{\rm gas}$) was observed to vary from $\approx10$ to $\approx100$~km~s$^{-1}$, but line widths $\gtrsim40$~km~s$^{-1}$ were found to be highly localised, while the bulk of the disc had $\sigma_{\rm gas}\lesssim30$~km~s$^{-1}$. Despite the fact that the spatial resolution of the combined 12-m plus ACA observations used in this work is about 3 times higher than those reported in \citet[][]{Ruffa19a}, the moment maps in Figures~\ref{fig:ngc612_mom0}, \ref{fig:ngc612_mom1}, and \ref{fig:ngc612_mom2} are overall in agreement with the features described above. In addition to these, however, a gap is clearly observed at the centre of the CO(2-1) distribution (Fig.\,~\ref{fig:ngc612_mom0}), suggesting a ring-like instead of a full-disc shape. Such central gas deficiency was unresolved in the lower-resolution CO data alone, but nevertheless tentatively inferred from their kinematic modelling. Accounting for the presence of a central hole was indeed required to best-fitting the observed CO surface brightness distribution \citep[][see also Section~\ref{sec:mod_results} and \ref{sec:discussion}]{Ruffa19b}. 

No evident Keplerian rise from material orbiting within the SMBH SOI can be identified in the major axis PVD of NGC\,0612 (Fig.\,~\ref{fig:ngc612_PVD}). This, on the other hand, shows a somewhat X-shaped structure (as visible from the velocity asymmetries at $\pm2\arcsec$), hinting at that typically observed in barred edge-on galaxies \citep[e.g.][]{Voort18}. This is discussed in detail in Section~\ref{sec:discussion}.  

\subsubsection*{NGC\,1574}
\begin{figure*}
\centering
\begin{subfigure}[t]{0.3\textheight}
\centering
 \caption{Moment zero}\label{fig:ngc1574_mom0_main}
\includegraphics[scale=0.5]{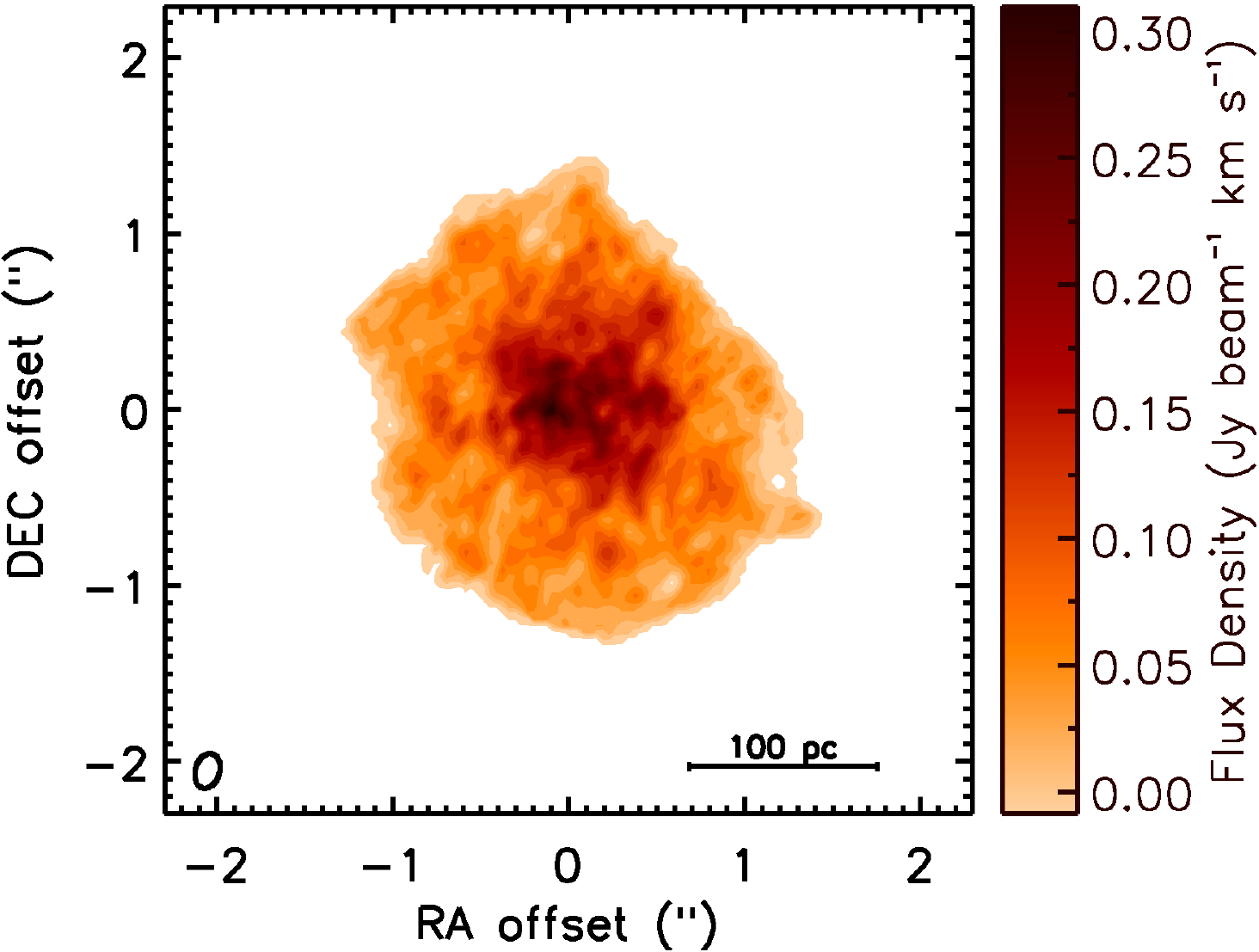}
\end{subfigure}
\hspace{2mm}
\begin{subfigure}[t]{0.3\textheight}
\centering
\caption{Moment one}\label{fig:ngc1574_mom1_main}
\includegraphics[scale=0.5]{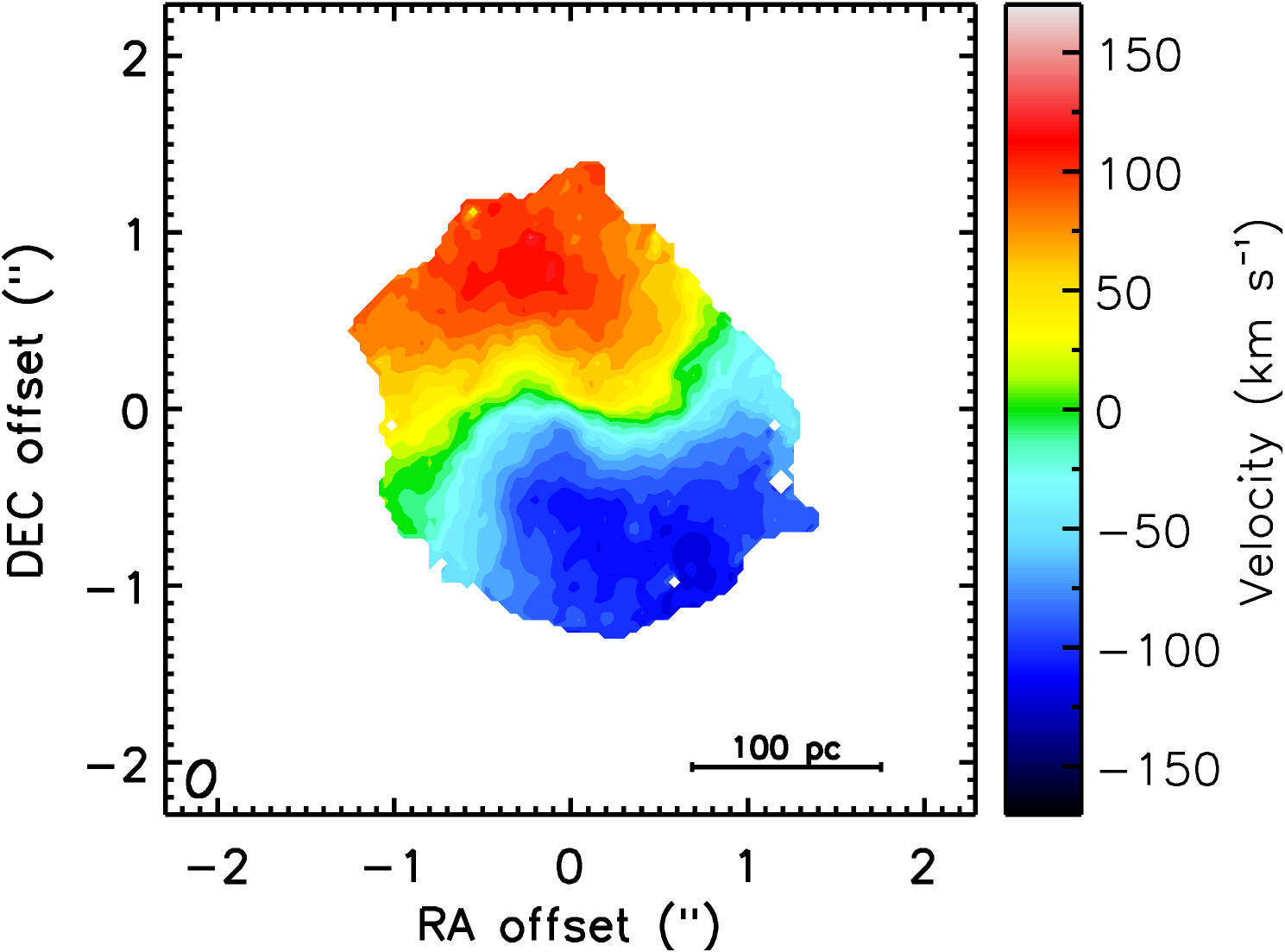}
\end{subfigure}

\medskip

\begin{subfigure}[t]{0.3\textheight}
\centering
\vspace{0pt}
\caption{Moment two}\label{fig:ngc1574_mom2_main}
\includegraphics[scale=0.5]{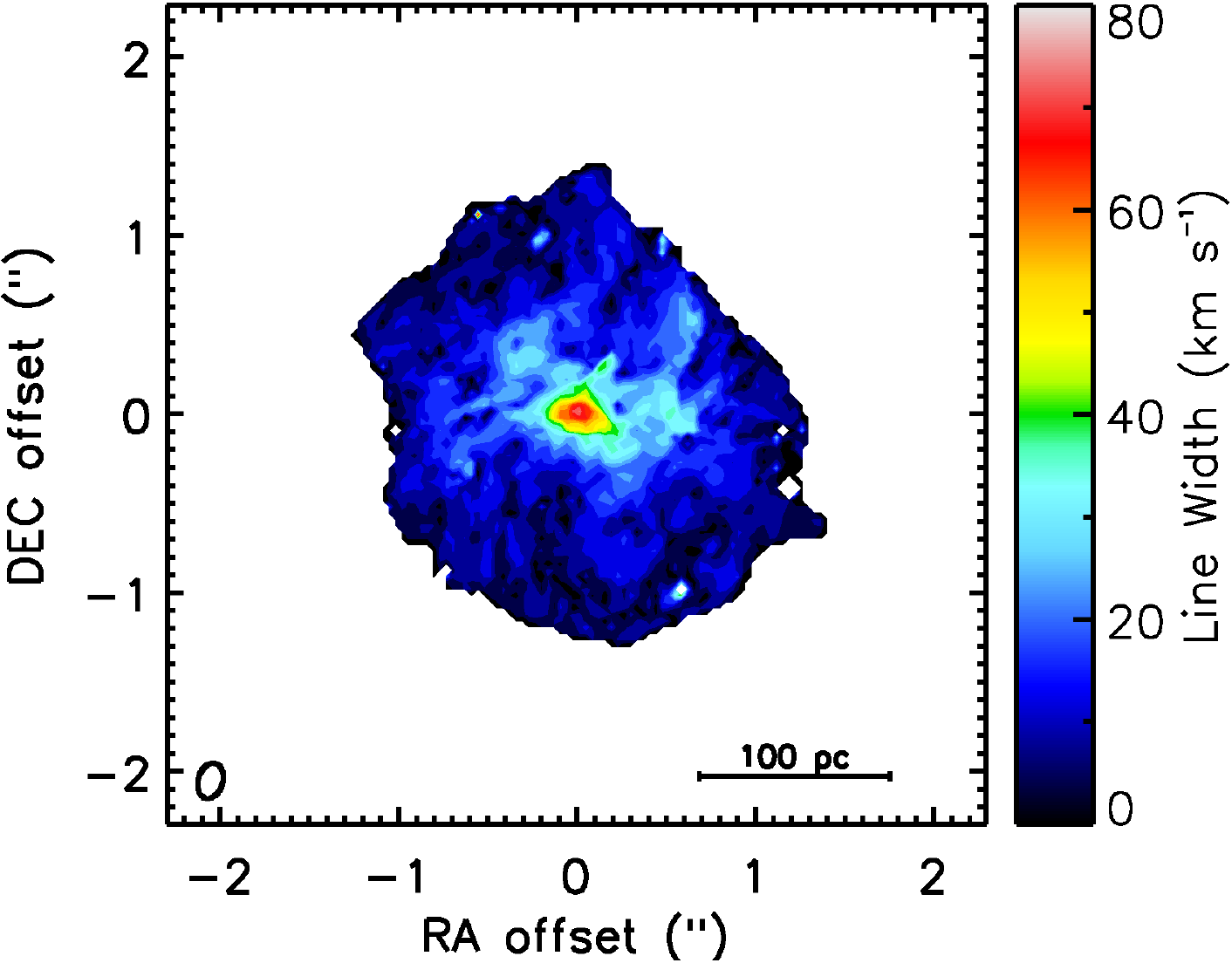}
\end{subfigure}
\begin{subfigure}[t]{0.3\textheight}
\centering
\vspace{0pt}
\caption{Major axis PVD}\label{fig:ngc1574_PVD_main}
\includegraphics[scale=0.48]{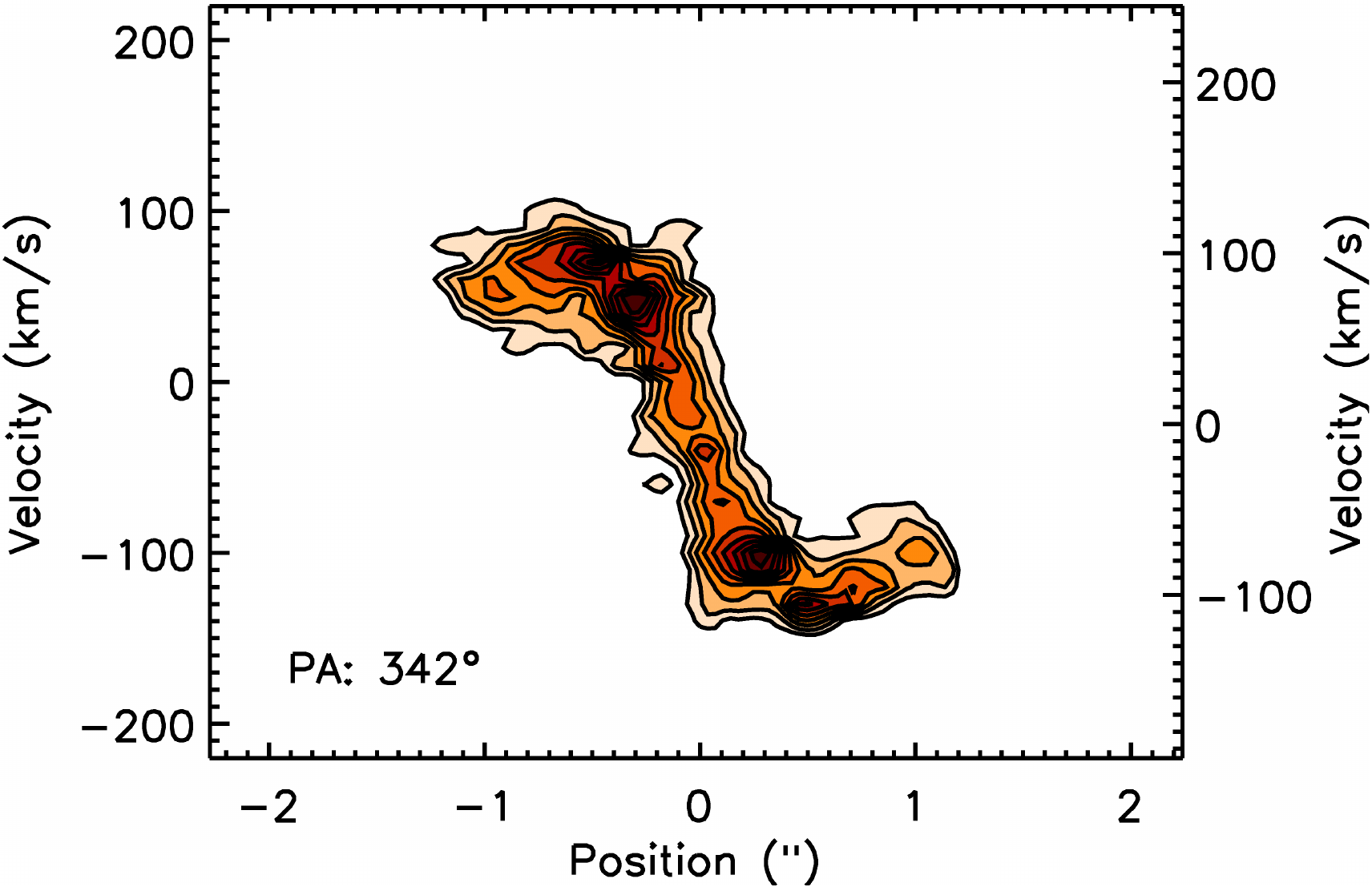}
\end{subfigure}
\caption{As in Figure~\ref{fig:NGC612}, but for NGC\,1574. We note that in this case the PVD slit has been orientated according to the PA value (indicated in the bottom-left corner of panel~\ref{fig:ngc1574_PVD_main}) encompassing the most central velocity structures.}\label{fig:NGC1574}
\end{figure*}

The integrated intensity map in Figure~\ref{fig:ngc1574_mom0_main} shows that the CO gas in NGC\,1574 is distributed in a disc, extending $\approx2.6\arcsec \times~2.2\arcsec$ ($\approx180 \times 150$~pc$^{2}$, in projection) around the optical nucleus of the galaxy (Figure~\ref{fig:NGC1574_optical}, right panel). The brightest CO features are concentrated within the central $\approx0.9\arcsec \times~0.9\arcsec$, with a peak at a RA offset of $\approx-0.3^{\prime \prime}$, surrounded by fainter emission outwards. 

The CO velocity pattern in Figure~\ref{fig:ngc1574_mom1_main} show that the gas is rotating, but with evident kinematic distortions (i.e.\,s-shaped iso-velocity contours) tracing the presence of unrelaxed substructures in the gas disc. These can be caused by either warps or non-circular motions, or a combination of both. Accurate 3D kinematic modelling can help differentiating between the two (see Section~\ref{sec:mod_results}). 

The line-of-sight velocity dispersion (Figure~\ref{fig:ngc1574_mom2_main}) is $\lesssim20$~km~s$^{-1}$ at the outskirts of the CO disc, indicating that the gas is dynamically cold in these areas. $\sigma_{\rm gas}$ then progressively increases towards the central regions, with the highest values ($\geq50$~km~s$^{-1}$) almost exclusively concentrated around the location of the line peak. In principle, such large $\sigma_{\rm gas}$ would imply that an ongoing perturbation (such as deviations from purely circular motions) is inducing turbulence within the gas clouds. Line-of-sight velocity dispersion, however, can be significantly inflated due to observational effects. Among these, beam smearing (i.e.\,contamination from partially-resolved or unresolved velocity gradients within the host galaxy) usually dominates in regions of large velocity gradients, such as the inner regions of galaxies. Full kinematic modelling is the only way to derive reliable estimates of the intrinsic $\sigma_{\rm gas}$ (see Section~\ref{sec:mod_results}). 

The major-axis PVD (Figure~\ref{fig:ngc1574_PVD_main}) clearly shows mild, symmetric velocity increases around the galaxy centre, which can be plausibly associated with the Keplerian upturn arising from material orbiting within the SMBH SOI \citep[e.g.][]{Davis17}. The signature appears slightly more prominent at positive velocities, albeit only in the faintest contour, but it is visible also at negative velocities. 

\subsubsection*{NGC\,4261}
\begin{figure*}
\centering
\begin{subfigure}[t]{0.3\textheight}
\centering
 \caption{Moment zero}\label{fig:ngc4261_mom0_main}
\includegraphics[scale=0.5]{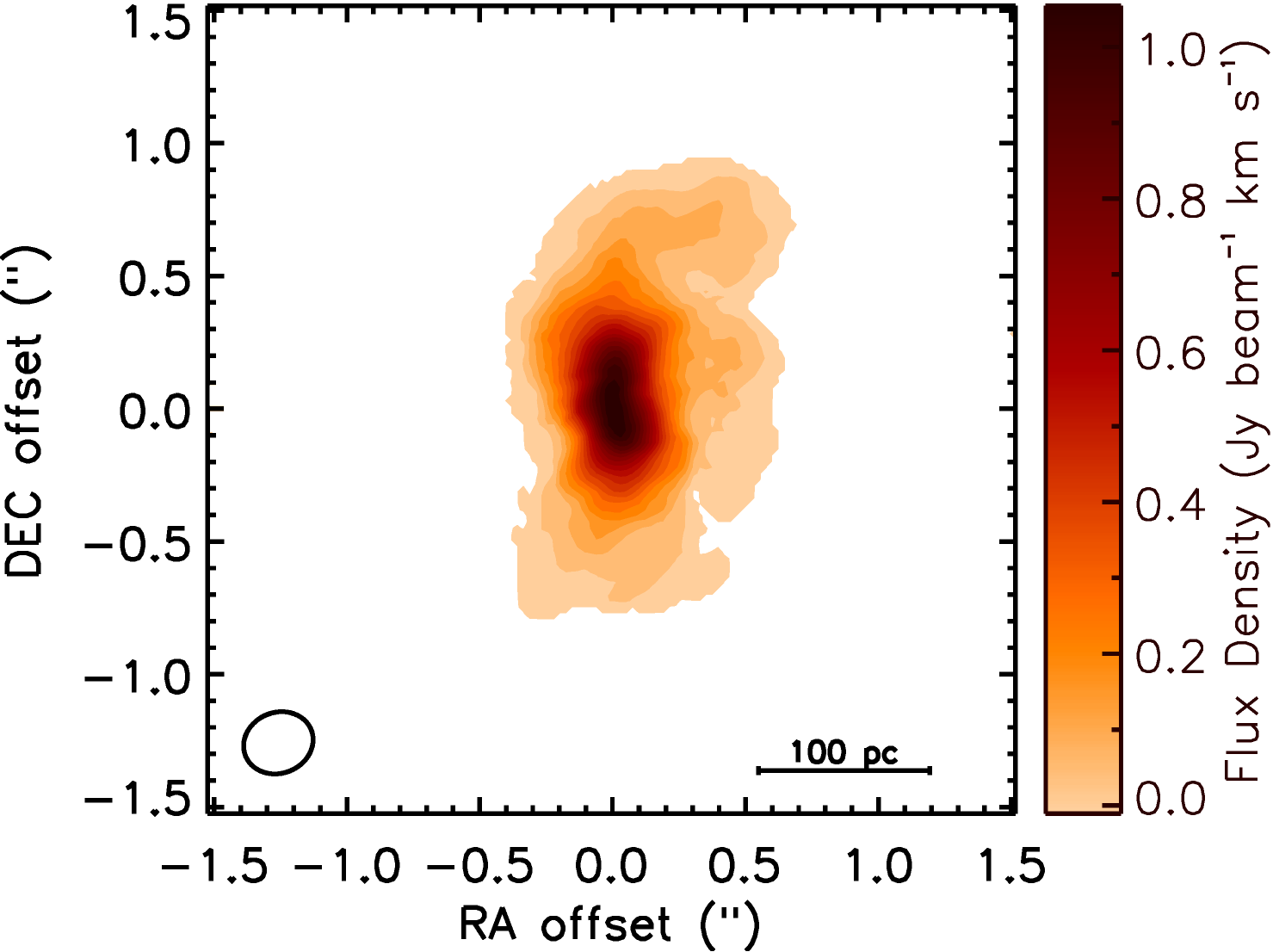}
\end{subfigure}
\hspace{4mm}
\begin{subfigure}[t]{0.3\textheight}
\centering
\caption{Moment one}\label{fig:ngc4261_mom1_main}
\includegraphics[scale=0.5]{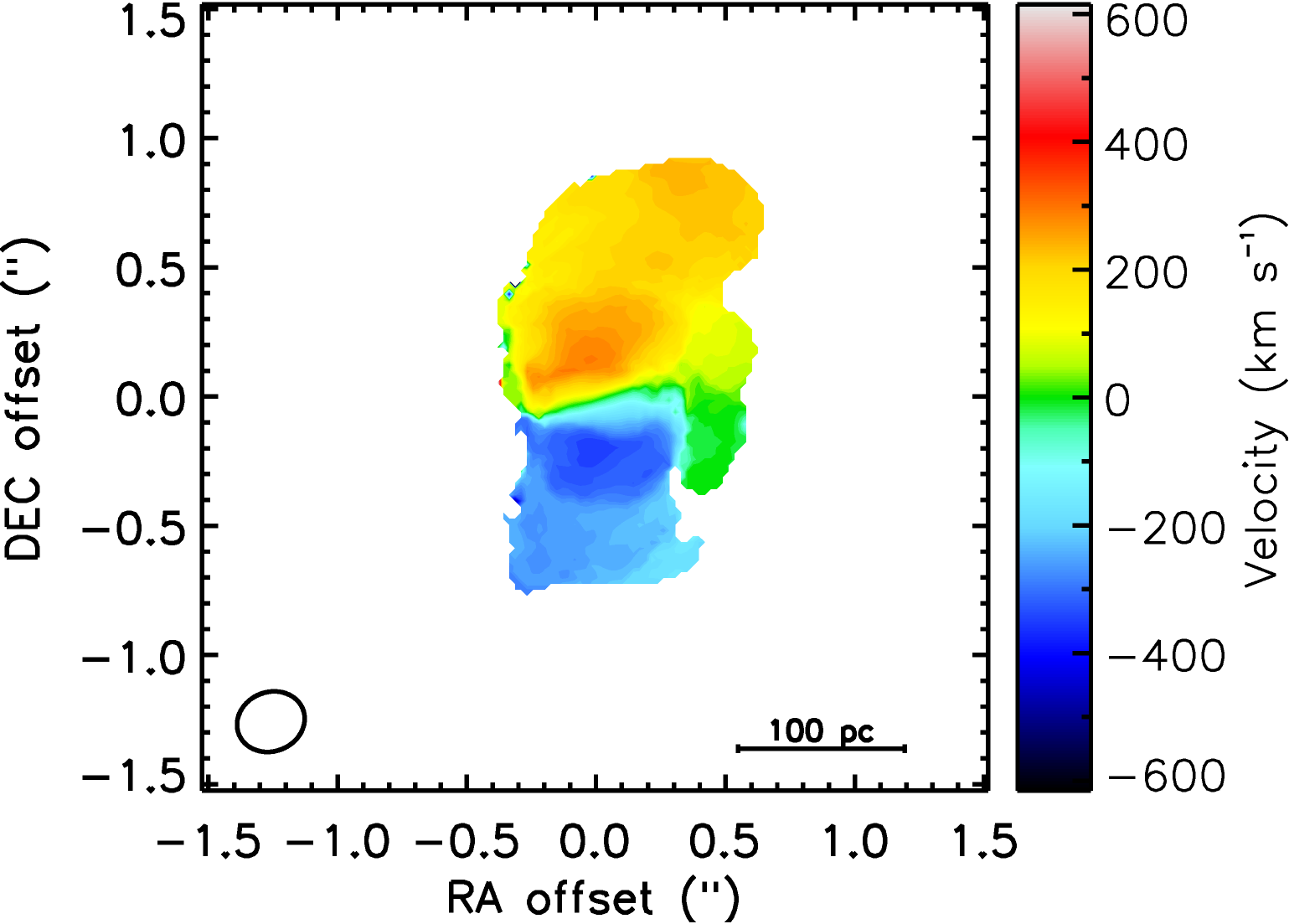}
\end{subfigure}

\medskip

\begin{subfigure}[t]{0.3\textheight}
\centering
\vspace{0pt}
\caption{Moment two}\label{fig:ngc4261_mom2_main}
\includegraphics[scale=0.5]{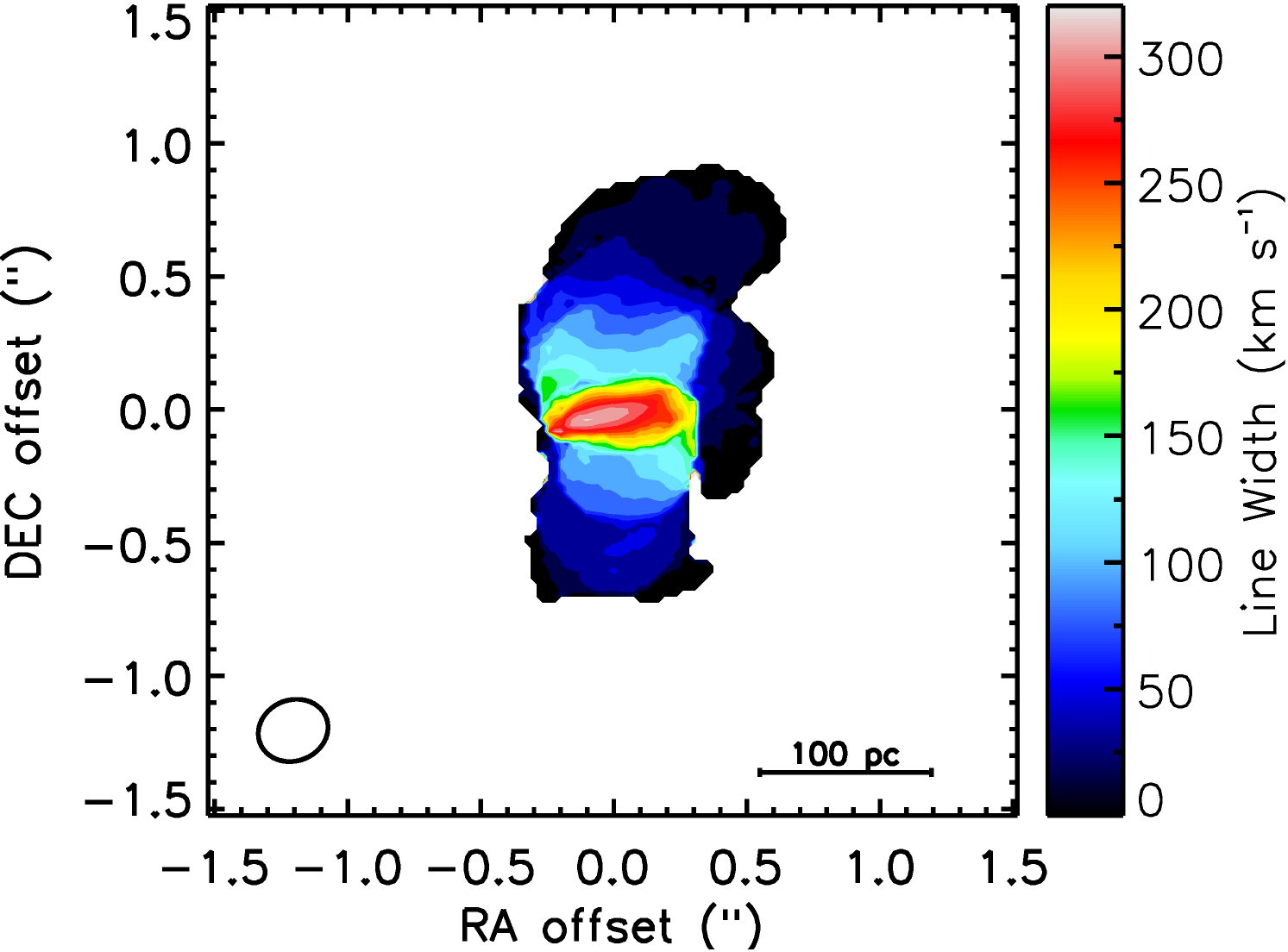}
\end{subfigure}
\hspace{4mm}
\begin{subfigure}[t]{0.3\textheight}
\centering
\vspace{0pt}
\caption{Major axis PVD}\label{fig:ngc4261_PVD}
\includegraphics[scale=0.48]{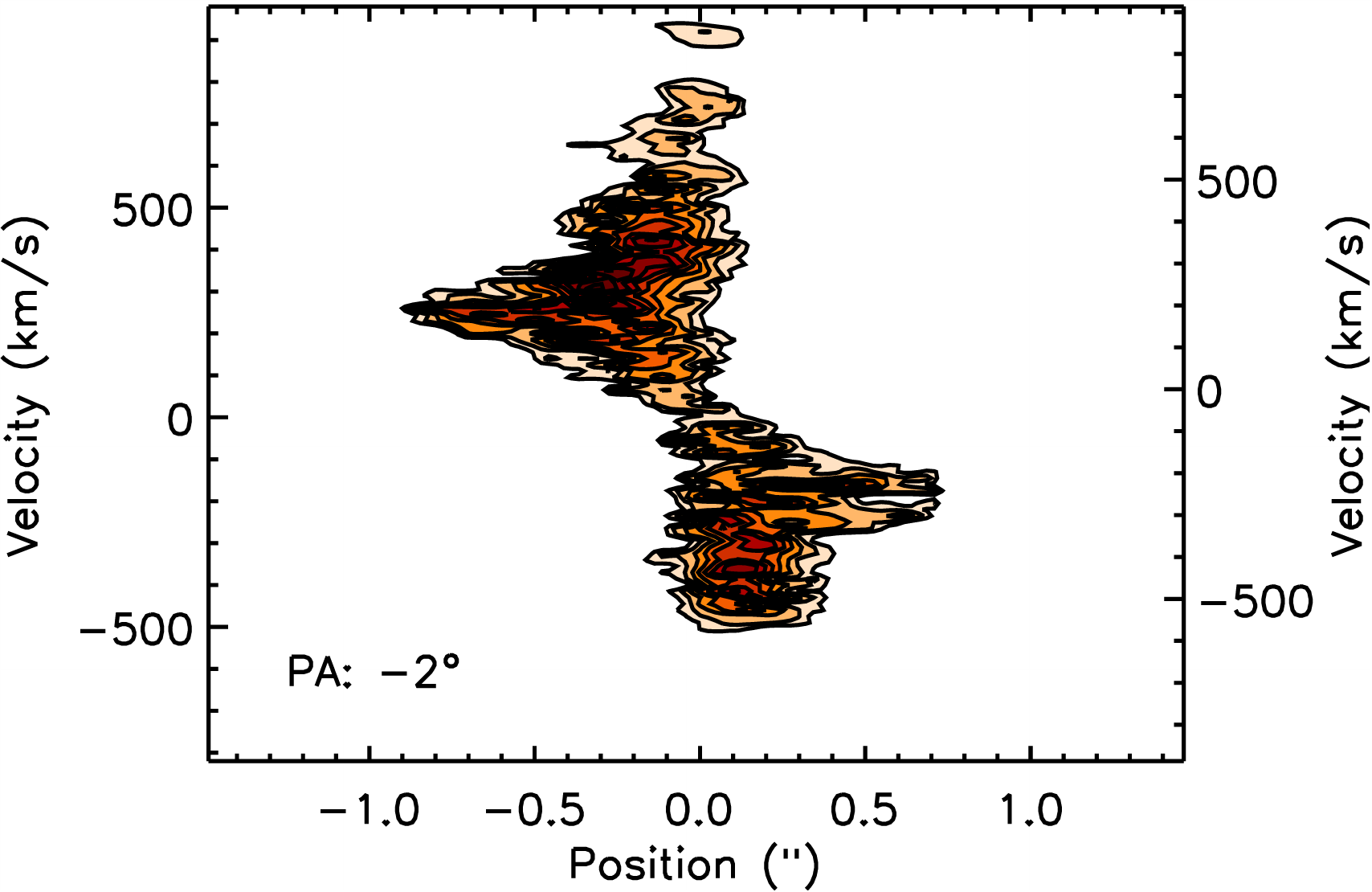}
\end{subfigure}
\caption{As in Figure~\ref{fig:NGC612}, but for NGC\,4261.}\label{fig:NGC4261}
\end{figure*}

The integrated intensity map in Figure~\ref{fig:ngc4261_mom0_main} shows that the CO gas in NGC\,4261 is distributed in a small-scale, disc-like structure, co-spatial with the nuclear dust disc (Figure~\ref{fig:NGC4261_optical}, right panel). The bulk of the CO emission arises from a bright centrally-concentrated region, surrounded by very faint emission extending up to $\approx1.7\arcsec$ in projection ($\approx250$~pc) along the major axis. The gas structure presents some asymmetries at its edges, where the major axis PA slightly changes orientation, suggesting the presence of a mild warp. 

Two velocity features can be clearly distinguished from the velocity map in Figure~\ref{fig:ngc4261_mom1_main}: one within ${\rm DEC~offsets}=\pm 0.4\arcsec$ from the centre characterised by very high velocities ($v\gtrsim400$~km~s$^{-1}$), and a slower one on larger scales. These components can be also clearly inferred from the major axis PVD in Figure~\ref{fig:ngc4261_PVD}, showing an abrupt velocity enhancement around the centre, followed by a steep decline and signs of a flat velocity component at increasing radii. 

The observed $\sigma_{\rm gas}$ (Figure~\ref{fig:ngc4261_mom2_main}) is $\lesssim40$~km~s$^{-1}$ at the outer edges of the gas distribution, and increases up to very large values ($\gtrsim300$~km~s$^{-1}$) towards the nucleus. In this case, however, the observed gas disc is only marginally resolved and viewed nearly edge-on (see Section~\ref{sec:mod_results}), therefore a combination of projection effects and beam smearing are likely strongly affecting the central $\sigma_{\rm gas}$ values.


\subsection{Line widths and profiles}\label{sec:line_profiles}
\begin{figure}
\centering
\begin{subfigure}[t]{0.3\textheight}
\caption{\textbf{NGC\,0612}}\label{fig:NGC612_spectrum}
\includegraphics[scale=0.33]{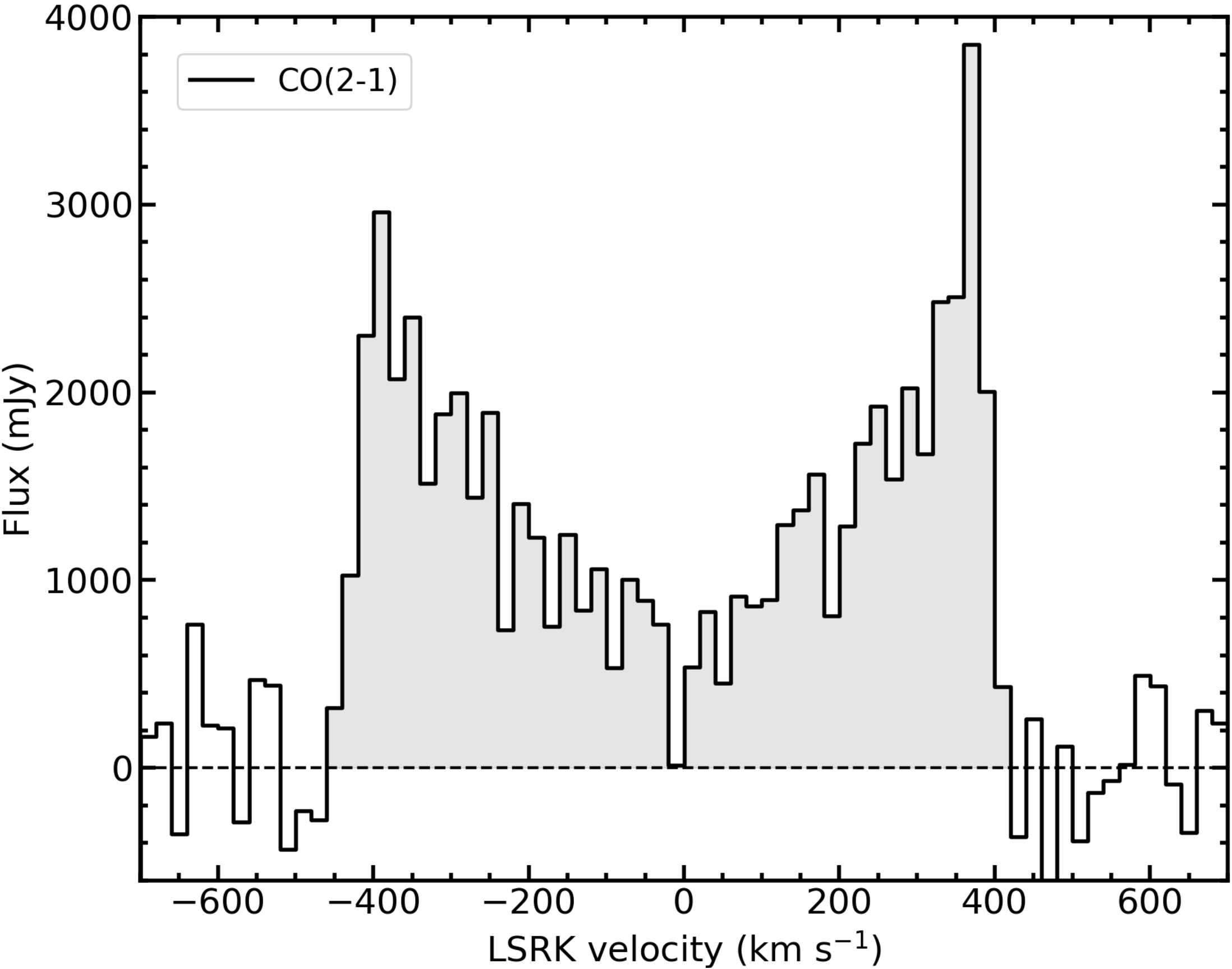}
\end{subfigure}
\medskip
\begin{subfigure}[t]{0.3\textheight}
\caption{\textbf{NGC\,1574}}\label{fig:NGC1574_spectrum}
\includegraphics[scale=0.33]{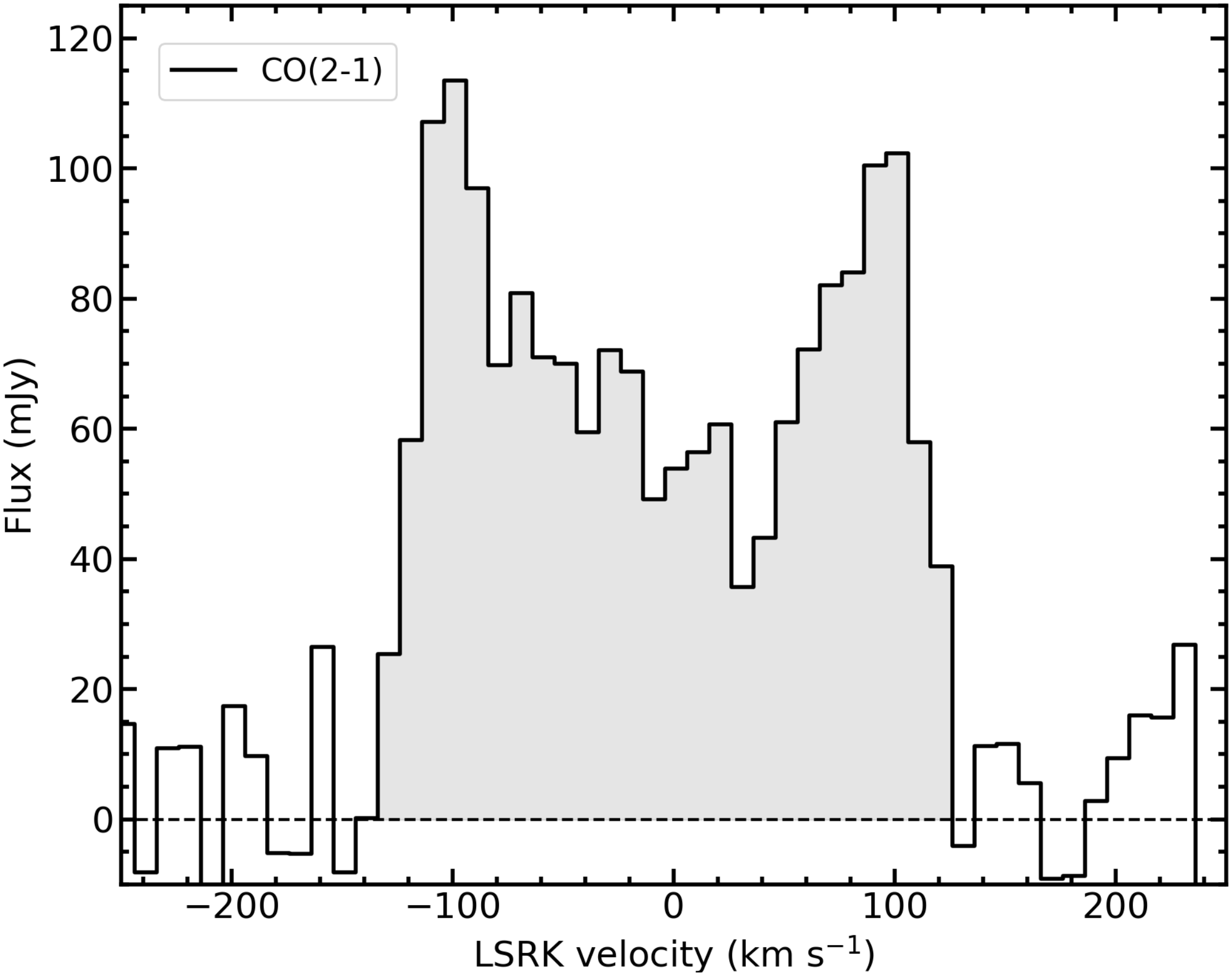}
\end{subfigure}
\vspace{0.2cm}
\medskip
\begin{subfigure}[t]{0.3\textheight}
\caption{\textbf{NGC\,4261}}\label{fig:NGC4261_spectum}
\centering
\includegraphics[scale=0.33]{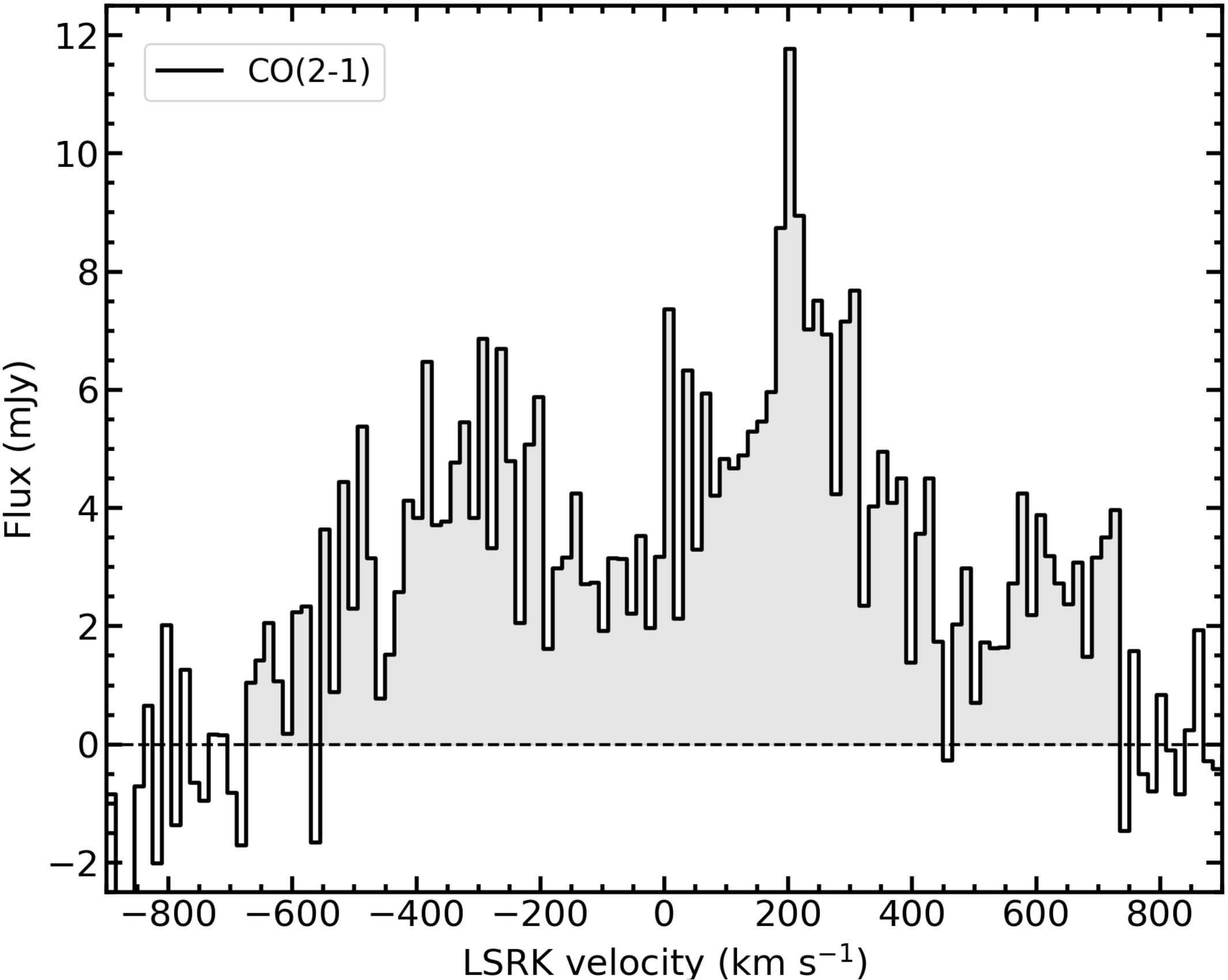}
\end{subfigure}
\caption[]{{\bf Top}: NGC\,0612 spectral profile extracted within a box of $4''\times24''$. {\bf Middle}: NGC\,1574 spectral profile extracted within a box of $2.8''\times2.8''$. {\bf Bottom}: NGC\,4261 spectral profile extracted within a box of $0.8''\times1.6''$. Each box includes all of the detected CO emission. In all the panels, the black dashed horizontal line indicates the zero flux level. The shaded regions highlight the spectral channels used to estimate the line FWZI (see Section~\ref{sec:line_profiles} for details). Velocities are measured in the source frame and the zero-point corresponds to the intensity-weighted centroid of the CO emission.}\label{fig:CO_spectra}
\end{figure}

The CO spectral profiles of the three targets are shown in Figure~\ref{fig:CO_spectra}, with velocity ranges narrowed around the detected line emission (while the total bandwidth extends up to $\sim 1300$~km~s$^{-1}$ in all the cases). These were extracted from the cleaned, un-smoothed data cubes within boxes enclosing all of the observed CO emission (the exact dimension of each box is indicated in the figure captions). 

The integrated spectral profiles of NGC\,0612 (Fig.\,\ref{fig:NGC612_spectrum}) and NGC\,1574 (Fig.\,\ref{fig:NGC1574_spectrum}) exhibit the well-defined double-horned shape expected from a rotating disc, with some asymmetries reflecting those in the gas distribution (Fig.\,\ref{fig:ngc612_mom0} and \ref{fig:ngc1574_mom0_main}). A central CO depression is also clearly visible in the NGC\,0612's spectrum (Fig.\,\ref{fig:NGC612_spectrum}), in agreement with the presence of a central gas deficiency. The spectrum of NGC\,4261 is more complex (Fig.\,\ref{fig:NGC4261_spectum}), but consistent with the kinematics observed in Figures~\ref{fig:ngc4261_mom1_main} and \ref{fig:ngc4261_PVD}. A rough double-peaked morphology can be identified between $\approx \pm400$~km~s$^{-1}$. This is surrounded by fainter structures extending up to $\approx \pm650$~km~s$^{-1}$, making the overall line profile very broad.

Line widths were measured as both full-width at zero-intensity (FWZI) and FWHM. The former was defined as the full velocity range covered by channels (identified interactively in the channel maps) with line intensities $\geq3\sigma$. These channels are highlighted by the grey shaded regions in Figure~\ref{fig:CO_spectra}. The CO integrated flux densities are calculated over this range. The FWHM was defined directly from the integrated spectra as the velocity difference between the two most distant channels from the line centre with intensities exceeding half of the line peak. 
The FWZI, FWHM and integrated flux densities are reported in Table~\ref{tab:line_parameters}.

\begin{table*}
\centering
\caption{Main CO line integrated parameters.}
\label{tab:line_parameters}
\begin{tabular}{l c c c c c c}
\hline
\multicolumn{1}{c}{ Target} &
\multicolumn{1}{c}{ Line FWHM }&
\multicolumn{1}{c}{ Line FWZI  }&
\multicolumn{1}{c}{ $S\textsubscript{CO}\Delta{\rm v}$}&
\multicolumn{1}{c}{ M\textsubscript{mol} } &
\multicolumn{1}{c}{ Size } \\
\multicolumn{1}{c}{  } &
\multicolumn{1}{c}{ (km~s$^{-1}$) } &
\multicolumn{1}{c}{ (km~s$^{-1}$) } &
\multicolumn{1}{c}{ (Jy km~s$^{-1}$) } &
\multicolumn{1}{c}{ (M$_{\rm \odot}$) } &
\multicolumn{1}{c}{ (kpc$^{2}$) } \\
\multicolumn{1}{c}{ (1) } &
\multicolumn{1}{c}{ (2) } &
\multicolumn{1}{c}{ (3) } &
\multicolumn{1}{c}{ (4) } &
\multicolumn{1}{c}{ (5) } &
\multicolumn{1}{c}{ (6) } \\
\hline
NGC\,0612   &   780   &  880  &  1209$\pm$121  &  $(1.0\pm0.1)\times10^{11}$  & 10.80$\times$1.60  \\
NGC\,1574  &  210  &  260  &  17.9$\pm$1.9  &  $(3.3\pm0.3)\times10^{7}$  & 0.18$\times$0.15  \\
NGC\,4261  &  540  &  1410  &  5.2$\pm$0.7  &  $(2.6\pm0.4)\times10^{7}$  & 0.25$\times$0.12  \\
\hline
\end{tabular}
\parbox[t]{1\textwidth}{ \textit{Notes.} $-$ Columns: (1) Target name. (2) Line FWHM. (3) Full velocity range covered by spectral channels with CO intensities $\geq$3$\sigma$ (FWZI, grey shaded region in Figure~\ref{fig:CO_spectra}). (4) Integrated CO flux density measured integrating numerically over the channels within the FWZI. The velocity ranges in columns (2) -- (4) are measured in the source frame. (5) Molecular gas mass derived using equation~\ref{eq:gas mass}. (6) Projected size of the observed CO distribution.}
\end{table*}

\subsection{Molecular gas mass}\label{sec:mol_masses}
We adopted the following relation to estimate the total molecular gas mass, including contributions from heavy elements \citep[$M_{\rm mol}$;][]{Bolatto13}:

\begin{multline}\label{eq:gas mass}
M_{\rm mol}=\dfrac{1.05\times10^{4}}{R_{\rm 21}}~\left(\dfrac{X_{\rm CO}}{2\times10^{20}~\dfrac{{\rm cm}^{-2}}{{\rm K~km~s}^{-1}}}\right) \\ \times \left(\dfrac{1}{1+z}\right)~\left(\dfrac{ S_{\rm CO}\Delta\nu}{{\rm Jy~km~s}^{-1}}\right)~\left(\dfrac{D_{{\rm L}}}{{\rm Mpc}}\right)^{2},
\end{multline}
where $S$\textsubscript{CO}$\Delta$v is the CO(1-0) integrated flux density, $R$\textsubscript{21} is the expected CO(2-1) to CO(1-0) flux density ratio, $z$ the galaxy redshift, $D$\textsubscript{L} the luminosity distance, and $X$\textsubscript{CO} the CO-to-H$_{2}$ conversion factor. Assuming a Milky Way $X$\textsubscript{CO} of $2 \times 10^{20}$\,cm$^{-2}$(K\,km\,s$^{-1}$)$^{-1}$ (shown to be appropriate for massive nearby ETGs; \citealp[see e.g.][]{Utomo15}), and $R$\textsubscript{21}=2.14 (for flux densities expressed in Jy~km~s$^{-1}$; see \citealp[][]{Ruffa22}), we estimated molecular masses ranging from $(2.6\pm0.4)\times10^{7}$ for NGC\,4261 to $(1.0\pm0.1)\times10^{11}$~M\textsubscript{$\odot$} for NGC\,0612 (see Table~\ref{tab:line_parameters}). We note that for this latter galaxy \citet{Ruffa19a} reported $M_{\rm mol}=(2.0\pm0.2)\times10^{10}$~M\textsubscript{$\odot$}, based only on the intermediate-resolution ALMA 12-m CO(2-1)
observations (Table~\ref{tab:ALMA observations summary}). This is about an order of magnitude lower than that estimated from the combined 12-m + ACA data used in this work. Such inconsistency can be largely ascribed to the differences in the measured CO(2-1) flux densities, being the one reported in \citet[][]{Ruffa19a} about 4.4 times lower than that listed in Table~\ref{tab:line_parameters}. These results clearly demonstrate how an incomplete uv–plane coverage can lead to resolve out a significant fraction of the total flux, especially in cases of extended, large-scale gas distribution like the one observed in NGC\,0612 (see also \citealp[][]{Leroy21}). A much smaller contribution in such a discrepancy has also to be ascribed to the slightly larger flux density ratio assumed for the previous calculation (i.e.\,$R$\textsubscript{21}$=2.32$; \citealp{Ruffa19a}).

\section{CO dynamical modelling}\label{sec:method}
The aim of this work is to obtain CO dynamical SMBH mass estimates. An accurate CO kinematic modelling has been thus carried out for our targets, adopting the same procedure extensively described in previous works of this series \citep[e.g.][see also \citealt{Ruffa19b,Ruffa22}]{Davis17,North19,Smith19}. We therefore provide only an outline of such procedure in the following and describe some relevant details on individual sources in Section~\ref{sec:mod_results}.

We analysed the CO kinematics using the forward-modelling approach implemented in the publicly-available {\sc Integrated Development Language} ({\sc IDL}) version of the \textsc{Kinematic Molecular Simulation} tool \citep[{\sc KinMS}\footnote{\url{https://github.com/TimothyADavis/KinMS}};][]{Davis13b}. This routine allows the user to input guesses for the gas distribution and kinematics, and constructs the corresponding mock data cube by calculating the line-of-sight projection of the circular velocity for a defined number (typically 10$^{5}$ - 10$^{6}$) of point-like sources representing the gas distribution. The {\sc IDL} \textsc{KinMS} routines are coupled with the Markov Chain Monte Carlo (MCMC) code (\textsc{KinMS\_MCMC}\footnote{\url{https://github.com/TimothyADavis/KinMS\_MCMC}}), that fits the data and outputs the full Bayesian posterior probability distribution, together with the best-fitting model parameters and their uncertainties.

As in previous WISDOM works, we assumed well-tested parametric functions to reproduce the observed gas distributions, fitting the parameters defining these functions to the data as part of the {\sc MCMC} procedure. Additional parameters needed to reproduce the gas distributions are those describing the disc geometry (i.e.\,PA and inclination), and the typical nuisance parameters (i.e.\,kinematic centre in RA, Dec and systemic velocity, and the total integrated flux).

To reproduce the observed gas kinematics we start from the assumption that the gas circular velocity mainly arises from the gravitational influence of the stars. We thus construct a stellar mass model to predict the circular velocity of the gas caused by the luminous matter, via the stellar mass-to-light ratio ($M/L$, whereby $v_{\rm circ}\propto\sqrt{M/L}$; see e.g.\,\citealp{DavisDermid17}). This usually allows us to obtain accurate $M_{\rm BH}$ estimates, once the contribution of visible matter to the observed gas kinematics is removed. We note that dark matter, while important on larger scales, is usually negligible in the nuclear regions of massive galaxies \citep[see e.g.][]{Cappellari13a}. 

We parametrise the luminous matter distribution from high-resolution optical images using the multi-Gaussian expansion (MGE; \citealp{Emsellem94,Cappellari02a}) procedure, adopting the fitting method and the \textsc{MGE\_FIT\_SECTORS} software package (\textsc{IDL} version)\footnote{Available from: \url{http://purl.org/cappellari/software}} of \citet[][]{Cappellari02a}. The circular velocity curve arising from this stellar mass model is then calculated using the \textsc{MGE\_CIRCULAR\_VELOCITY} function (\textsc{IDL} version)\footnote{\url{http://purl.org/cappellari/software}, part of the Jeans Anisotropic MGE (JAM) dynamical modelling package of \citet{Cappellari08}.}. In this procedure, the two-dimensional MGE model components are deprojected (at a given inclination) to a three-dimensional mass distribution, calculating the gravitational potential, and thus the circular velocity arising from the luminous matter. Such circular velocity is then multiplied element-wise by the square root of the assumed $M/L$, and a point mass representing the SMBH is added in the centre. These latter are included as additional free parameters in the \textsc{MCMC} fit. The velocity curve obtained in output is then automatically used by the \textsc{KinMS} tool to model the observed CO kinematics.

The \textsc{KinMS\_MCMC} procedure used to fit the simulated cubes utilises Gibbs sampling and adaptive stepping to explore the parameter space. The posterior distribution of each model is described by the log-likelihood function $\ln {\rm P} = - \chi^{2}/2$. As extensively explained in \citet{Smith19}, we re-scale the uncertainty in the cube by a factor of $(2N)^{0.25}$, where $N$ is the number of pixels with detected emission in the mask, as defined in Section \ref{sec:obs}. This ensures that the uncertainty in the $\chi^2$ statistics for high N does not lead to unrealistically small \textit{systematic} uncertainties in our final fit values \citep[see also][]{BoschVen09,Mitzkus17}. In principle, a further correction could be included for the fact that our pixel grid oversamples the beam, and thus nearby pixels are spatially correlated. However, applying this correction (e.g.\,via a covariance matrix) is computationally expensive, and the effect much smaller than the uncertainty scaling described above \citep[see][for an extensive discussion on this issue]{Davis17}. We thus neglect the covariance correction here.

\begin{table*}
\begin{center}
\caption{Best-fitting parameters with formal uncertainties from the MCMC fits of the three targets.}
\label{tab:MCMC-params}
\begin{tabular*}{0.8\textwidth}{@{\extracolsep{\fill}}lcccc} 
\hline
Parameter & Search range & Best fit & $1\sigma$ uncertainty & $3\sigma$ uncertainty \\
\hline
\hline
\multicolumn{5}{c}{ {\bf NGC\,612} } \\
\hline
$\log$(M$_{\rm BH,AT}$) ($M_{\odot}$) & 6 - 11 & $<9.51$ & $-$ & $-$ \\
$\log$(M$_{\rm BH,S}$) ($M_{\odot}$) & 6 - 11 & $<9.38$ & $-$ & $-$ \\
$\log$(M$_{\rm BH,BD}$) ($M_{\odot}$) & 6 - 11 & $<9.19$ & $-$ & $-$ \\
\hline
{\bf Molecular gas disc} & & & & \\
PA ($^\circ$) & \multicolumn{4}{c}{ 182.81 (fixed) }\\ 
Inc ($^\circ$) & \multicolumn{4}{c}{ 80.82 (fixed) } \\ 
$\sigma_{\rm gas}$ (km s$^{-1}$) & 10 - 100 & 65.71 & $-2.63, +2.18$ & $-6.23, +6.98$ \\
Gaussian mean $\mu$ (\arcsec) & 0 - 10 & 4.89 & $-0.14, +0.26$ & $-0.68, +0.66$ \\
Gaussian width $\sigma_{\rm gauss}$ (\arcsec) & 0 - 10 & 3.82 & $-0.19, +0.10$ & $-0.48, +0.49$ \\
$R_{\rm hole}$ (\arcsec) & 0.01 - 5 & $<0.94$ & $-$ & $-$ \\
\hline
{\bf Nuisance parameters} & & & & \\
Integrated intensity (Jy~km~s$^{-1}$) & 100 - 800 & 567.85 & $-20.67, +28.01$ & $-68.71, +64.21$ \\
Centre X offset (\arcsec) & \multicolumn{4}{c}{ -0.01 (fixed) } \\
Centre Y offset (\arcsec) & \multicolumn{4}{c}{ 0.40 (fixed) }\\
Centre velocity offset (km s$^{-1}$) & \multicolumn{4}{c}{ -59.40 (fixed) }\\
\hline
\hline
\multicolumn{5}{c}{ {\bf NGC\,1574} } \\
\hline
{\bf Mass model} & & & & \\
$\log$(M$_{\rm BH}$) ($M_{\odot}$) & 6 - 10 & 8.00 & $-0.09, +0.07$ & $-0.24, +0.19$ \\
Stellar $M/L$ ($M_{\odot}/L_{\odot,{\rm F606W}}$) & 1 - 10 & 7.33 & $-0.83, +0.92$ & $-2.24, +1.97$ \\
\hline
{\bf Molecular gas disc} & & & & \\
PA$_{\rm 0}$ ($^\circ$) & 330 - 360 & 342 & $\pm 0.02$ & $\pm 0.05$ \\ 
PA$_{\rm 1}$ ($^\circ$) & 385 - 415 & 392 & $\pm 0.01$ & $\pm 0.04$ \\
Inc ($^\circ$) & 24 - 34 & 26.75 & $-2.44, +0.89$ & $-2.44, +5.56$ \\ 
$\sigma_{\rm gas}$ (km s$^{-1}$) & 5 - 40 & 15.50 & $\pm 1.00$ & $\pm 3.60$ \\
$R_{\rm disc}$ (\arcsec) & 0.1 - 1.5 & 0.56 & $-0.02, +0.03$ & $-0.07, +0.09$ \\ 
\hline
{\bf Nuisance parameters} & & & & \\
Integrated intensity (Jy~km~s$^{-1}$) & 5 - 30 & 15.47 & $-0.96, +1.62$ & $-3.54, +3.56$ \\
Centre X offset (\arcsec) & -1 - 1 & -0.02 & $-0.01, +0.02$ & $-0.03, +0.04$ \\
Centre Y offset (\arcsec) & -1 - 1 & -0.04 & $\pm 0.01$ & $\pm 0.02$ \\
Centre velocity offset (km s$^{-1}$) & -50 - 50 & -26.28 & $-0.73, +1.15$ & $-2.61, +2.56$ \\
\hline 
\hline
\multicolumn{5}{c}{ {\bf NGC\,4261} } \\
\hline
{\bf Mass model} & & & & \\
$\log$(M$_{\rm BH}$) ($M_{\odot}$) & 7 - 10 & 9.21 & $\pm 0.01$ & $\pm 0.05$ \\
Stellar $M/L$ ($M_{\odot}/L_{\odot,{\rm F160W}}$)$^{*}$ & 0.05 - 8 & 1.28 & $-0.86, +0.2$ & $-1.14, +1.98$\\
\hline
{\bf Molecular gas disc} & & & & \\
PA ($^\circ$) & -10 - 10 & -2.10 & $-1.63, +1.56$ & $-4.82, +5.82$ \\ 
Inc ($^\circ$) & \multicolumn{4}{c}{ 60.80 (fixed) } \\
$\sigma_{\rm gas}$ (km s$^{-1}$) & 10 - 150 & 61.71 & $-5.11, +5.18$ & $-$ \\ 
$R_{\rm disc}$ (\arcsec) & 0.05 - 1 & 0.19 & $\pm 0.01$ & $\pm 0.03$ \\ 
\hline
{\bf Nuisance parameters} & & & & \\
Integrated intensity (Jy~km~s$^{-1}$) & 0.5 - 8 & 4.63 & $-0.30, +0.19$ & $\pm 0.68$ \\
Centre X offset (\arcsec) & -1 - 1 & -0.01 & $\pm 0.01$ & $\pm 0.02$ \\
Centre Y offset (\arcsec) & -1 - 1 & -0.02 & $\pm 0.01$ & $\pm 0.02$ \\
Centre velocity offset (km s$^{-1}$) & -20 - 80 & 54.61 & $-2.60, +0.76$ & $-5.96, +5.23$ \\
\hline
\end{tabular*}
\parbox[t]{0.8\textwidth}{{\textit{Note:} The X and Y offsets are measured with respect to the image phasecenter, which usually coincides with the location of the unresolved core continuum emission. The velocity offsets are measured with respect to the central channel of the cube, and their best-fitting values thus define the difference with respect to the galaxy systemic velocity (reported in Table~\ref{tab:general}).\\
$^{*}$Based on the H-band MGE solutions reported in \citet[][]{Boizelle21}. See the text for details.}}
\end{center}
\end{table*}

To ensure the kinematic fitting process converges, we set reasonable priors (provided in Table~\ref{tab:MCMC-params}) for the parameters that are left free to vary in the fit. Initially, the step size in each fit is adaptively scaled to ensure a minimum acceptance fraction and the chain converges. Once the \textsc{MCMC} chains converged, we re-run the entire chain for an additional $10^{5}$ steps with a fixed step size to produce the full final posterior probability distribution. For each model parameter these probability surfaces are then marginalised over to produce a best-fit value (median of the marginalised posterior distribution) and associated 68 and 99\% confidence levels (CL).

\subsection{Modelling and results on individual sources}\label{sec:mod_results}
\subsubsection*{NGC 0612}
As mentioned in Section~\ref{sec:targets}, a full \textsc{KinMS} modelling of the CO disc in NGC\,0612 has been previously presented by \citet[][]{Ruffa19b} using only the intermediate-resolution ALMA 12-m observation. We thus repeated the kinematic modelling starting from the results obtained in such previous work, and concentrated our efforts here on the attempt of obtaining an $M_{\rm BH}$ estimate exploiting the $\approx3$ times higher-resolution of the combined ALMA data used in this work.

Ruffa et al.\, previously found that the gas surface brightness profile of NGC\,0612 is well reproduced by a Gaussian function with a central hole. The same parametrisation turned out to be the best way to reproduce the observed gas distribution also in the higher-resolution data, although such a simple modelling necessarily misses some of the small-scale, asymmetric features visible in Figure~\ref{fig:ngc612_mom0}. This, however, does not affect the analysis of the gas kinematics and thus our conclusions. The Gaussian surface brightness profile added a total of three free parameters in our \textsc{KinMS\_MCMC} fit: the Gaussian centre ($\mu_{\rm gauss}$), width ($\sigma_{\rm gauss}$), and the radius of the central hole ($R_{\rm hole}$). The $PA$ and inclination (${\rm Inc}$) were assumed to be constant throughout the disc, and the gas to be in purely circular motions.

Differently from \citet{Ruffa19b}, in this work we attempted to carry out an MGE modelling of the NGC\,0612's luminous stellar matter. To this aim, we used the HST WFC3 F160W image shown in Figure~\ref{fig:NGC612_optical}. The prominent dust distribution present in the equatorial plane of the galaxy, however, leads to a substantial extinction even at such infrared wavelengths (the longest available), and can strongly affect the derived stellar light profile. We attempted to mitigate such negative effect by fully masking in our MGE modelling the dust lane extending in the North-South direction to the East of the nucleus. Although there is strong dust obscuration encompassing also the nuclear regions, we accurately avoided to mask those areas to prevent issues on the location of the galaxy centre (the first fundamental step of the MGE fitting). The resulting best-fitting MGE model (not shown here) has been then fed to {\sc KinMS} to calculate the gas circular velocity curve, adding two free parameters to the {\sc KinMS\_{MCMC}} fit: a constant $M/L$ and the SMBH mass. We also included a velocity dispersion parameter ($\sigma_{\rm gas}$). Unless otherwise required (see e.g.\,the case of NGC\,4261), $\sigma_{\rm gas}$ is assumed to be spatially-constant, as our aim is to simply obtain a reference average value for the intrinsic gas velocity dispersion. The overall model had a total of 12 parameters (including the nuisance parameters), which were all initially left free to vary in the first \textsc{MCMC} fit. 

This model, while providing an overall good fit to the gas distribution and geometry, failed to adequately reproduce the observed gas kinematics, indicating an issue in the prediction of the gas circular velocity curve. This has likely to be ascribed to the prominent dust lane and the consequent extension of the masked region, making challenging to obtain a reliable MGE modelling of the stellar surface brightness profile from the innermost to the outermost regions of the gas disc. We then decided to adopt a different approach in this case, that is to model the observed CO kinematics using meaningful parametric functions. We performed the fit using the latest \textsc{Python} version of the \textsc{KinMS} routines, the \textsc{KinMS\_fitter}\footnote{\url{https://github.com/TimothyADavis/KinMS\_fitter}}, which allows to use a number of built-in functions describing the observed gas velocity profile. We tested three different models: i) one assuming that the CO rotation curve follows an arctangent function (AT), that is the simplest functional form that has been shown to provide a good fit to most rotation curves on $\sim$kpc scales \citep[see e.g.][this is also the same parametrisation adopted for the NGC\,0612's velocity curve in \citealp{Ruffa19b}]{Swinbank12,Voort18}; ii) one assuming kinematics arising from a Sersic stellar density profile (S), iii) and one assuming a bulge/disc stellar density profile (BD; this is implemented in \textsc{KinMS} as a summation of two Sersic profiles with indices 1 and 4, respectively). In all the cases, a Keplerian velocity component was also added to account for the inner gas kinematics and obtain a SMBH mass estimate. The observed gas distribution and velocity dispersion were modelled as described above. 
We note that at this stage we fixed the RA, DEC and velocity offsets, along with the PA and inclination to their previous best-fitting values (the latter two also perfectly in agreement with those found in \citealt[][]{Ruffa19b}) to reduce the dimensionality of the model and speed-up the computational time (which is of the order of days in this case, due to the very high resolution of the combined data and the large size of the CO disc). For the same reason, we restricted our CO modelling within a central area of $12\arcsec \times 12\arcsec$ in size.

\begin{figure*}
\begin{subfigure}[t]{0.3\textheight}
\caption{}\label{fig:NGC612_model_PVD}
\includegraphics[scale=0.52]{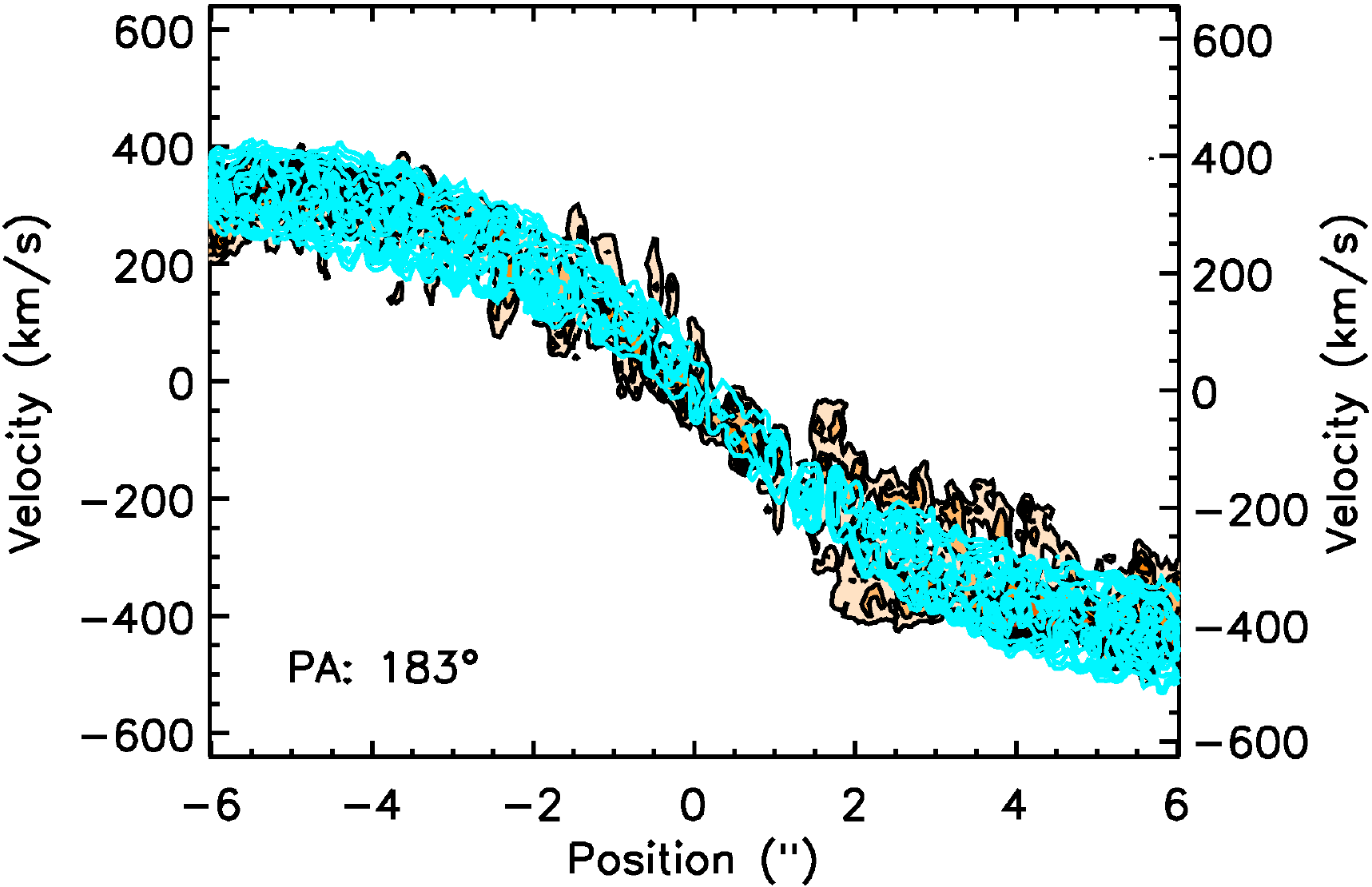}
\end{subfigure}
\hspace{20mm}
\begin{subfigure}[t]{0.3\textheight}
\caption{}\label{fig:N612_BHmass_comp}
\includegraphics[scale=0.33]{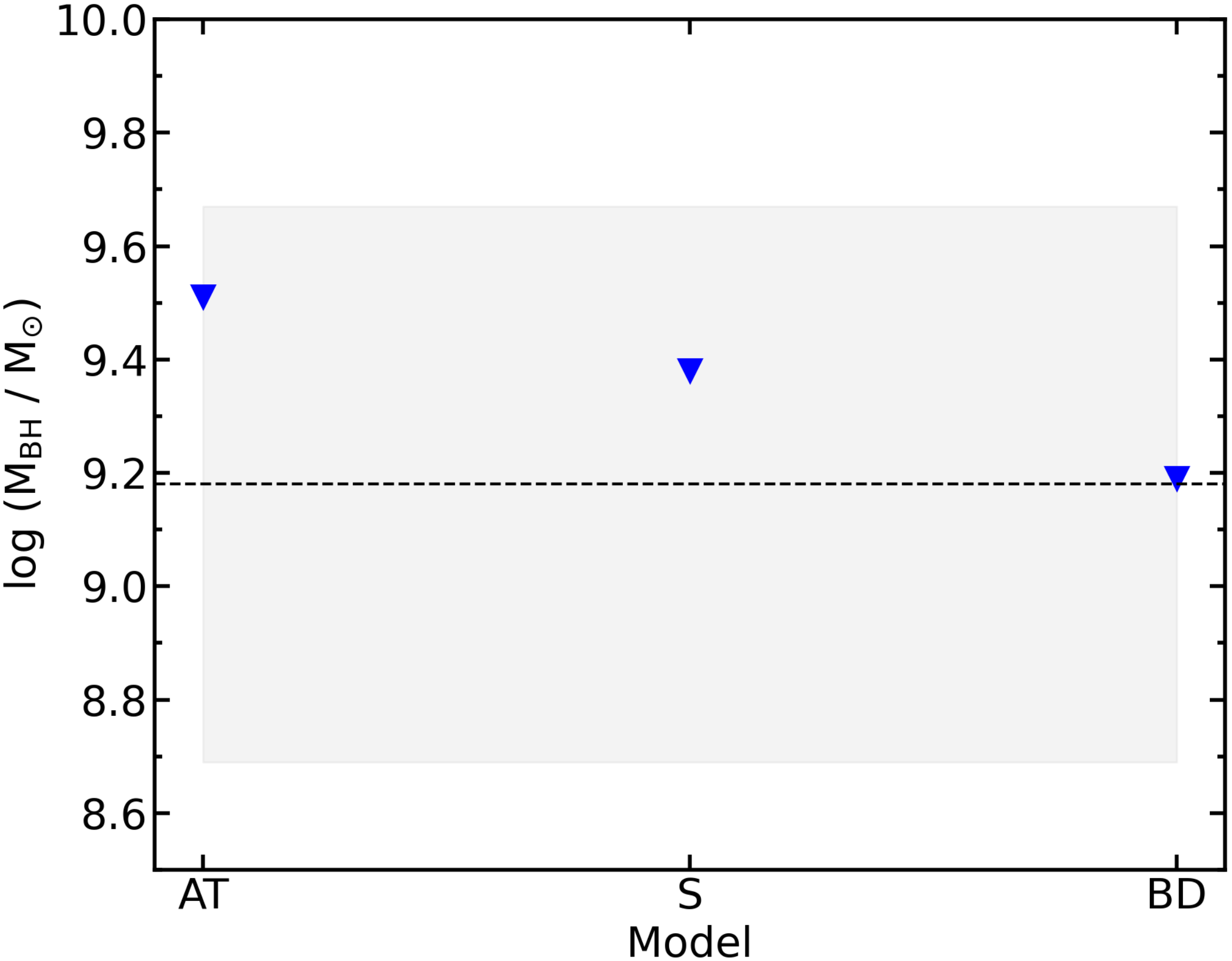}
\end{subfigure}
\caption[]{{\bf Left panel:} Major axis PVD of NGC\,0612 extracted within the fitting area (i.e.\,$12\arcsec \times 12\arcsec)$ and with the contours from the best-fitting Sersic model overlaid in cyan. {\bf Right panel:} $M_{\rm BH}$ upper limits obtained for each of the three kinematic models tested on NGC\,0612: one assuming that the CO rotation curve follows an arctangent function (AT), one assuming that the CO kinematics arises from a Sersic stellar density profile (S), and one assuming a bulge/disc stellar density profile (BD). The black dashed horizontal line indicates the value of $M_{\rm BH}$ expected from the $M_{\rm BH}-\sigma_{\rm e}$ relation of \citet{Vandenbosch16} as calculated Section~\ref{sec:targets}. The gray shaded area illustrates the scatter of such relation.}\label{fig:N612_mod_results}.
\end{figure*}

None of the three tested models provide a particularly good fit to the observed CO velocity curve, although all of them acceptably reproduce the overall kinematics (as shown by the maps in Fig.\,\ref{fig:NGC612_moments_all_appendix}). As illustrative case, we show in the left panel of Figure~\ref{fig:N612_mod_results} the major axis PVD of NGC\,0612 (within the fitting area) with the contours from the best-fitting Sersic model overlaid. Most importantly, the SMBH mass remains unconstrained in all the cases, with a conservative overall upper limit of $M_{\rm BH}\lesssim3.2\times10^{9}$~M$_{\odot}$ (corresponding to the largest among the three obtained). All the three upper limits, however, are in agreement with the prediction from the $M_{\rm BH}-\sigma_{\rm \filledstar}$ relation, and well within its scatter (see Fig.\,\ref{fig:N612_mod_results}, right panel). We discuss these result in Sections~\ref{sec:NGC612_discuss} and \ref{sec:msigma_discuss}.

Since our inability of placing a robust constraint on $M_{\rm BH}$ turned out to be essentially independent of the adopted mass/kinematic model and none of those tested can be elected as the most reliable, in Table~\ref{tab:MCMC-params} we just provide the search ranges and obtained $M_{\rm BH}$ upper limits for each of the three attempts (instead of the full list of parameters describing each model). We also list the search ranges, best-fitting values and $1(3)\sigma$ uncertainties of the nuisance parameters and those describing the gas distribution and geometry that are in common to all the three attempts, reporting in particular the values obtained for the Sersic model (similar results have been obtained in the other two cases). All these parameters are mostly in agreement with those obtained by \citet[][once re-scaled for the central $12\arcsec \times 12\arcsec$ area]{Ruffa19b}, and thus not further discussed here. The corner plots and one-dimensional marginalisation of relevant, non-nuisance parameters of the Sersic model (but common to all those tested) are shown in Figure~\ref{fig:NGC612_conts}.

\subsubsection*{NGC\,1574}
At the resolution achieved with the combined ALMA observations (see Table~\ref{tab:line_images}), we found that the CO surface brightness distribution in NGC\,1574 (Fig.\,\ref{fig:ngc1574_mom0_main}) is fully consistent with an axisymmetric thin exponential disc (a form that has been shown to be appropriate for the gas in many ETGs; \citealp[e.g.][]{Davis13b}). This has been parametrised as
\begin{eqnarray}
\Sigma_{\rm disc} \propto 
	e^{-\dfrac{r}{R_{\rm disc}}},  
\end{eqnarray}
where $\Sigma_{\rm disc}$ is the surface brightness at radius $r$ from the centre, and R$_{\rm disc}$ is the exponential disc scale length, which was left as a free parameter in our \textsc{KinMS\_MCMC} fit. We initially fit the inclination and PA as single values throughout the disc, and assumed the gas to be in purely circular motions.

\begin{figure}
\centering
\includegraphics[scale=0.5]{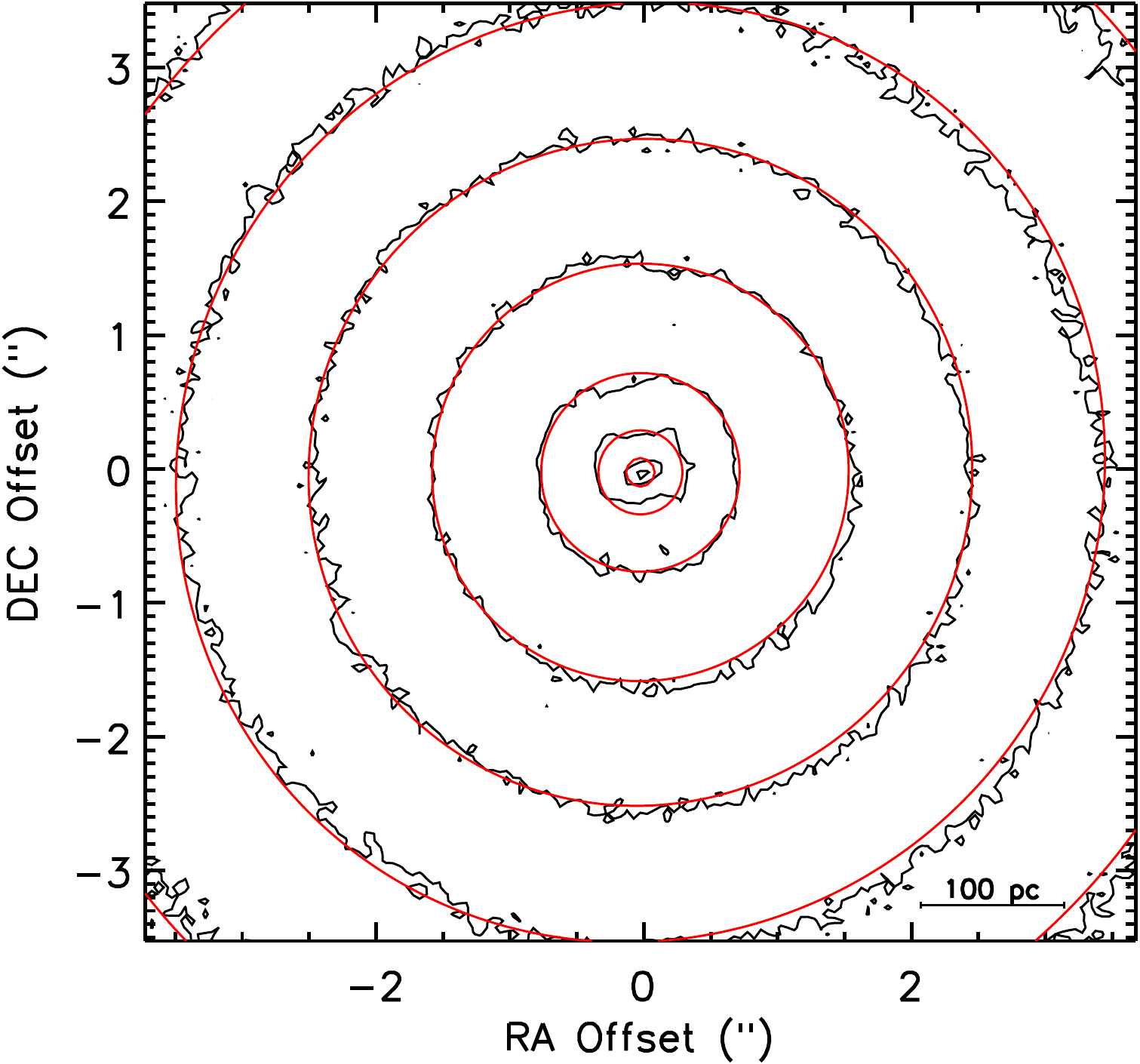}
\caption[]{MGE model of NGC\,1574 (red contours), overlaid on the HST optical (F606W) image (black contours). The panel shows a zoom-in in the galaxy regions relevant for our analysis (see the text for details).}\label{fig:NGC1574_MGE_model}
\end{figure}

The MGE model of the stellar light distribution was constructed using the HST WFPC2 F606W image shown in Figure~\ref{fig:NGC1574_optical} (the longest wavelength available, to minimise the effect of dust extinction). A zoom-in of the resulting MGE fitting in the areas relevant for our analysis (i.e.\,those over which the CO disc extends) is shown in Figure~\ref{fig:NGC1574_MGE_model}, the values of each best-fitting Gaussian are listed in Table~\ref{tab:NGC1574_MGE_parameters}. The circular velocity curve arising from this mass model has been then predicted in {\textsc KinMS}, with a constant $M/L$ ratio and the SMBH treated as point mass. We also included in this model a spatially-constant velocity dispersion parameter ($\sigma_{\rm gas}$). 

As expected, while well reproducing gas distribution, the simple model described above failed to reproduce the velocity distortions visible in Figure~\ref{fig:ngc1574_mom1_main}, leaving very large residuals in the data-model moment one map. As mentioned in Section~\ref{sec:cube_analysis}, such s-shaped iso-velocity contours may result from either warps or non-circular motions, as both phenomena produce similar kinematic features. In NGC\,1574, however, we found that they are best reproduced by a PA warp. We tested different (arbitrary) functional forms to add a radial PA variation to the model described above, and found that the observed trend was best reproduced by performing a linear interpolation between an inner and an outer PA value (both left free to vary in the fit). 

\begin{figure*}
\centering
\includegraphics[scale=0.6]{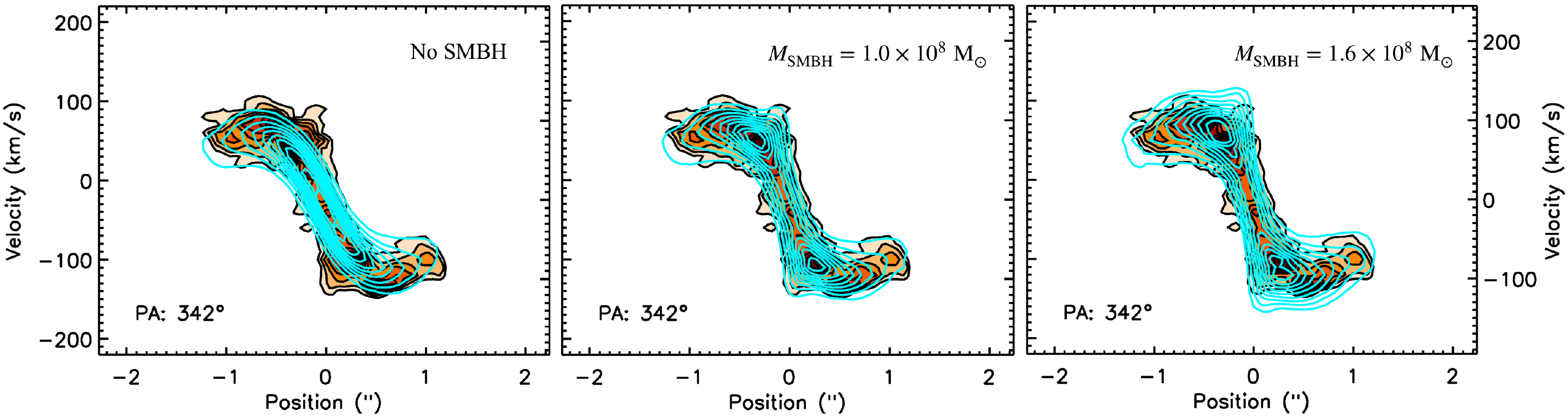}
\caption[]{Observed PVD along the kinematic major axis of NGC\,1574 with different models overlaid in cyan contours. The left panel shows a model without a SMBH, the central panel with the best-fitting SMBH and the right one with an overly large SMBH (see the text for details). \label{fig:NGC1574_PVDs_model}}
\end{figure*}

The final adopted model had 11 free parameters: the gas integrated flux density, lower and upper PA ($PA_{0}$ and $PA_{1}$), inclination angle ($Inc$), RA, DEC and velocity centres, exponential disc scale length ($R_{\rm disc}$), gas velocity dispersion ($\sigma_{\rm gas}$), SMBH mass ($M_{\rm BH}$), and $M/L$ (see Table~\ref{tab:MCMC-params}). As usual, we first run an initial fit with about $10^{3}$ steps to evaluate the goodness of the overall model and ensure the chain converged. At this stage, we found the inclination parameter settled at a best-fitting value of $29^{\circ}$, although with relatively large uncertainties (of the order of $\pm5^{\circ}$). We then re-run the entire chain for additional $10^{5}$ steps. As we increased the number of steps, however, the inclination parameter tended to converge towards very small values (i.e.\,$<19^{\circ}$), leading - in turn - to an implausibly large $M/L$ ratio and decreasing the overall goodness of the fit. From the optical dust distribution, assuming that this is intrinsically circular, we estimated an inclination of $\approx27^{\circ}$, reasonably in agreement with the value reported above. Since CO is typically co-spatial with dust in ETGs \citep[e.g.][]{Alatalo13,Ruffa19a}, we assume that also the CO disc inclination should be consistent with this value. In our final fit, we thus set a Gaussian prior for the inclination with mean equal to our initial best-fitting value and width of $10^{\circ}$. This provided a much physically-meaningful constraint on the $M/L$, but necessarily introduced larger uncertainties on both the $M/L$ and SMBH mass estimate (see Section~\ref{sec:NGC1574_discuss} for a discussion on this issue). Nevertheless, we found strong evidence for the presence of a SMBH at the centre of NGC\,1574, with a best-fitting $M_{\rm BH}=(1.0\pm0.2)\times10^{8}$~M$_{\odot}$ (1$\sigma$ uncertainty). The major axis PVD with the best-fitting model contours overlaid is shown in the middle panel of Figure~\ref{fig:NGC1574_PVDs_model}. For comparison, we also overlay the contours of a model PVD obtained removing the SMBH (Fig.\,\ref{fig:NGC1574_PVDs_model}, left panel), and of one obtained setting an overly large SMBH mass (Fig.\,\ref{fig:NGC1574_PVDs_model}, right panel). These plots clearly show the need for a central dark object with the best-fitting mass to fully account for the gas kinematics at small radii.

\begin{figure}
\centering
\includegraphics[scale=0.3]{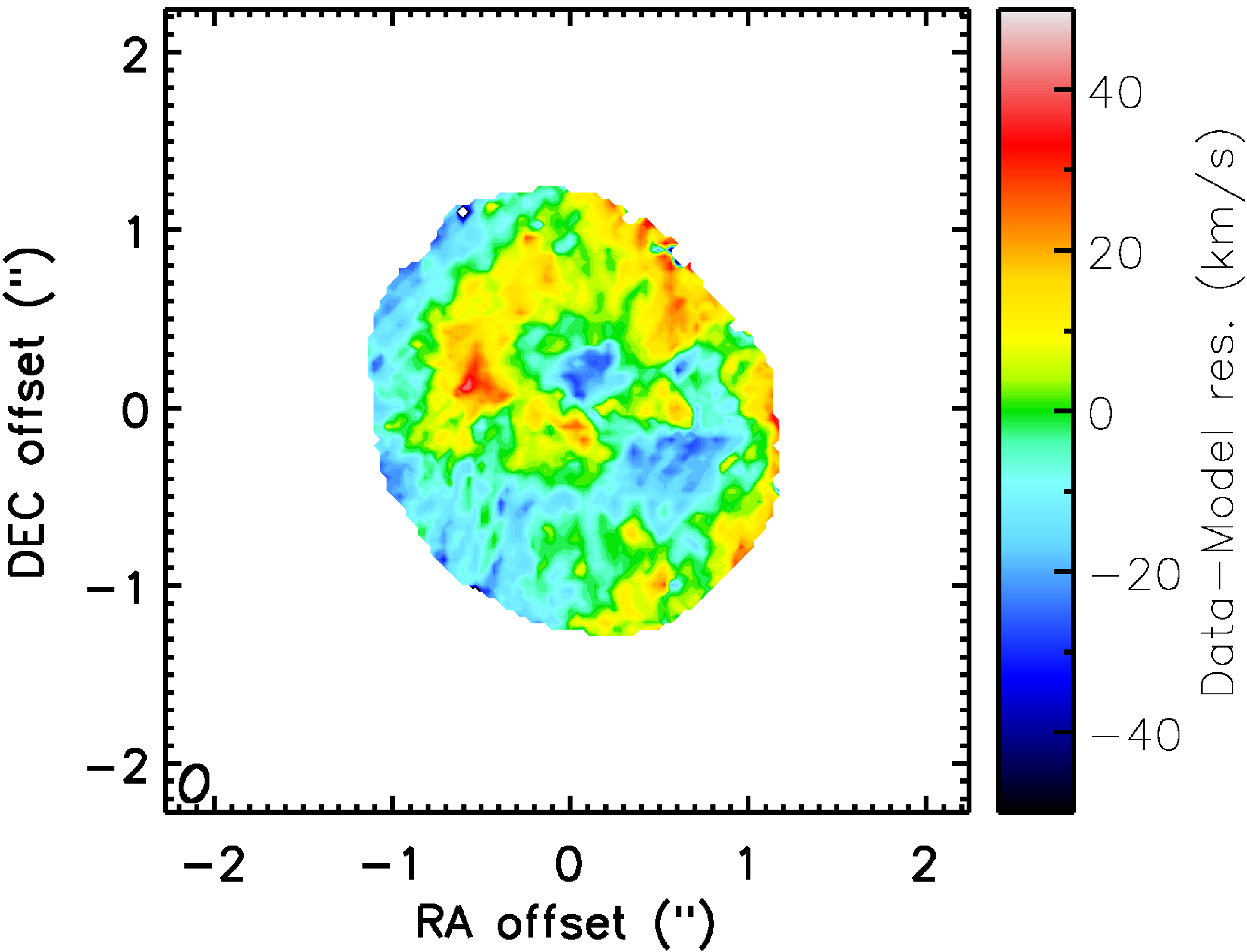}
\caption[]{Residual (data - model) mean velocity map of NGC\,1574. The synthesized beam is shown in the bottom-left corner. The bar to the right show the colour scale in km~s$^{-1}$. East is to the left and North to the top.}\label{fig:NGC1574_mode_res}
\end{figure}

The best-fitting and formal uncertainties of each model parameter are listed in Table~\ref{tab:MCMC-params}. The corner plots and one-dimensional marginalisation of the non-nuisance model parameters are shown in Figure~\ref{fig:NGC1574_conts}. As demonstrated by the maps in Figure~\ref{fig:NGC1574_moments_all}, the adopted model provides a very good fit to the data. In particular, by comparing Figures~\ref{fig:ngc1574_mom1} and \ref{fig:ngc1574_mom1_mod}, it is clear that the PA warp included in our model well reproduces most of the asymmetries observed in the rotation pattern. Nevertheless, some residuals are observed in the data-model velocity map (Fig.\,\ref{fig:NGC1574_mode_res}), with average velocities around $\approx\pm20$~km~s$^{-1}$ (corresponding to about $\pm40$~km~s$^{-1}$ deprojected\footnote{Observed velocities are projected along the line of sight. Intrinsic values can be estimated by correcting for the inclination of the gas distribution, i.e.\,$v_{\rm deproj}=v_{\rm obs}/\sin{i}$.}). This suggests that the adopted model is not a complete representation of the gas kinematics, and may point towards the presence of non-circular motions. We discuss these results in Section~\ref{sec:NGC1574_discuss}.

\subsubsection*{NGC\,4261}
As for NGC\,1574, we adopted the simple exponential disc parametrisation to reproduce the gas distribution observed in NGC\,4261. As anticipated in Section~\ref{sec:targets}, a CO dynamical SMBH mass measurement in this object has been recently reported by \citet[][]{Boizelle21}. To allow a coherent comparison with their results, we thus adopted the MGE solutions (constructed from an H-band mosaic) reported in Table~1 of \citet[][]{Boizelle21}, and reproduced in our Table~\ref{tab:NGC4261_MGE_param}. We fitted the disc inclination as single value throughout the disc and assumed the gas to be in purely circular motions. To account for the putative presence of a PA warp (see Fig.\,\ref{fig:NGC4261} and Sect.\,\ref{sec:cube_analysis}), we attempted a fit allowing the PA to vary with radius. The PA variation was implemented testing various functional forms (e.g.\,linear, exponential, etc.\,), but in all the cases the best-fitting model turned out to be consistent with no PA variation. This suggest that - if present - the warp is likely very mild and have only a minor effect on the observed CO disc geometry. We thus adopted a spatially-constant PA parameter. 
The gas velocity dispersion, $\sigma_{\rm gas}$, was also initially assumed to be constant throughout the disc. 

The final adopted model had 10 free parameters: the gas integrated flux density, PA and inclination ($Inc$), RA, DEC and velocity centres, exponential disc scale length ($R_{\rm disc}$),  the gas velocity dispersion ($\sigma_{\rm gas}$), the SMBH mass ($M_{\rm BH}$), and the $M/L$ (see Table~\ref{tab:MCMC-params}). We note that, while left free to vary within the whole physical range, the inclination parameter remained poorly constrained in our fitting, varying from very low to very high values in subsequent iterations. Since our aim here is just to cross-check an existing CO dynamical $M_{\rm BH}$, in our final model we fixed the inclination to the best-fitting value reported by \citet[][]{Boizelle21}.

\begin{figure*}
\centering
\includegraphics[scale=0.6]{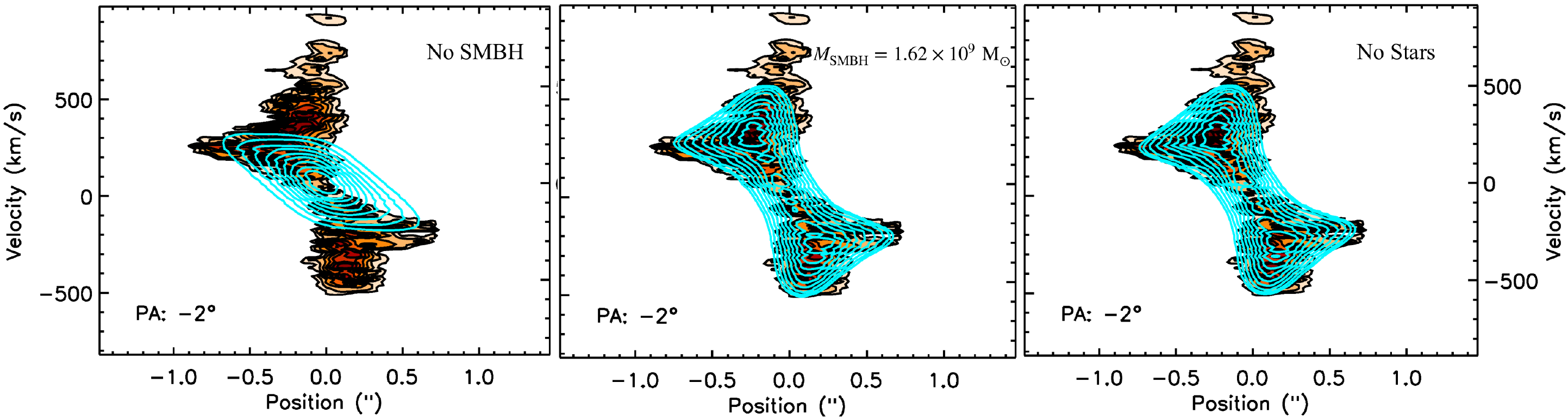}
\caption[]{Observed PVD along the kinematic major axis of NGC\,4261 with different models overlaid in cyan contours. The left panel shows a model without a SMBH, the central panel with the best-fitting SMBH and the right one a model with the $M/L$ ratio set to zero (see the text for details). \label{fig:NGC4261_PVDs_model}}
\end{figure*}

The best-fitting and formal uncertainties of each model parameter are listed in Table~\ref{tab:MCMC-params}. A comparison between the full set of the data moment maps and those extracted from the best-fitting simulated data cube is provided in Figure~\ref{fig:NGC4261_moments_all}. The corner plots and one-dimensional marginalisation of the non-nuisance model parameters for NGC\,4261 are shown in Figure~\ref{fig:NGC4261_conts}. We obtained a best-fitting SMBH mass of $(1.62{\pm 0.04})\times10^{9}$~M$_{\odot}$ (1$\sigma$ uncertainty), in agreement (within their respective errors) with the value reported by \citet[][]{Boizelle21}. The major axis PVD with the best-fitting model contours overlaid is shown in the middle panel of Figure~\ref{fig:NGC4261_PVDs_model}. Although the presence of an high-mass SMBH at the centre of this object is already well established, it is interesting to consider the same PVD with overlaid the contours of a model without a SMBH (Fig.\,\ref{fig:NGC4261_PVDs_model}, left panel), and of one obtained removing the contribution of luminous matter (Fig.\,\ref{fig:NGC1574_PVDs_model}, right panel). These plots, together with the very low and poorly constrained $M/L$ of $1.28^{+1.98}_{-1.14}$ obtained at the best fit (see Tab.\,\ref{tab:MCMC-params} and Fig.\,\ref{fig:NGC4261_conts}), clearly show that the observed CO kinematics is dominated by the SMBH gravitational influence.

We note that, while testing different scenarios to reproduce the observed gas kinematics, \citet[][]{Boizelle21} found some evidence for an exponentially-decaying gas velocity dispersion in NGC\,4261, with a central peak of 162~km~s$^{-1}$ and a much lower value of $\approx20$~km~s$^{-1}$ towards the edges of the gas disc. For comparison, we thus performed another fit allowing the velocity dispersion to vary with radius as
\begin{eqnarray}
\sigma_{\rm gas} (R) = \sigma_{0} + \sigma_{1}
	e^{-\dfrac{r}{R_{\rm t}}},  
\end{eqnarray}
where $R_{\rm t}$ is the scale length of the velocity dispersion, and $\sigma_{0}$ and $\sigma_{1}$ parametrise the radial variation. At the best fit, we obtain  $\sigma_{0}\approx143$~km~s$^{-1}$, $\sigma_{1}\approx21$~km~s$^{-1}$ and $R_{\rm t}\approx0.41''$, in agreement with what found by \citet[][]{Boizelle21}. These values, however, remained fairly unconstrained in our modelling, with posterior probability distributions that resulted quite flat within the search range. Furthermore, allowing for such a $\sigma_{\rm gas}$ gradient does not provide any evident improvement in our reproduction of the CO kinematics, and does not make any difference in the best-fitting SMBH mass value. We thus conclude that a velocity dispersion gradient is not required to reproduce the observations. Nevertheless, we found the best-fitting $\sigma_{\rm gas}$ to be $\approx60$~km~s$^{-1}$, quite large with respect to that typically expected (i.e.\,$\sigma_{\rm gas} \lesssim20$~km~s$^{-1}$; \citealp[e.g.][]{Davis17}). This may indicate that an ongoing perturbation (such as deviations from purely circular motions) is partly affecting the dynamical state of the gas. This scenario seems in agreement with the presence of a very high velocity structure (with $v>200$~km~s$^{-1}$) at the centre of the major axis PVD which is not reproduced by our model assuming gas in purely circular motions (Fig.\,\ref{fig:NGC4261_PVDs_model}). A discussion of the obtained results, including speculation on the origin of such high-velocity feature, is provided in Section~\ref{sec:NGC4261_discuss}.

\section{Discussion}\label{sec:discussion}

\subsection{SMBH mass measurement: main uncertainties}
The typical errors associated with a SMBH mass derived through the molecular gas technique are extensively discussed in previous papers of this series \citep[e.g.][]{Onishi17,Davis17,Davis18,North19,Smith19}. Each WISDOM paper builds from the previous ones and focuses on the sources of uncertainty that are relevant for the targeted object(s). In the following, we discuss the uncertainties on the SMBH mass estimates specific for each galaxy studied in this work. 

We note that in all the cases additional uncertainties on the best-fitting parameters may arise neglecting the possible contribution of the molecular gas itself (in addition to that of the stars and the SMBH) to the gravitational potential in the areas interested by the kinematic modelling. To ensure that this is not the case for our targets, we calculated the total mass enclosed within a certain radius (assuming spherical symmetry, $M_{\rm enc} = v^{2}(r)r/G$, where $v(r)$ is the de-projected rotational velocity at radius $r$ and $G$ is the gravitational constant). The radius has been set to be the maximum CO radius for NGC\,1574 and NGC\,4261 (that is $1.2\arcsec$ and $0.8\arcsec$, respectively), and $6\arcsec$ for NGC\,0612 (as $12\arcsec \times 12\arcsec$ is the central area within which we restricted the kinematic modelling; see Section~\ref{sec:mod_results}). We then compared the molecular gas mass (that is $M_{\rm mol}$ listed in Table~\ref{tab:line_parameters} for NGC\,1574 and NGC\,4261) with the total mass enclosed within those areas, and found that the contribution of $M_{\rm mol}$ is totally negligible for both NGC\,1574 and NGC\,4261 (maximum 5\% of the total enclosed mass). The same is not true for NGC\,0612, where $M_{\rm mol}$ within $12\arcsec \times 12\arcsec$ is $\approx7.8\times10^{10}$~M$_{\odot}$ and thus $\approx$50\% of the total enclosed mass. We tested the impact that such gas mass might have on our results by running the {\sc KinMS} models described in Section~\ref{sec:mod_results} both including and excluding the gravitational potential of the gas in the calculation of the circular velocity, and found that the best-fitting SMBH mass upper-limits remain essentially unchanged in all the cases. Therefore, we consider the results obtained for NGC\,0612 robust against this uncertainty.

We also note that all the specific uncertainties described below are usually rather small when compared to that introduced by the distance measurement. Dynamical SMBH mass estimates are indeed systematically affected by the assumed galaxy distance, with $M_{\rm BH} \propto D$. The distances for our targets have been assumed as described in Section~\ref{sec:targets}, and have typical formal uncertainties of the order of $\approx 10\%$. Since this is a simple normalisation, however, we adopt standard practice and do not include it in the systematic uncertainties of our $M_{\rm BH}$ estimates.

\subsubsection{NGC\,0612}\label{sec:NGC612_discuss}
The analysis of the combined CO(2-1) observations of NGC\,0612 presented in this paper allows us to confirm and complement the results obtained from the kinematic modelling of the sole intermediate-resolution data presented by \citet[][]{Ruffa19b}, and to place some constraints on the mass of the SMBH at the centre of this object.

Despite the angular resolution of the combined ALMA data analysed in this work allowing us to well resolve the predicted $R_{\rm SOI}$ (see Tab.\,\ref{tab:line_images} and Sect.\,\ref{sec:targets}), we do not detect any clear Keplerian rise around the centre of the CO velocity curve (Fig.~\ref{fig:ngc612_PVD}). The well-known radio activity of this galaxy provides strong evidence for the presence of an active SMBH at its centre \citep[e.g.][]{Ruffa19a}, therefore arguably ruling out the possibility that the nuclear potential could be simply not dominated by a SMBH. The most likely explanation for not detecting a clear sign of the presence of a central dark object here is that there is a CO deficiency in the vicinity of the SMBH. As anticipated in Sections~\ref{sec:mom_maps} and \ref{sec:mod_results}, the presence of a central hole in the CO disc of NGC\,0612 was already tentatively inferred by \citet{Ruffa19b}, reporting a best-fitting R$_{\rm hole}\gtrsim 0.3'' \approx180$~pc. The CO disc in NGC\,0612 is viewed almost edge-on (with $Inc=81^{\circ}$; Tab.\,\ref{tab:MCMC-params}), hence projection effects (i.e.\,line-of-sight passing through material at many different radii) make it challenging to obtain a robust estimate of R$_{\rm hole}$ also from the higher-resolution data presented in this paper. Nevertheless, a central CO cavity can be now clearly inferred by-eye in Figure~\ref{fig:NGC612} (see also Fig.\,\ref{fig:NGC612_moments_all_appendix}), and we estimate a best-fitting upper limit of $R_{\rm hole}\lesssim0.94\arcsec$. Combining the two result, we conclude that the size of the central CO hole is likely significantly larger than the predicted $R_{\rm SOI}$ ($\approx0.13''$ or 77~pc; see Sect.\,\ref{sec:targets}).

Cavities at the centre of the CO distribution are starting to be observed in a growing number of galaxies detected at high resolution with ALMA \citep[e.g.][]{Onishi17,Davis18,Combes19,Smith19,Garcia19,Ruffa19a,Smith21a}. One possibility to explain such central gas deficiency is that some mechanism dissociates the molecular gas or prevents it from forming (or accumulating) in the very centre of the galaxy. The dissociation of molecules can occur due to both AGN feedback and/or UV radiation from young stars. Both mechanisms are plausible in NGC\,0612, as it hosts an AGN with large-scale jets and form stars at a rate of 7~M$_{\odot}$~yr$^{-1}$ (see Sect.\,\ref{sec:targets}). When an AGN is present, it is also possible that there is molecular gas at the centre, but this is excited towards higher-$J$ transitions by X-ray radiation from the accretion disc or by interaction with the radio jets. When one of these mechanisms is ongoing, usually the central gap is filled when observing CO lines with $J_{\rm upper}\geq3$ \citep[e.g.][]{Wada18,Combes19,Ruffa22} or dense molecular gas tracers (such as HCO$^{+}$; e.g.\,\citealt[][]{Imanishi18}). 

Dynamical effects can also affect the distribution and survival of molecular gas in the nuclear regions of galaxies. For instance, the strong shear or tidal acceleration expected near the SMBH can disrupt gas clouds, where the molecular gas usually forms and survives thanks to its self-shielding from photo-dissociating radiation. Non-axisymmetric gravitational instabilities induced by stellar bars can also give rise to nuclear gaps, relying on bar-induced gravitational torques that force the gas outwards to the inner Lindblad resonance (ILR), or inwards on to the central SMBH \citep[e.g.][]{Combes01}. Bars in edge-on galaxies leave also their kinematical signature in molecular gas velocity curves, showing characteristic X-shapes or sharp edges near the turnover radius. These are indicative of the presence of two velocity components along the line of sight: an inner rapidly rising component associated with gas located within the ILR, and an outer slowly rising component associated with gas on nearly circular orbits beyond co-rotation \citep[e.g.][]{Alatalo13,Combes13}. The major axis PVD in Figure~\ref{fig:ngc612_PVD} shows a weak X-shape, tempting us to speculate on bar-induced gravitational instabilities giving rise to the central CO hole. However, follow-up observations of NGC\,0612 targeting different molecules and/or transitions of the same species are necessary to prove that the central cavity is truly devoid of gas, and thus draw solid conclusions on its formation.

Similar holes (and thus no clear Keplerian upturns) have been already observed also in other objects of the WISDOM sample \citep[see e.g.][]{Onishi17,Davis18,Smith19,Smith21a}. These works nevertheless demonstrated that accurate estimates of the SMBH mass can be obtained even in cases like these, provided that $R_{\rm SOI}$ is sufficiently large compared to $R_{\rm hole}$ and a high-accuracy stellar mass model can be obtained from optical imaging. This latter condition is necessary to ensure a reliable calculation of the velocity enhancement compared to that expected from the stellar mass alone. The prominent dust structure in NGC\,0612, however, makes its optical stellar light highly extincted even at infrared wavelengths (see Fig.\,\ref{fig:NGC612_optical} and Sect.\,\ref{sec:mod_results}), preventing us from obtaining a reliable MGE model and thus a robust prediction of the molecular gas kinematics. Although mostly physically sensible, our arbitrary parametrisations of the stellar mass density and - in turn - of the gas kinematics are necessarily less accurate than a full MGE photometric analysis. This, combined with the presence of a central CO void with a much larger radius than $R_{\rm SOI}$, likely explain our inability to put solid constraints on the NGC\,0612's SMBH mass, which is found to have a (conservative) upper limit of $M_{\rm BH}\lesssim3.2\times10^{9}$~M$_{\odot}$ (see also Sect.\,\ref{sec:msigma_discuss}).

\subsubsection{NGC\,1574}\label{sec:NGC1574_discuss}
We have presented ALMA+ACA CO(2-1) observations of NGC\,1574, showing the presence of a nearly face-on molecular gas disc concentrated within the inner $\approx200$~pc of the galaxy (Fig.\,\ref{fig:NGC1574} and Sect.\,\ref{sec:mom_maps}). These data allow us to provide for the first time evidence for the presence of a massive dark object at the centre of NGC\,1574, and obtain the first measure of its mass, $M_{\rm BH}=(1.0\pm0.2)\times10^{8}$~M$_{\odot}$ (1$\sigma$ uncertainty). The plots in Figure~\ref{fig:NGC1574_PVDs_model} clearly demonstrate the robustness of our finding (see Sect.\,\ref{sec:mod_results}). There are, however, a number of systematic uncertainties that must be taken into account in such a $M_{\rm BH}$ estimate. 

For nearly face-on objects such as NGC\,1574, the inclination constitutes a main source of uncertainty. This remains fairly unconstrained in our best-fitting model, with a sample truncated by the lower bound of the search range (see Sect.\,\ref{sec:mod_results} and Fig.~\ref{fig:NGC1574_conts}). The difficulty in determining the inclination in cases like this and its effect on the SMBH mass estimation have been extensively discussed in a previous paper of this series \citep[][]{Smith19}. We provide a brief summary in the following. 

Since we only observe the line-of-sight projection of the gas velocities (i.e.\,$v_{\rm obs} \sin{i}$), SMBH mass uncertainties are strictly correlated to those of the inclination (with $M_{\rm BH} \propto 1/\sin^{2}{i}$). In cases like NGC\,1574, the inclination uncertainties thus typically dominates the error budget of the estimated SMBH mass, and induce large uncertainties also on the $M/L$. The posterior probability distributions in Figure~\ref{fig:NGC1574_conts} indeed show a clear degeneracy between inclination and $M_{\rm BH}$ or $M/L$, and a covariance between $M_{\rm BH}$ and $M/L$ (although this latter is exaggerated by plotting linear against logarithmic scales). Contrarily to what expected, the latter is a positive correlation, and it is also clear from the contours in Figure~\ref{fig:NGC1574_conts} that the $M/L$ remains poorly constrained at the best-fit. All these uncertainties are likely a consequence of the $M_{\rm BH}$–Inc and $M/L$–Inc correlations.
We nevertheless note that, despite such large inclination-related uncertainties, $M_{\rm BH}$ varies by only $\pm0.3$~dex within the inclination range (as demonstrated by the narrow $3\sigma$ confidence interval in Fig.\,\ref{fig:NGC1574_conts}), making us confident of our estimation.

Other potential uncertainties may arise from assumptions on the dynamical state of the gas. The best-fitting average velocity dispersion is consistent with the bulk of the gas being dynamically cold ($\sigma_{\rm gas} = 15.5\pm3.6$~km~s$^{-1}$; Tab.\,\ref{tab:MCMC-params}). This also indicates that the CO disc is mostly rotationally supported with $v_{\rm rot}/\sigma_{\rm gas} =14$, where $v_{\rm rot}$ is the de-projected rotation velocity of the gas in the nearly flat portion of the rotation curve (see Fig.~\ref{fig:ngc1574_PVD_main}) and $v_{\rm rot}/\sigma_{\rm gas} \geq 10$ are typically associated to relaxed molecular gas discs in the nearby Universe \citep[see e.g.][]{Wisnioski15}. The evident distortions in the rotation pattern turned out to be well described by the presence of a PA warp (Sect.\,\ref{sec:mod_results} and Fig.\,\ref{fig:NGC1574_moments_all}). In addition to that, however, non-circular motions may be present. The map in Figure~\ref{fig:NGC1574_mode_res} indeed shows some velocity residuals with peaks of $\approx\pm60$~km~s$^{-1}$ (de-projected). Some non-circular motions can occur in non-axisymmetric potentials induced by stellar bars (such as that hosted in the central regions of NGC\,1574; see Sect.\,\ref{sec:targets}), giving rise to a shock front along the bar edges and leading to an inward/outward flow of material that can also potentially fuelling the central SMBH (see also \citealt[][]{Ruffa19b}). It is however worth noting here that, while well reproducing most of the observed velocity asymmetries, our simple parametrisation of the PA warp is completely arbitrary and the model is axi-symmetric by construction. Velocity residuals like those observed in Figure~\ref{fig:NGC1574_mode_res} may thus also arise from the small discrepancies between the data and the model that can be introduced by these two factors. Nevertheless, the clear detection of symmetric Keplerian upturns at the centre of the CO velocity curve, the relatively low formal uncertainties of our $M_{\rm BH}$ determination, and the fact that the amplitude of radial motions - if present - would be low compared to that of the underlying circular velocity ($\approx25$\% at their peak), indicate that our $M_{\rm BH}$ determination is robust against all these uncertainties.

\subsubsection{NGC\,4261}\label{sec:NGC4261_discuss}
We carried out the CO dynamical modelling in NGC\,4261 with the primary aim of exploiting the higher resolution and sensitivity of our combined ALMA+ACA dataset to cross-check (using also a different modelling technique) the CO dynamical $M_{\rm BH}$ estimate recently reported by \citet[][]{Boizelle21}. At the best fit, we obtain $(1.62{\pm 0.04})\times10^{9}$~M$_{\odot}$ (1$\sigma$ uncertainty), in agreement with such previous work (reporting $M_{\rm BH}=1.67\times10^{9}$~M$_{\odot}$), and much larger than that previously obtained from ionised gas dynamical modelling (i.e.\,$M_{\rm BH}=5\times10^{8}$~M$_{\odot}$; \citealt[][]{Ferrarese96,Humphrey09}). We refer the reader to \citet[][]{Boizelle21} for a detailed discussion on the origin of the discrepancy between recent and earlier measurements and on the general uncertainties of the CO dynamical modelling in NGC\,4261, and focus here on discussing the uncertainties emerged from our modelling. 

We find weak constraints on the H-band $M/L$ with a best-fitting value of $1.28^{+1.98}_{-1.14}$ ($3\sigma$ uncertainties). The posterior probability distribution of this parameter is asymmetric (Fig.~\ref{fig:NGC4261_conts}) and truncated by the lower bound of the $M/L$ ratio prior (which is essentially zero). This, in principle, could introduce large uncertainties in the derived $M_{\rm BH}$, as expected from the covariance between the two parameters (equivalent to the conservation of total dynamical mass; see \citealt[][]{Smith19} for a detailed discussion on this issue). It is clear, however, from the plots in Figures~\ref{fig:NGC4261_conts} and \ref{fig:NGC4261_PVDs_model} that the effect of such co-variance is very small and the SMBH mass determination is largely independent of it. Thanks to the high spatial resolution of our data, we fully resolve $R_{\rm SOI}$ (see Sect.\,\ref{sec:targets}) and thus unambiguously detect the symmetric Keplerian velocity features due to the presence of a central massive dark object. This, together with the evident low stellar contribution at the probed scales, makes our SMBH mass almost independent of the luminous mass model and - in turn - leads to a $M_{\rm BH}$ determined with relatively small uncertainties (as demonstrated by the Gaussian and narrow distribution in Fig.\,\ref{fig:NGC4261_conts}). We nevertheless note that for NGC\,4261 an accurate r-band stellar $M/L=8.5$ (5\% error) has been reported by \citet{Cappellari13b}. It is therefore interesting to verify how such value compares with our H-band CO estimate in the central galaxy regions. To this aim, we used an r-band (AB system) SDSS image \citep{York00} of NGC\,4261 in combination with an H-band (Vega system) 2MASS image \citep{Skrutskie06} to measure a colour $r-H=2.85$ within a $r=10''$ circular aperture centred on the galaxy nucleus. We then adopted the solar luminosity $M_{\odot,H}=3.37$~mag given in Table~1 of \citet{Boizelle21} and the $M_{\odot,r}=4.64$~mag from Table~1 of \citet{Cappellari13a}. Plugging these values into Equation (22) of \citet{Cappellari16}, we accurately converted the stellar $\lg(M/L)_{\rm stars}=0.820$ ($M_{\odot}/L_{\odot,r}$) (6\% uncertainty) given in \citet{Cappellari13b} into an H-band value $M/L=1.54\pm0.09$ ($M_{\odot}/L_{\odot,H}$). This value agrees within $1\sigma$ with our H-band estimate from CO dynamical modelling.

\begin{figure*}
\centering
\includegraphics[scale=0.55]{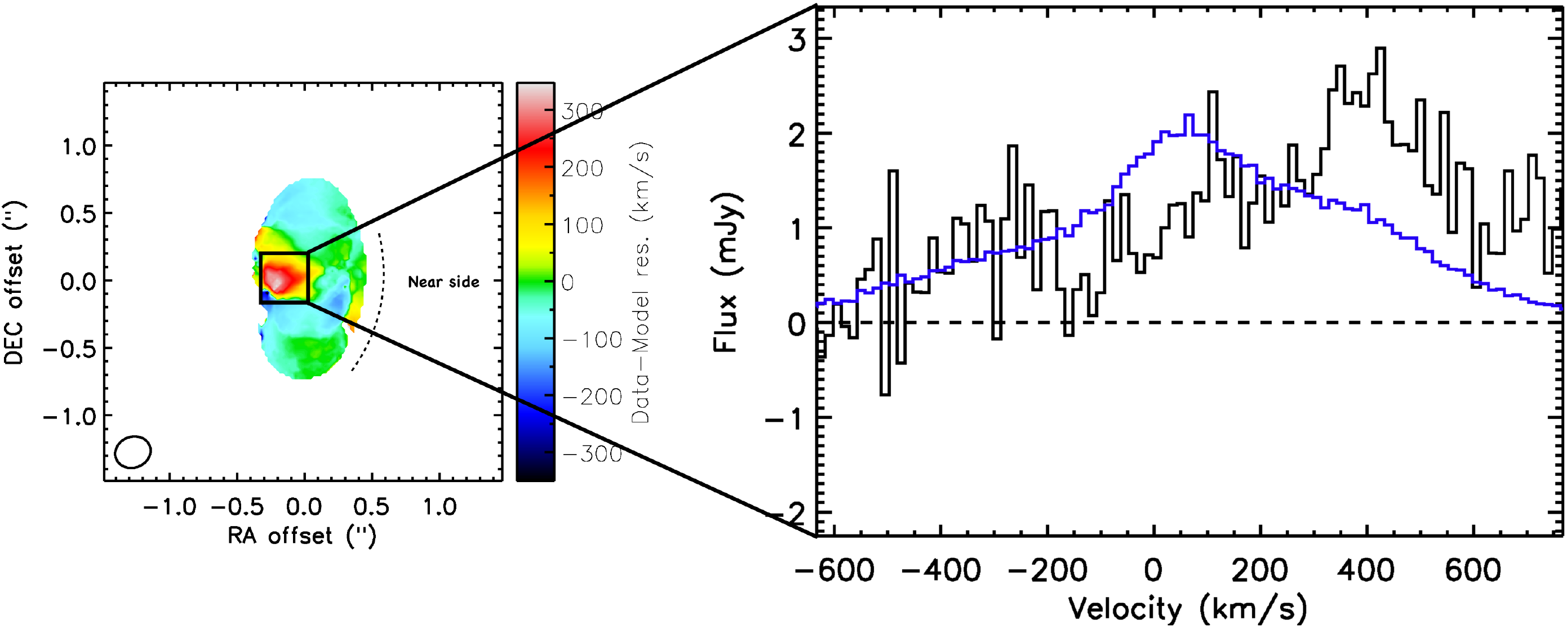}
\caption[]{{\bf Left panel:} Residual (data - model) mean velocity map of the CO in NGC\,4261. The curved black dashed line marks the near side of the CO disc as inferred from dust extinction in Figure~\ref{fig:NGC4261_optical}. The synthesized beam is shown in the bottom-left corner of the panel. The bar to the right show the colour scale in km~s$^{-1}$. East is to the left and North to the top. {\bf Right panel:} CO(2-1) spectral profile extracted from the cleaned data cube within the box illustrated on the left panel and enclosing the region of the most significant data-model velocity residuals. The corresponding spectral profile from our best-fitting model is overlaid in blue. The black dashed horizontal line indicates the zero flux level.}\label{fig:NGC4261_mod_res}
\end{figure*}

As discussed for NGC\,1574, other potential uncertainties arise from the assumptions that the molecular gas is dynamically cold and rotating only on circular orbits. The best-fitting average value of the intrinsic gas velocity dispersion is consistently large in this case ($\sigma_{\rm gas}\approx60$~km~s$^{-1}$), but this is clearly dominated by the increase observed in the very inner regions of the gas distribution (Fig.\,\ref{fig:ngc4261_mom2_main}). This may indicate that an ongoing perturbation (such as deviations from purely circular motions) may partly affect the dynamical state of the gas in those areas, inducing turbulence within the gas clouds. This scenario seems in agreement with the presence of very high velocity structures (with $v>200$~km~s$^{-1}$) at the centre of the observed major axis PVD (positive velocities) which are not reproduced by our model (Fig.\,\ref{fig:NGC4261_PVDs_model}). Very high residuals can be also observed at the same location in the data-model velocity map (Fig.\,\ref{fig:NGC4261_mod_res}, left panel). As a further check, we then extracted the spectral profile from the cleaned CO data cube within the region enclosing such residual velocities. This is illustrated in Figure~\ref{fig:NGC4261_mod_res}, with overlaid the corresponding spectral profile from our best-fitting model. It is clear from this plot that there is an additional, high-velocity spectral component that is not reproduced by our model. Similar features are usually related to local gas flows and indicate departures from pure circular rotation \citep[see e.g.][]{Dominguez20,North21,Ruffa22}. By comparing the velocities of the broad kinematic component with the near side of the disc (as inferred from dust extinction), we suggest that such component may trace a circumnuclear gas outflow. However, the clear dominance of the SMBH in the central regions (see Fig.\,\ref{fig:NGC4261_PVDs_model}) and the very low formal uncertainties of our $M_{\rm BH}$ determination, suggest that such putative non-circular motions are unlikely to significantly affect the derived SMBH mass.

Similar deviations from purely circular motions in gas distributions can be induced by various mechanisms. Among these, there are jet-ISM interactions on (sub-)kpc scales \citep[e.g.][]{Combes13,Garcia14,Oosterloo17,Ruffa20,Ruffa22}. A growing number of both observations and simulations indeed show that small scale jet–gas interactions can impact the gas distribution (by - e.g.\,- locally compressing or fragmenting gas clouds) and produce turbulent cocoons of shocked gas that can be accelerated up to $\approx$1000~km~s$^{-1}$ over a wide range of directions with respect to the radio jet axis (which in this case is nearly perpendicular to the gas disc; e.g.\,\citealt[][]{Temi22}). This process usually gives rise to features similar to those seen in the PVD of NGC\,4261 \citep[see e.g.][and references therein]{Oosterloo17,Mukherjee18a,Mukherjee18b,Murthy19,Zovaro19,Ruffa22}. Higher-resolution observations of both the radio jet and molecular gas components in NGC\,4261 would be needed, however, to accurately test this hypothesis and draw solid conclusions on this aspect. It is nevertheless interesting to note that disturbances such as those found in the CO kinematics of NGC\,4261 are not commonly seen in the cold gas components of local, non-jetted ETGs (see e.g.\,\citealt[][]{Ruffa19a} for a detailed discussion on this aspect). In particular, the emerging picture is that CO discs in LERGs such as NGC\,4261 are systematically more disturbed than in radio-quiet objects \citep[e.g.][]{Boizelle17,Ruffa19a,Ruffa19b,North19,North21}. This provides additional hints on the impact of jet-induced AGN feedback onto the surrounding ISM and - more in general - on the interplay between the central active SMBH and its host galaxy.

\subsection{The $M_{\rm BH}-\sigma_{\rm \filledstar}$ relation: comparison to literature}\label{sec:msigma_discuss}
\begin{figure}
\includegraphics[scale=0.39]{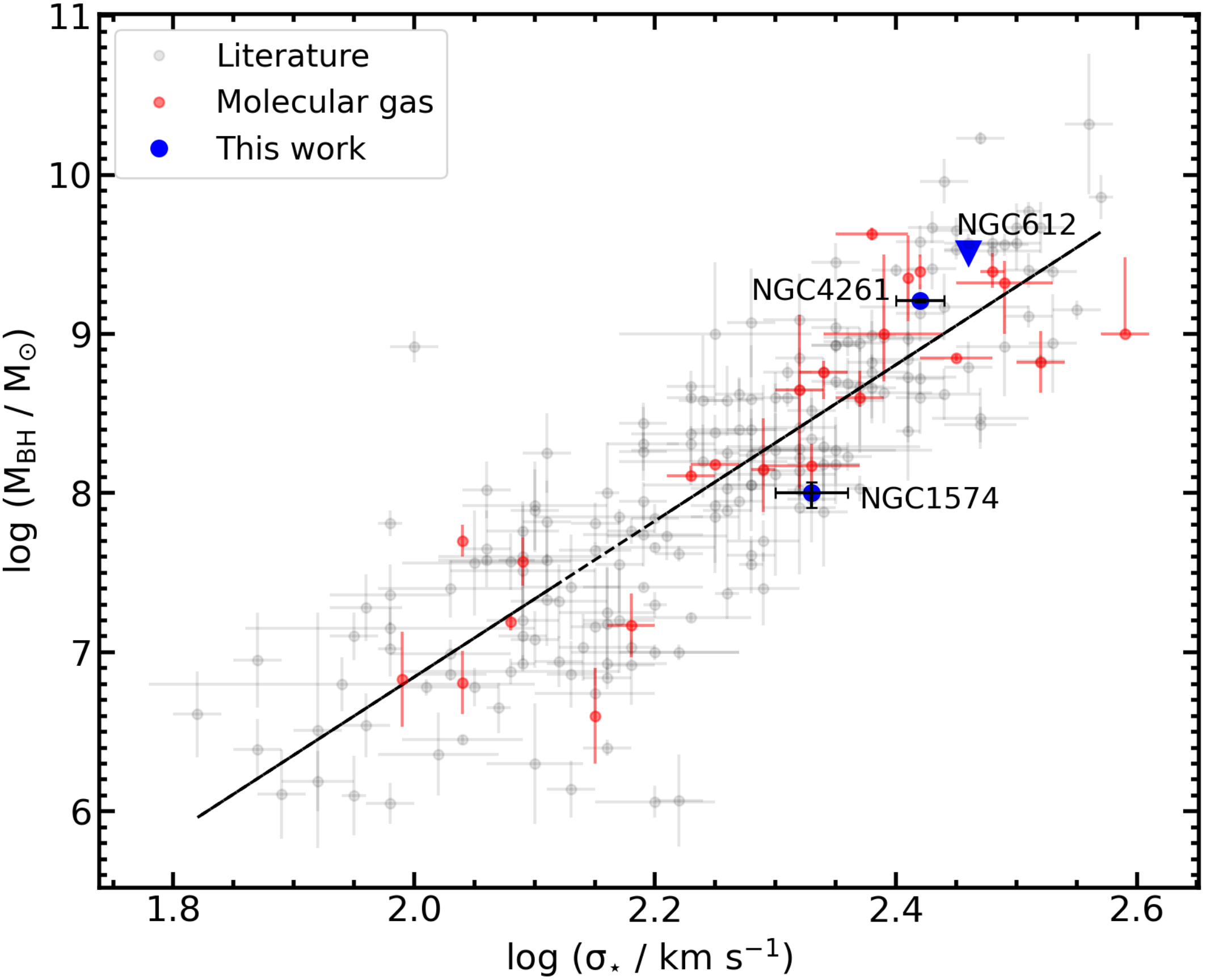}
\caption[]{$M_{\rm BH}-\sigma_{\rm \filledstar}$ relation from the compilation of \citet[][]{Vandenbosch16} (grey data points and black dashed line). The BH mass measurements/upper limit obtained in this work for NGC0612, NGC\,1574 and NGC\,4261 are shown as large blue data points (as highlighted also by the corresponding text labels), while other SMBH mass measurements obtained using the molecular gas technique are plotted in red. For CO dynamical SMBH measurements, the error bars correspond to $1\sigma$ uncertainties. See the text for details.}\label{fig:M_sigma}
\end{figure}

To investigate how our measurements compare to the overall $M_{\rm BH}-\sigma_{\rm \filledstar}$ relation, in Figure~\ref{fig:M_sigma} we show the location of the objects studied in this work on the BH mass measurements and best-fitting line from the compilation of \citet[][]{Vandenbosch16}. Large blue circles/triangle are used to highlight our BH mass estimates for NGC\,0612, NGC\,1574 and NGC\,4261. Other existing SMBH mass measurements from studies of the molecular gas kinematics are shown as red circles \citep[and taken from][]{Davis13a,Onishi15,Barth16,Onishi17,Davis17,Davis18,Smith19,Combes19,Nagai19,North19,Boizelle19,Ruffa19b,Nguyen20,Boizelle21,Smith21b,Cohn21,Kabasares22}. 

The upper limit on NGC\,0612 and the estimate on NGC\,4261 lie slightly above the best-fitting $M_{\rm BH}-\sigma_{\rm \filledstar}$ relation of \citet[][]{Vandenbosch16}, but well within the scatter. NGC\,1574 is also still located within the scatter, but at the lower edge of the relation. If correct, this would indicate that - at the assumed $\sigma_{\rm \filledstar}$ - the BH at the centre of NGC\,1574 is under-massive with respect to what expected (by about a factor of two; see Sect.\,\ref{sec:targets} and Tab.\,\ref{tab:MCMC-params}). The relatively large error bars on the assumed $\sigma_{\rm \filledstar}$ prevent us from drawing solid conclusions from this comparison. Nevertheless, on a pure speculative basis, it is interesting to note that similar results would suggest differences in the relative BH-host galaxy interplay for ETGs harbouring an AGN (such as NGC\,0612 and NGC\,4261) and those in which instead the SMBH has not yet switched on (such as NGC\,1574). This - in turn - would provide essential information on the role of AGN feeding/feedback processes in the overall BH-host galaxy co-evolution.

\section{Summary}\label{sec:conclusions}
As part of the WISDOM series, we present an accurate CO kinematic modelling in the three nearby ETGs NGC\,0612, NGC\,1574 and NGC\,4261 using ALMA+ACA observations of the $^{12}$CO(2-1) line with spatial resolutions of 14-58~pc. The primary aim of this work was to obtain molecular gas dynamical measurements of the SMBH mass in these objects. The main results can be summarised as follows:
\begin{itemize}
\item In NGC\,0612, we detect a nearly edge-on, clumpy disc of molecular gas, extending
up to $\approx10$~kpc along its major axis. At the centre of this gas distribution a clear gap is observed with an estimated radius $R_{\rm hole}\lesssim0.94\arcsec$, thus likely much larger than the predicted $R_{\rm SOI}$ ($\approx0.13\arcsec$). This would explain why - despite this latter being well resolved at the angular resolution of our ALMA+ACA data - we do not detect any clear sign of the presence of a dark massive object at the centre of NGC\,0612. The disc is found to be mostly regularly rotating, with signs of a mild warp on large scales. Using a publicly-available HST WFC3 F160W image, we attempted to construct a stellar mass model (from which the gas circular velocity can be predicted). The prominent dust distribution present in the equatorial plane of the galaxy, however, leads to a substantial extinction even at such infrared wavelengths, preventing us from obtaining a robust prediction of the CO kinematics. After testing different arbitrarily-chosen parametric functions, we were able to acceptably reproduce the observed velocity and set a conservative upper limit of $M_{\rm BH}\lesssim3.2\times10^{9}$~M$_{\odot}$. This is in agreement with predictions from the $M_{\rm BH}-\sigma_{\rm \filledstar}$ relation ($M_{\rm BH}\approx1.5 \times 10^{9}$~M$_{\odot}$) and well within its scatter. 
\item In NGC\,1574, the CO gas is distributed in a disc, extending $<200$~pc around the optical nucleus of the galaxy. The line-of-sight CO velocity shows that the gas is rotating, but with evident kinematic distortions (i.e.\,s-shaped iso-velocity contours) that we found arising mainly from a position angle warp. In addition, low-amplitude non-circular motions might also be present ($\leq20$\% of the underlying circular velocity). Mild, symmetric velocity increases are clearly detected around the centre of the gas velocity curve, and can be plausibly associated with the Keplerian upturn arising from material orbiting within the SMBH SOI. This, together with an accurate prediction of the gas circular velocity caused by the luminous matter, allowed us to obtain the first measure of the SMBH mass in NGC\,1574, $M_{\rm BH}=(1.0\pm0.2)\times10^{8}$~M$_{\odot}$ (1$\sigma$ uncertainty). Although with some uncertainties, this value places NGC\,1574 at the lower edges of the $M_{\rm BH}-\sigma_{\rm \filledstar}$ relation, but still within its scatter. 
\item In NGC\,4261, the CO gas is distributed in a small-scale, disc-like structure, extending up to $\approx250$~pc along its major axis. The mean line-of-sight velocity and major axis velocity curve clearly indicate that the central gas kinematics is dominated by the SMBH gravitational influence, allowing us to determine an accurate SMBH mass of $M_{\rm BH}=(1.62{\pm 0.04})\times10^{9}$~M$_{\odot}$. This is fully consistent with a previous CO dynamical SMBH mass estimate in this object, and allows us to perform a crucial cross check between the different methodologies developing in this field. Signs of non-circular motions (likely outflow) are also identified in the central CO regions of this object. The derived $M_{\rm BH}$ places NGC\,4261 slightly above the $M_{\rm BH}-\sigma_{\rm \filledstar}$ relation, but well within its scatter.
\end{itemize}

\section*{Acknowledgements}
IR and TAD acknowledge support from the UK Science and Technology Facilities Council through grants ST/S00033X/1 and ST/W000830/1. TGW acknowledges funding from the European Research Council (ERC) under the European Union’s Horizon 2020 research and innovation programme (grant agreement No. 694343). This paper makes use of the following ALMA data: ADS/JAO.ALMA\#[2015.1.00419.S], \#[2015.1.01572.S], \#[2016.2.00046.S], \#[2016.2.00053.S], \#[2017.1.00301.S], \#[2017.1.00904.S], and \#[2018.1.00397.S]. ALMA is a partnership of ESO (representing its member states), NSF (USA) and NINS (Japan), together with NRC (Canada), NSC and ASIAA (Taiwan), and KASI (Republic of Korea), in cooperation with the Republic of Chile. The Joint ALMA Observatory is operated by ESO, AUI/NRAO and NAOJ. The National Radio Astronomy Observatory is a facility of the National Science Foundation operated under cooperative agreement by Associated Universities, Inc. This paper has also made use of the NASA/IPAC Extragalactic Database (NED) which is operated by the Jet Propulsion Laboratory, California Institute of Technology under contract with NASA. This research used the facilities of the Canadian Astronomy Data Centre operated by the National Research Council of Canada with the support of the Canadian Space Agency. We acknowledge the usage of the HyperLeda database (\url{http://leda.univ-lyon1.fr}). This research made also use of {\tt Astropy} \citep[][]{Astropy13,Astropy18}, {\tt Matplotlib} \citep[][]{Hunter07}, and {\tt NumPy} \citep[][]{Walt11,Harris20}. 

\section*{Data Availability}
The ALMA data used in this article are available to download at the ALMA archive (\url{https://almascience.nrao.edu/asax/}). The calibrated data, final data products and original plots generated for the research study underlying this article will be shared upon reasonable request to the first author.



\bibliographystyle{mnras}
\bibliography{mybibliography} 




\appendix

\section{Moment maps}\label{sec:mom_maps_appendix}

We provide here the full set of observed and best-fitting mock moment maps for the three galaxies studied in this work. We note that for NGC\,0612 the plotted area is restricted to the one within which the CO modelling has been carried out (i.e.\,the inner $12\arcsec\times12\arcsec$; see Section~\ref{sec:mod_results} for details).

\begin{figure*}
\begin{subfigure}[t]{0.3\textheight}
\centering
 \caption{Data moment zero}\label{fig:ngc612_mom0_appendix}
\includegraphics[scale=0.52]{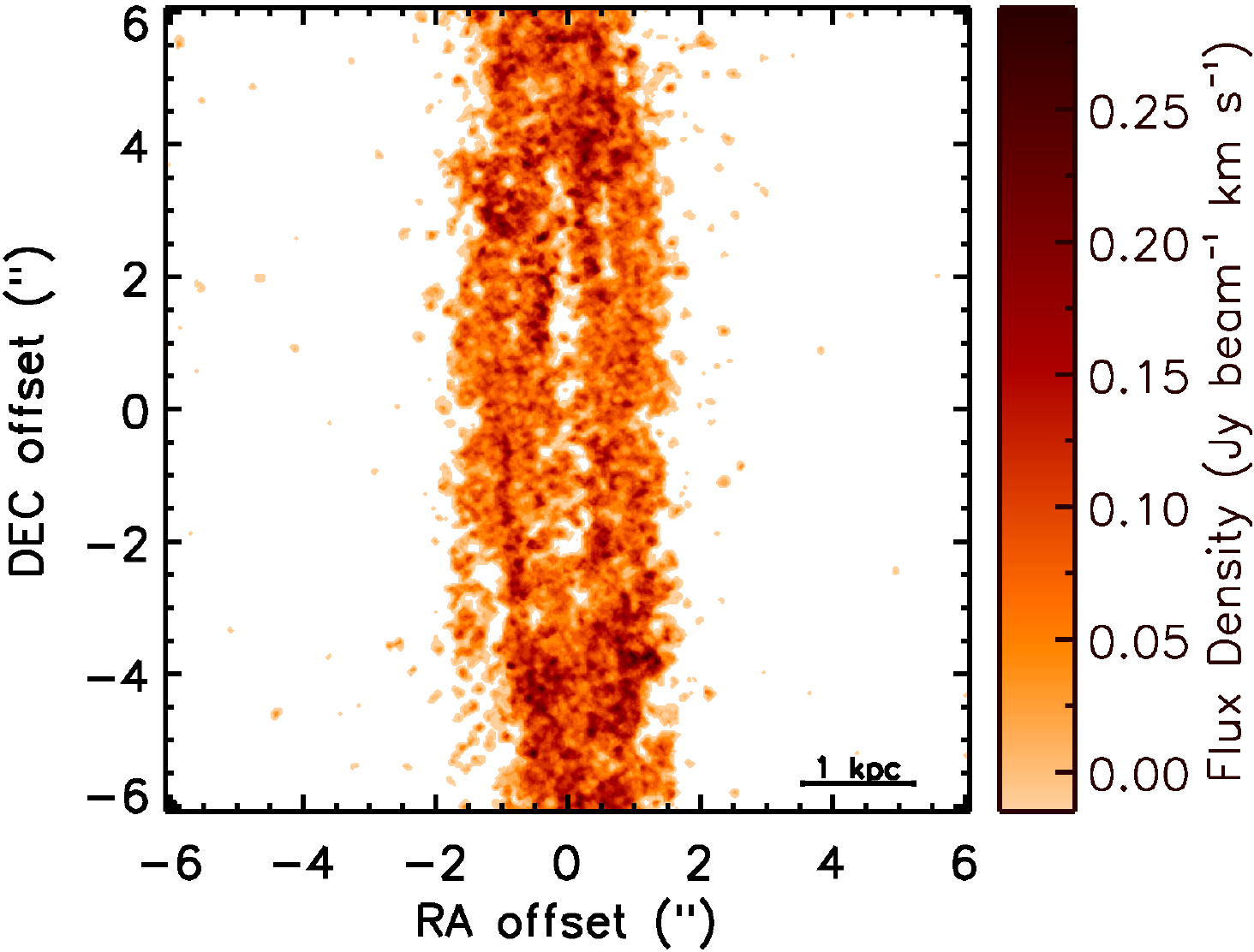}
\end{subfigure}
\hspace{5mm}
\begin{subfigure}[t]{0.3\textheight}
\centering
\caption{Model moment zero}\label{fig:ngc612_mom0_mod_appendix}
\includegraphics[scale=0.52]{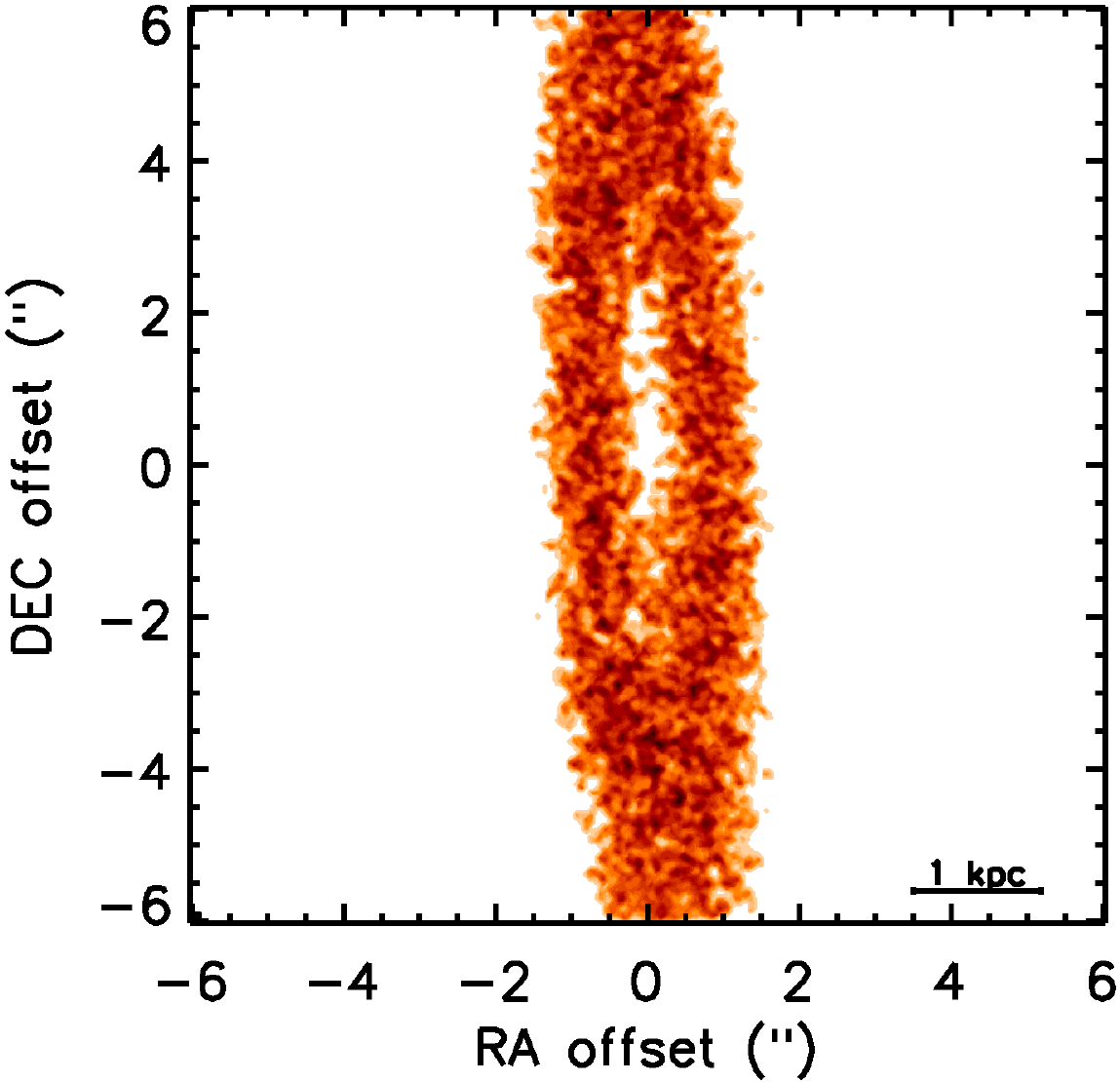}
\end{subfigure}

\medskip

\begin{subfigure}[t]{0.3\textheight}
\centering
 \caption{Data moment one}\label{fig:ngc612_mom1_appendix}
\includegraphics[scale=0.52]{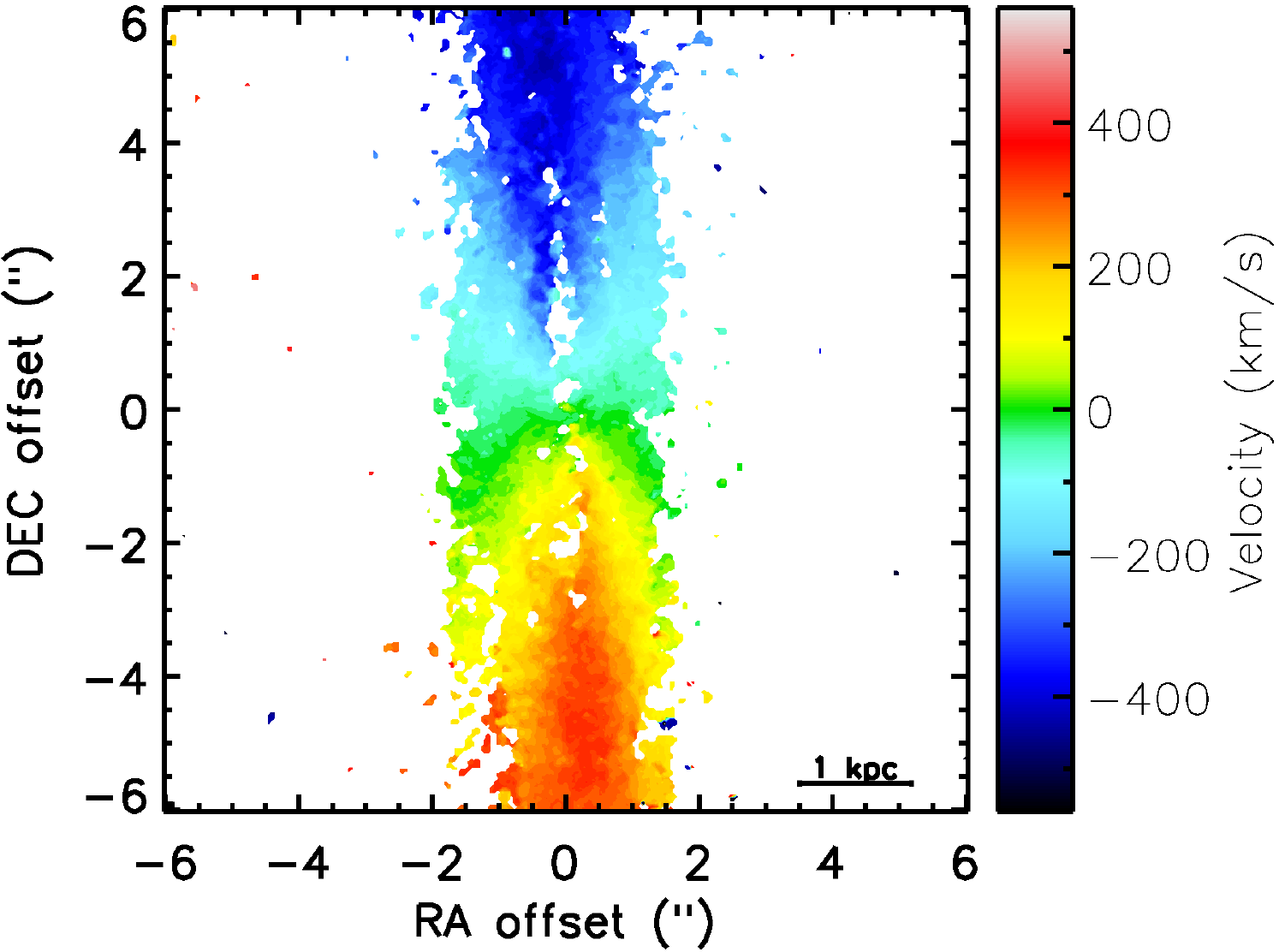}
\end{subfigure}
\hspace{6mm}
\begin{subfigure}[t]{0.3\textheight}
\centering
\caption{Model moment one}\label{fig:ngc612_mom1_mod_appendix}
\includegraphics[scale=0.33]{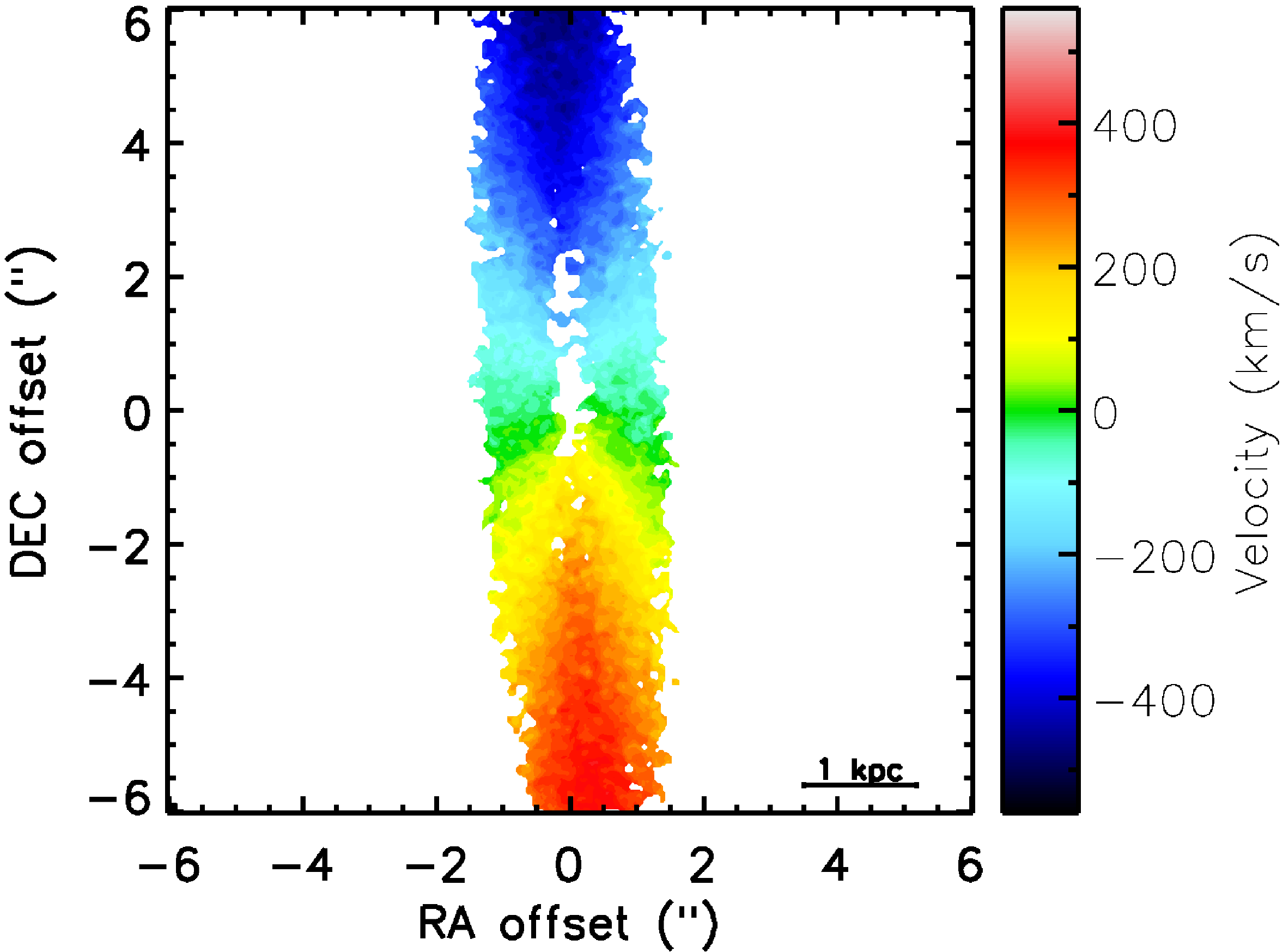}
\end{subfigure}

\medskip

\begin{subfigure}[t]{0.3\textheight}
\centering
\caption{Data moment two}\label{fig:ngc612_mom2_appendix}
\includegraphics[scale=0.52]{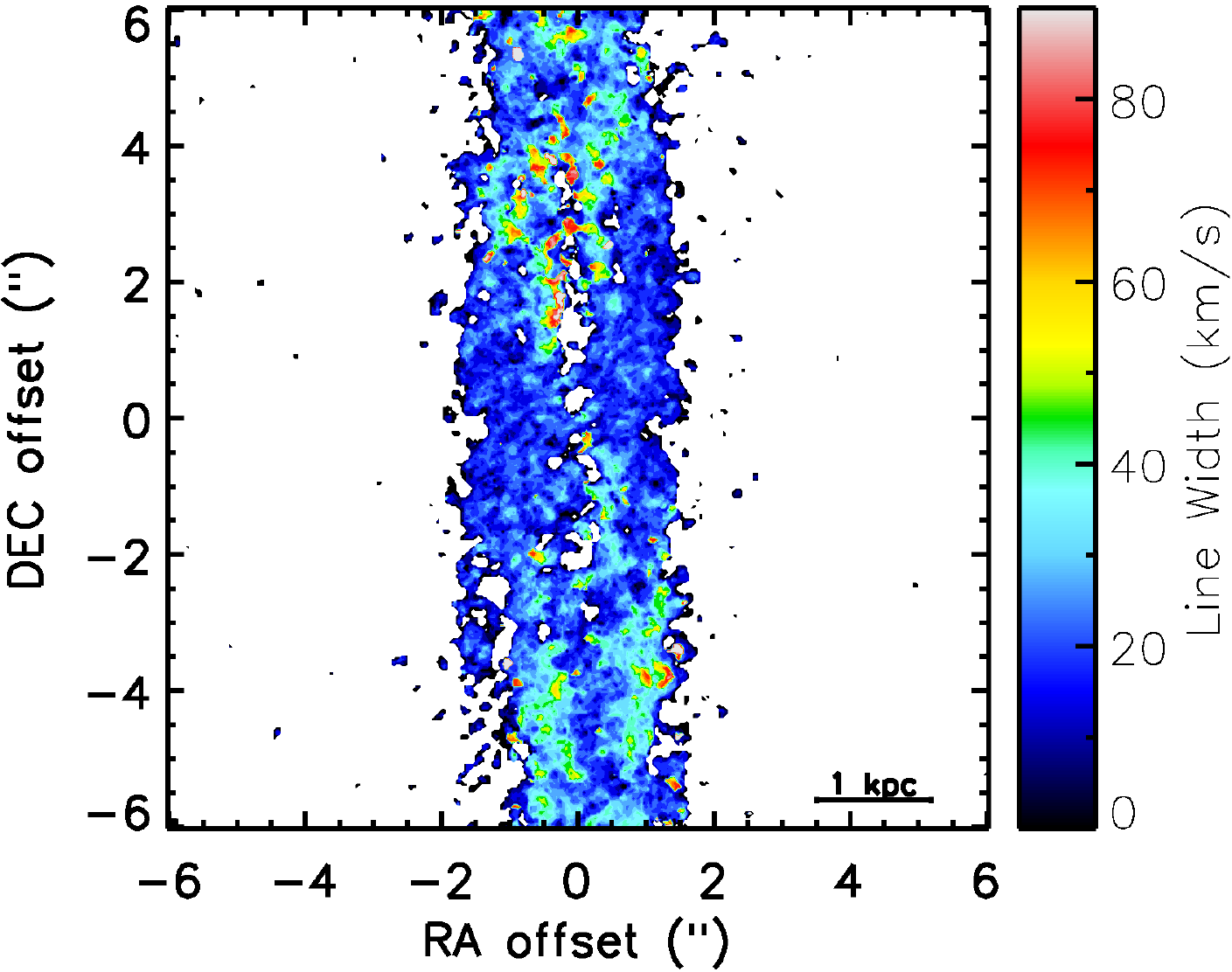}
\end{subfigure}
\hspace{5mm}
\begin{subfigure}[t]{0.3\textheight}
\centering
\caption{Model moment two}\label{fig:ngc612_mom2_mod_appendix}
\includegraphics[scale=0.52]{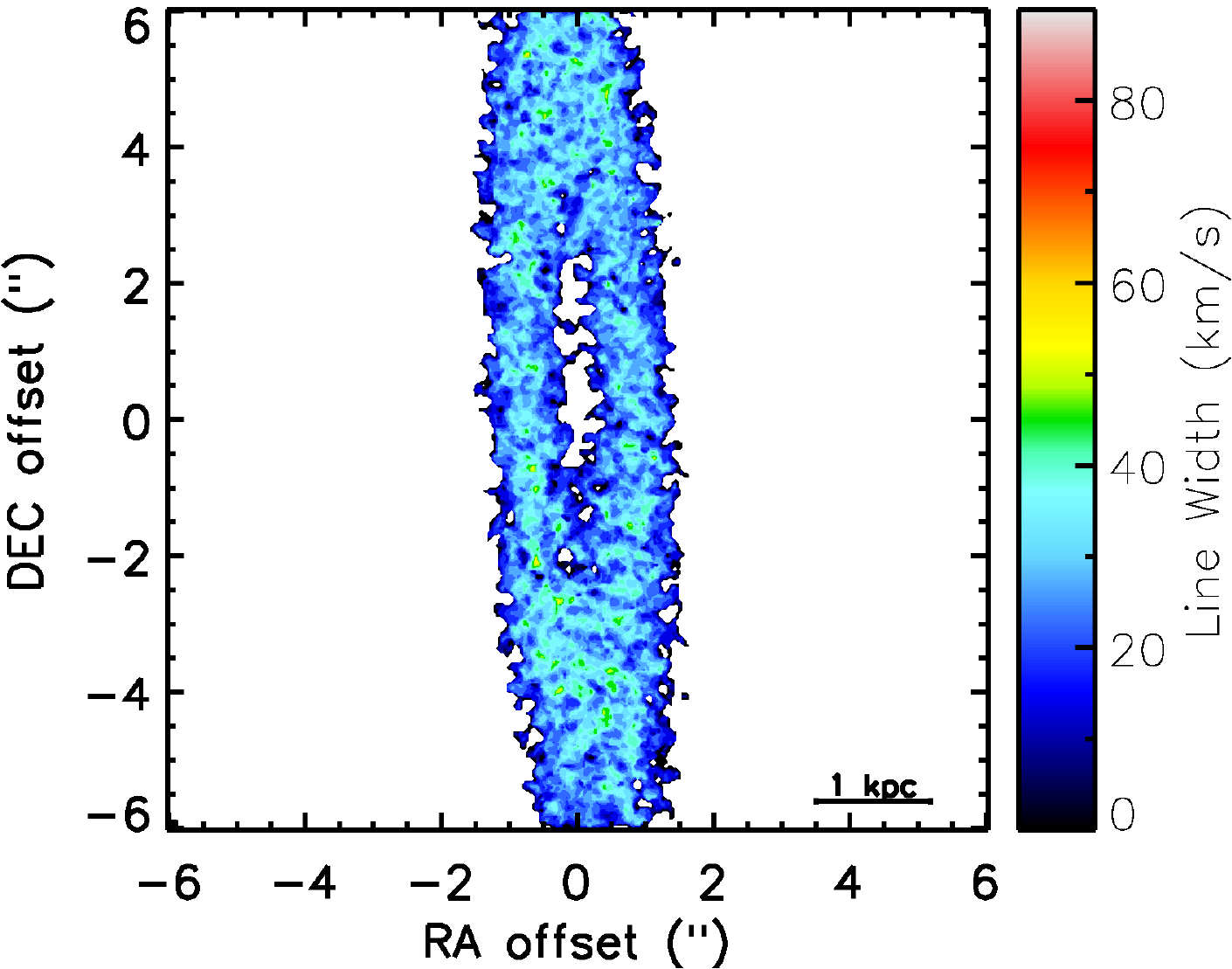}
\end{subfigure}
\caption{CO(2-1) integrated intensity (moment 0), mean line-of-sight velocity (moment 1) and velocity dispersion (moment 2) maps of NGC\,0612. The moments extracted from the observed data cubes are in the left-hand panels, while those extracted from the simulated data cubes of one of the three tested models (i.e.\,the Sersic model) are in the right-hand panels.}\label{fig:NGC612_moments_all_appendix}
\end{figure*}

\begin{figure*}
\begin{subfigure}[t]{0.3\textheight}
\centering
 \caption{Data moment zero}\label{fig:ngc1574_mom0}
\includegraphics[scale=0.52]{ngc1574_mom0.pdf}
\end{subfigure}
\hspace{10mm}
\begin{subfigure}[t]{0.3\textheight}
\centering
\caption{Model moment zero}\label{fig:ngc1574_mom0_mod}
\includegraphics[scale=0.3]{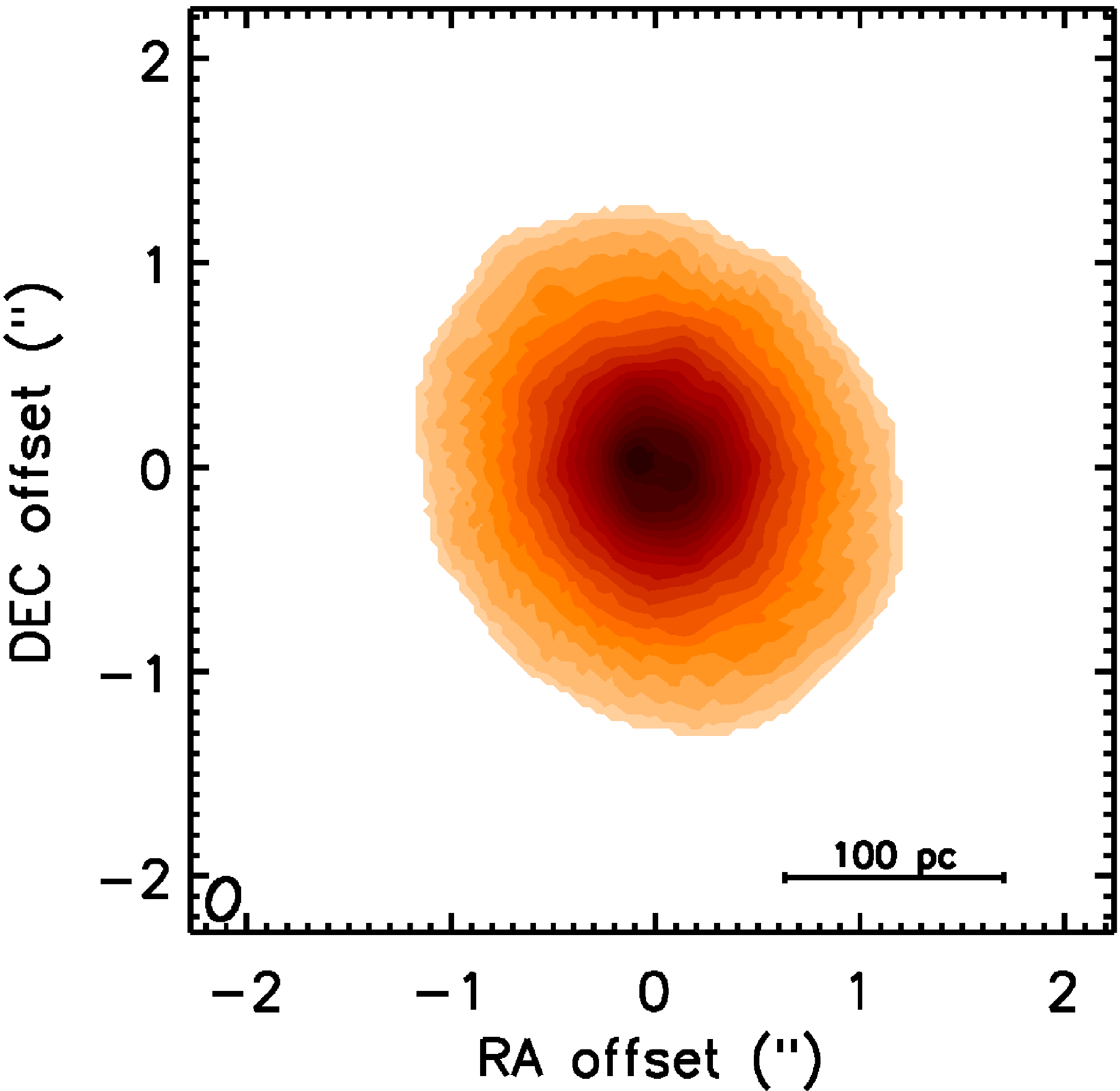}
\end{subfigure}

\medskip

\begin{subfigure}[t]{0.3\textheight}
\centering
 \caption{Data moment one}\label{fig:ngc1574_mom1}
\includegraphics[scale=0.52]{ngc1574_mom1.pdf}
\end{subfigure}
\hspace{12mm}
\begin{subfigure}[t]{0.3\textheight}
\centering
\caption{Model moment one}\label{fig:ngc1574_mom1_mod}
\includegraphics[scale=0.3]{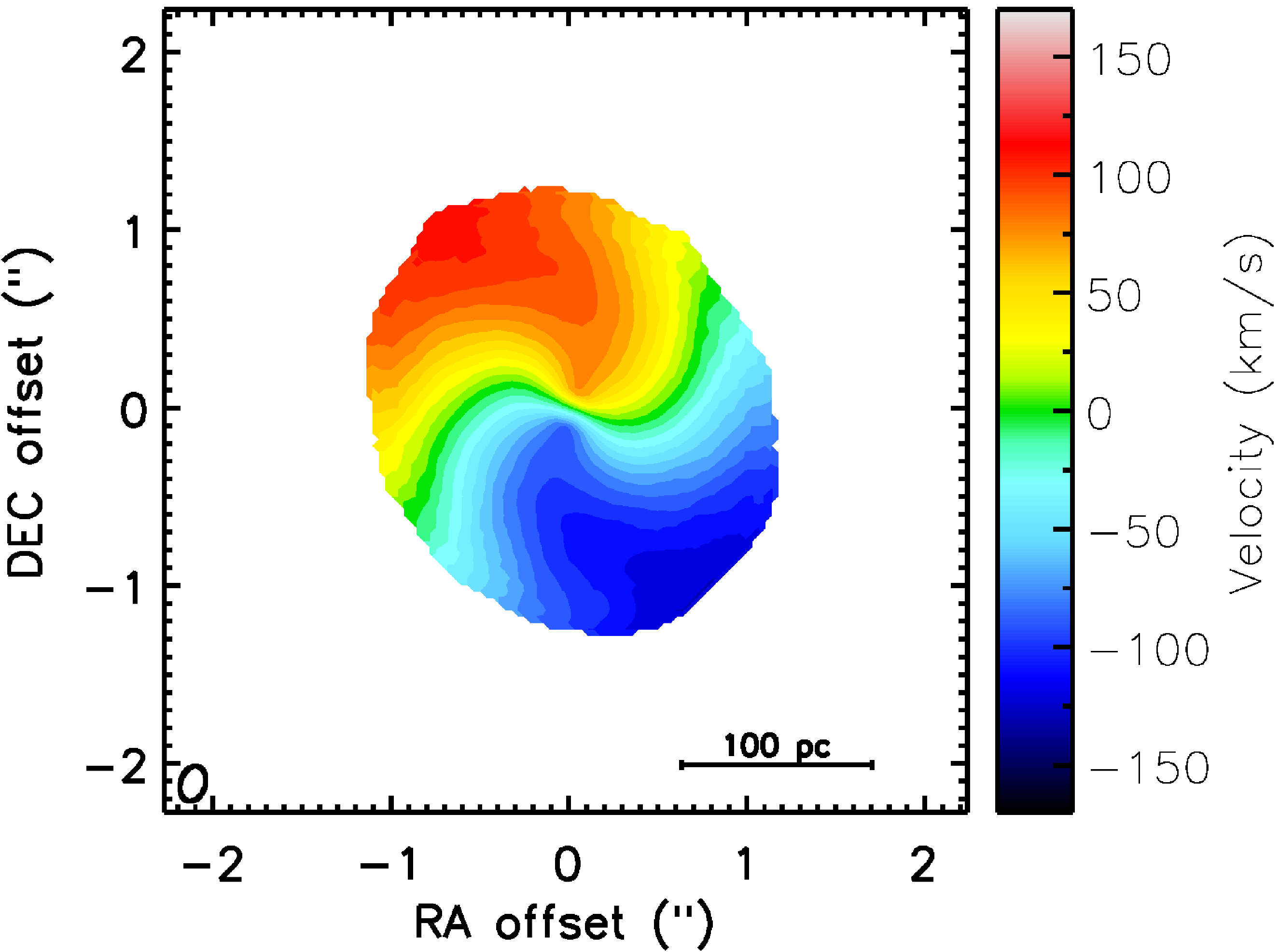}
\end{subfigure}

\medskip

\begin{subfigure}[t]{0.3\textheight}
\centering
\vspace{0pt}
\caption{Data moment two}\label{fig:ngc1574_mom2}
\includegraphics[scale=0.52]{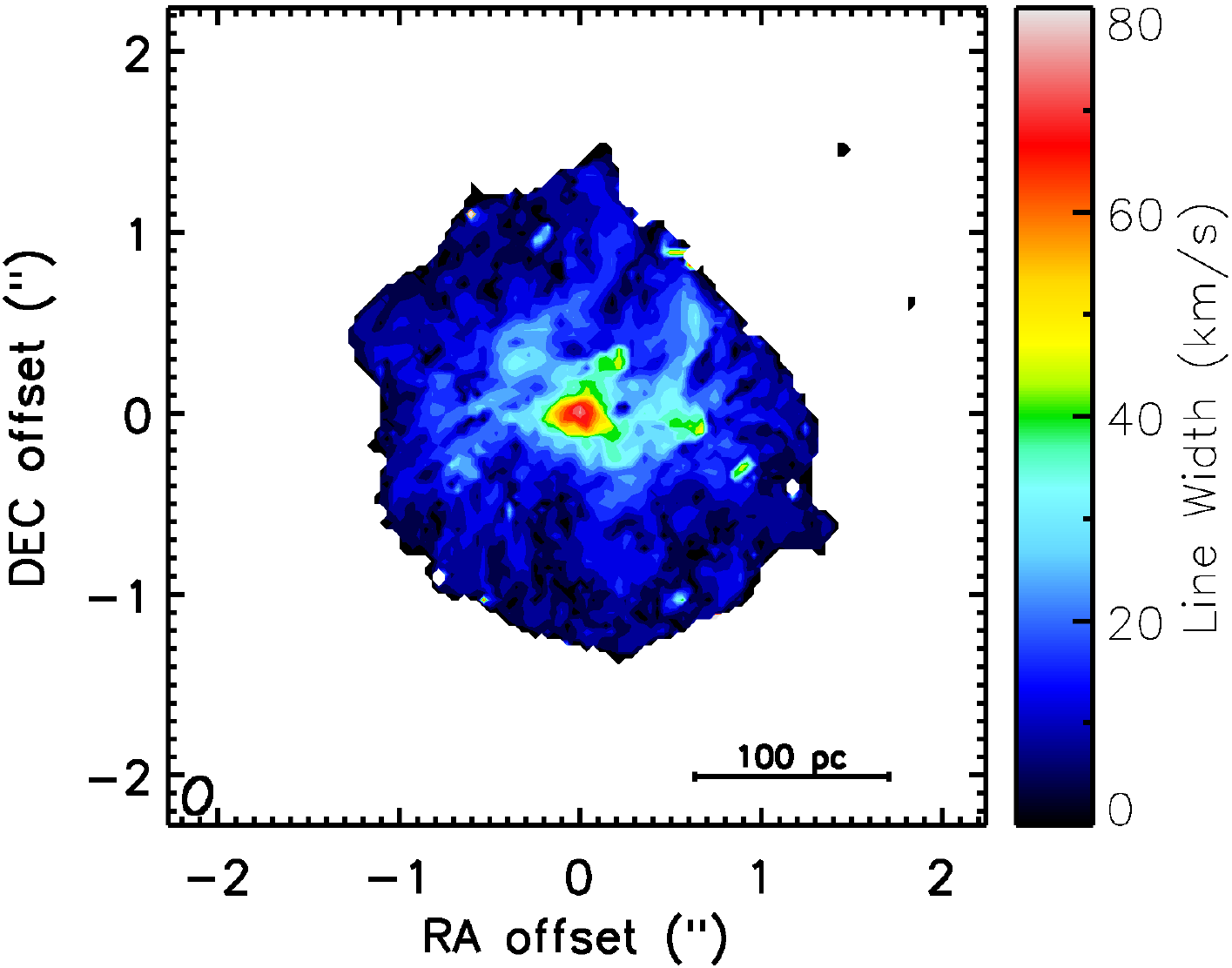}
\end{subfigure}
\hspace{10mm}
\begin{subfigure}[t]{0.3\textheight}
\centering
\vspace{0pt}
\caption{Model moment two}\label{fig:ngc1574_mom2_mod}
\includegraphics[scale=0.3]{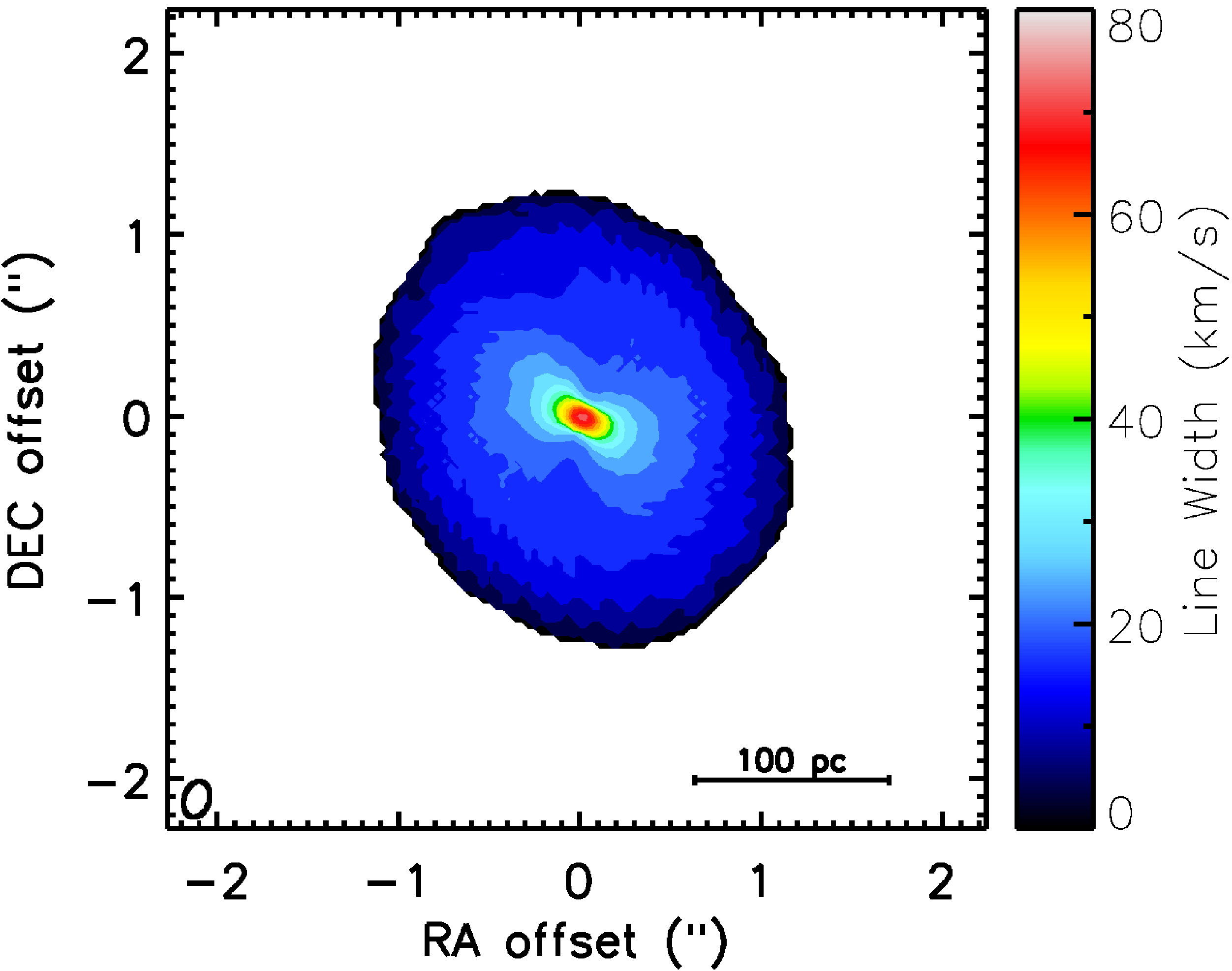}
\end{subfigure}
\caption{Same as Figure~\ref{fig:NGC612_moments_all_appendix}, but for NGC\,1574.}\label{fig:NGC1574_moments_all}
\end{figure*}

\begin{figure*}
\begin{subfigure}[t]{0.3\textheight}
\centering
 \caption{Data moment zero}\label{fig:ngc4261_mom0}
\includegraphics[scale=0.52]{ngc4261_mom0.pdf}
\end{subfigure}
\hspace{10mm}
\begin{subfigure}[t]{0.3\textheight}
\centering
\caption{Model moment zero}\label{fig:ngc4261_mom0_mod}
\includegraphics[scale=0.52]{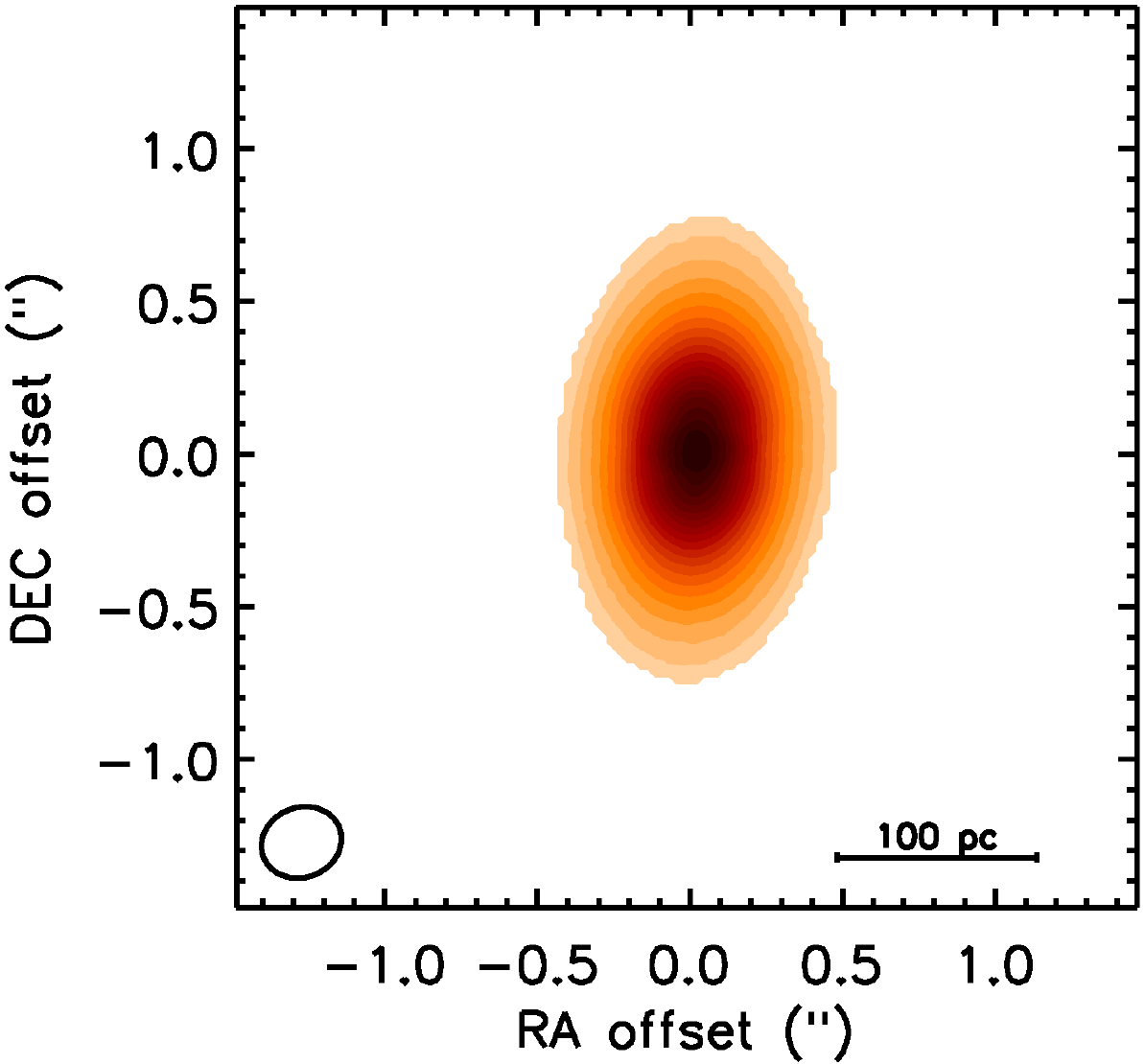}
\end{subfigure}

\medskip

\begin{subfigure}[t]{0.3\textheight}
\centering
 \caption{Data moment one}\label{fig:ngc4261_mom1}
\includegraphics[scale=0.52]{ngc4261_mom1.pdf}
\end{subfigure}
\hspace{12mm}
\begin{subfigure}[t]{0.3\textheight}
\centering
\caption{Model moment one}\label{fig:ngc4261_mom1_mod}
\includegraphics[scale=0.52]{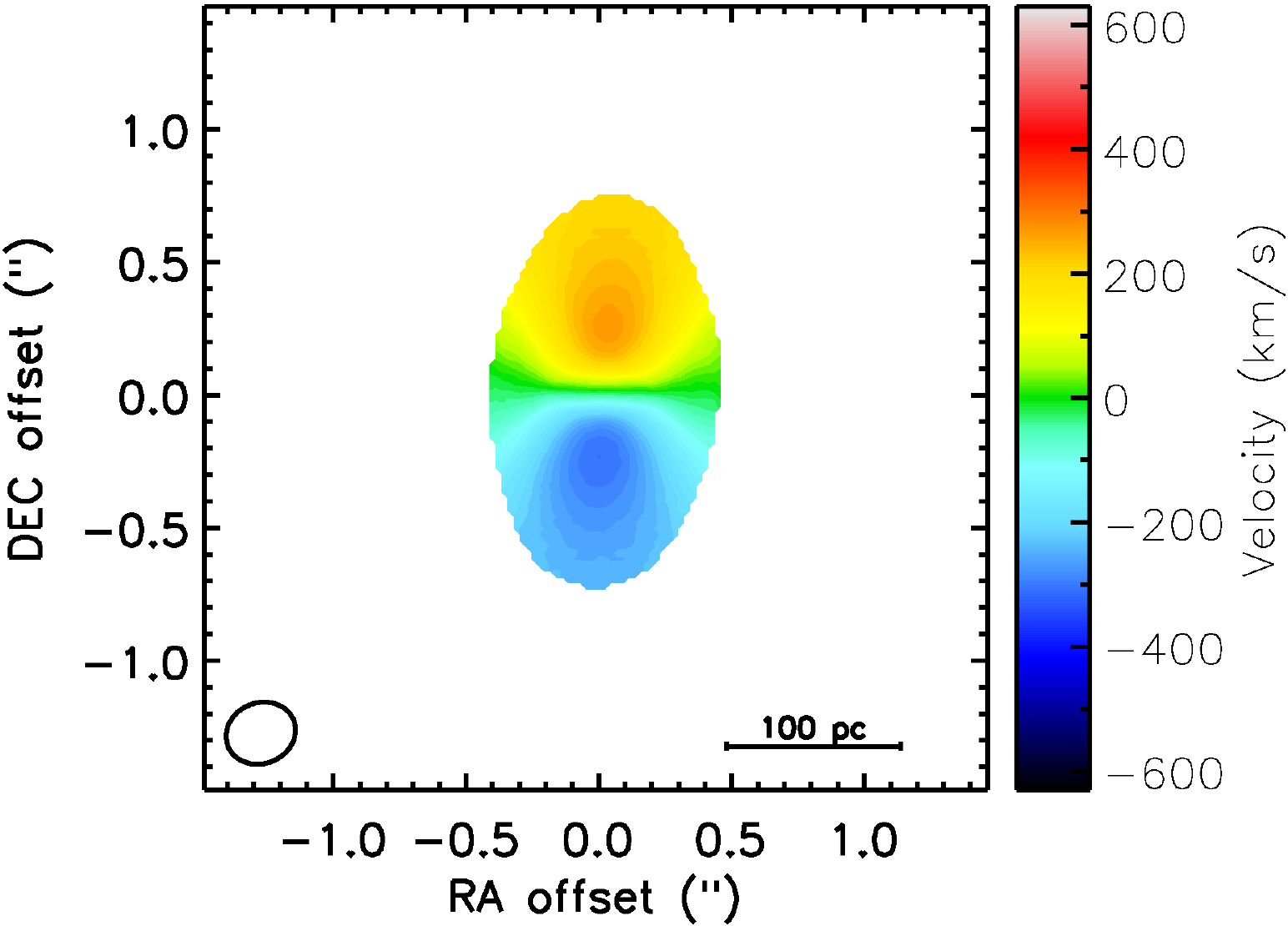}
\end{subfigure}

\medskip

\begin{subfigure}[t]{0.3\textheight}
\centering
\vspace{0pt}
\caption{Data moment two}\label{fig:ngc4261_mom2}
\includegraphics[scale=0.52]{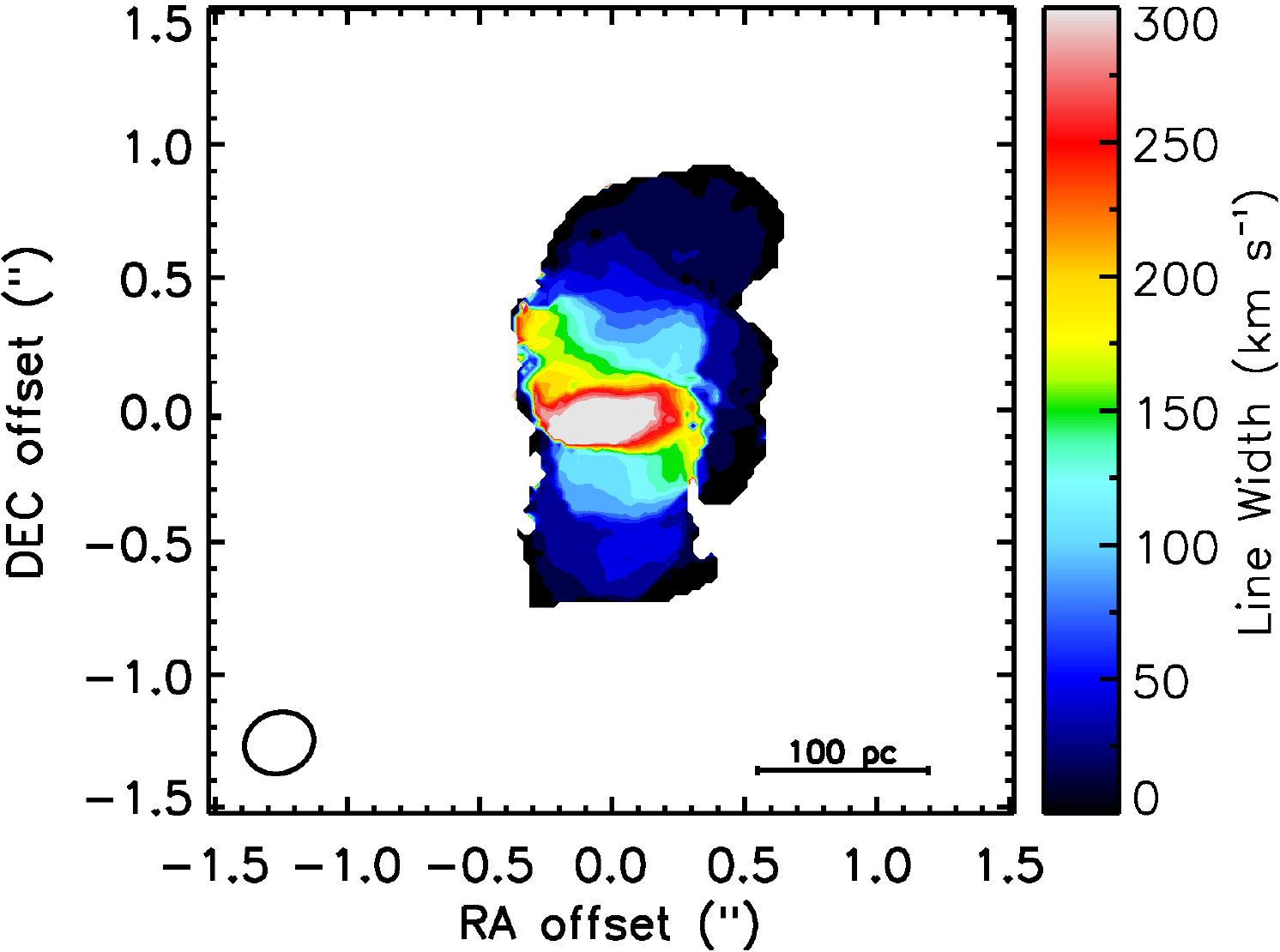}
\end{subfigure}
\hspace{10mm}
\begin{subfigure}[t]{0.3\textheight}
\centering
\vspace{0pt}
\caption{Model moment two}\label{fig:ngc4261_mom2_mod}
\includegraphics[scale=0.52]{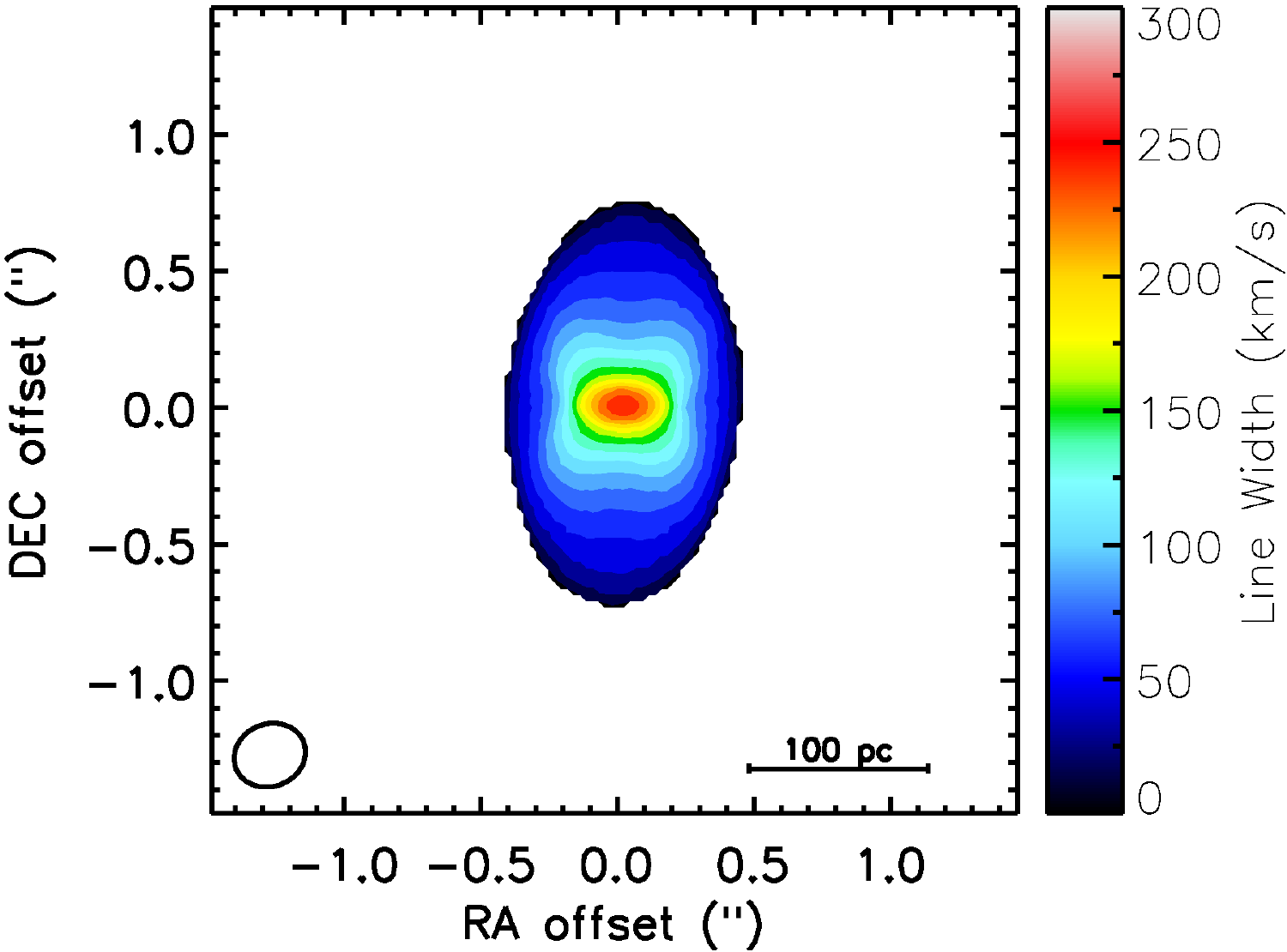}
\end{subfigure}
\caption{Same as Figure~\ref{fig:NGC612_moments_all_appendix}, but for NGC\,4261.}\label{fig:NGC4261_moments_all}
\end{figure*}

\section{Corner plots}\label{sec:corner_plots_appendix}
We show here the corner plots of the best-fitting models of all the targets.

\begin{figure*}
\centering
\includegraphics[scale=0.6]{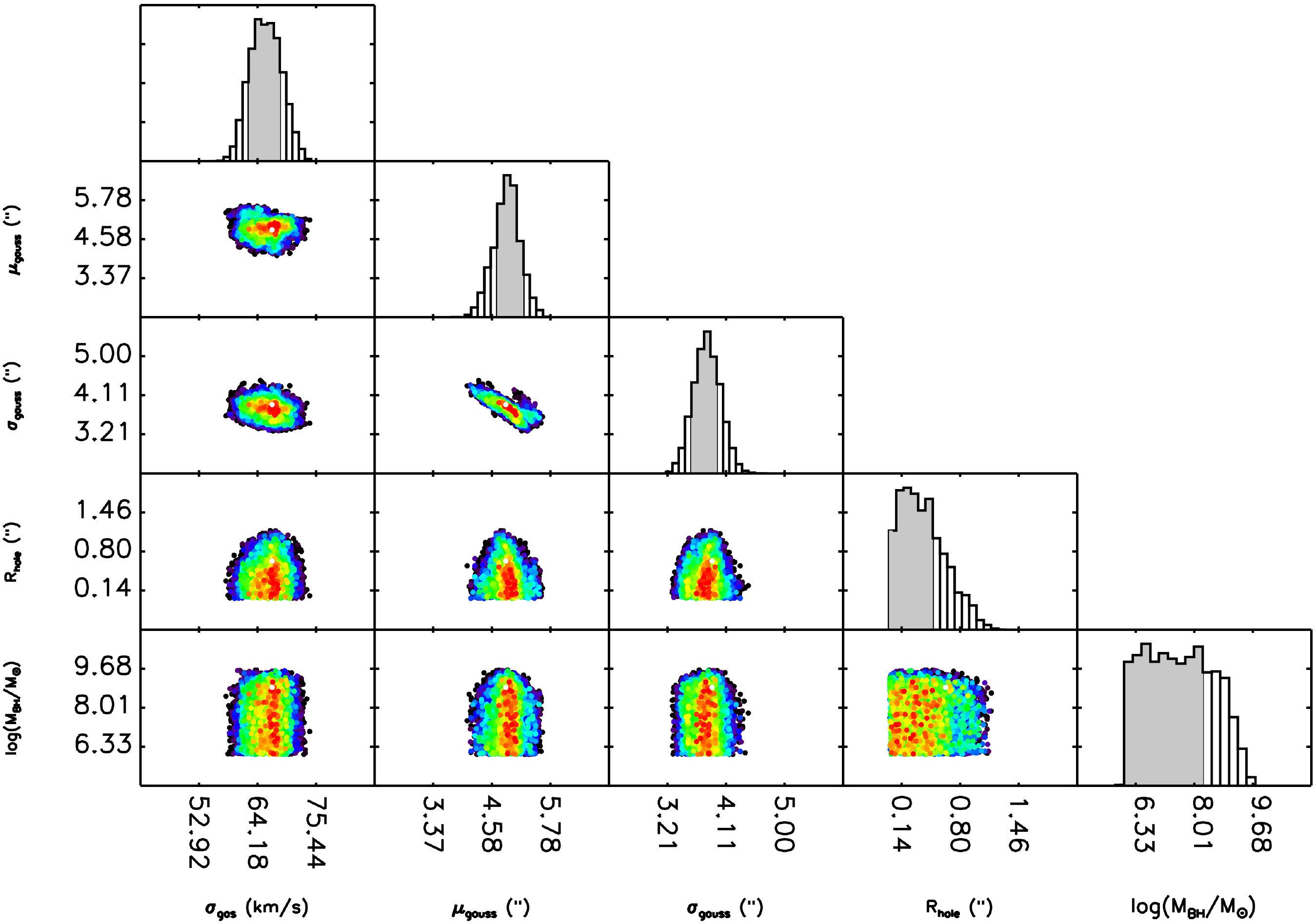}
\caption[]{Corner plots showing covariances between relevant (i.e.\,non-nuisance) parameters of one of the three tested models for NGC\,0612 (i.e.\,the Sersic model; see the text for details). The colours represent increasing confidence intervals from 68.3\% (red, 1$\sigma$) to 99.7\% (blue, 3$\sigma$). The white dots show the $\chi^{2}_{\rm min}$ values. Histograms show the one-dimensional marginalised posterior distribution of each model parameter. The grey shaded regions indicate the 68\% (1$\sigma$) confidence intervals. \label{fig:NGC612_conts}}
\end{figure*}

\begin{figure*}
\centering
\includegraphics[scale=0.6]{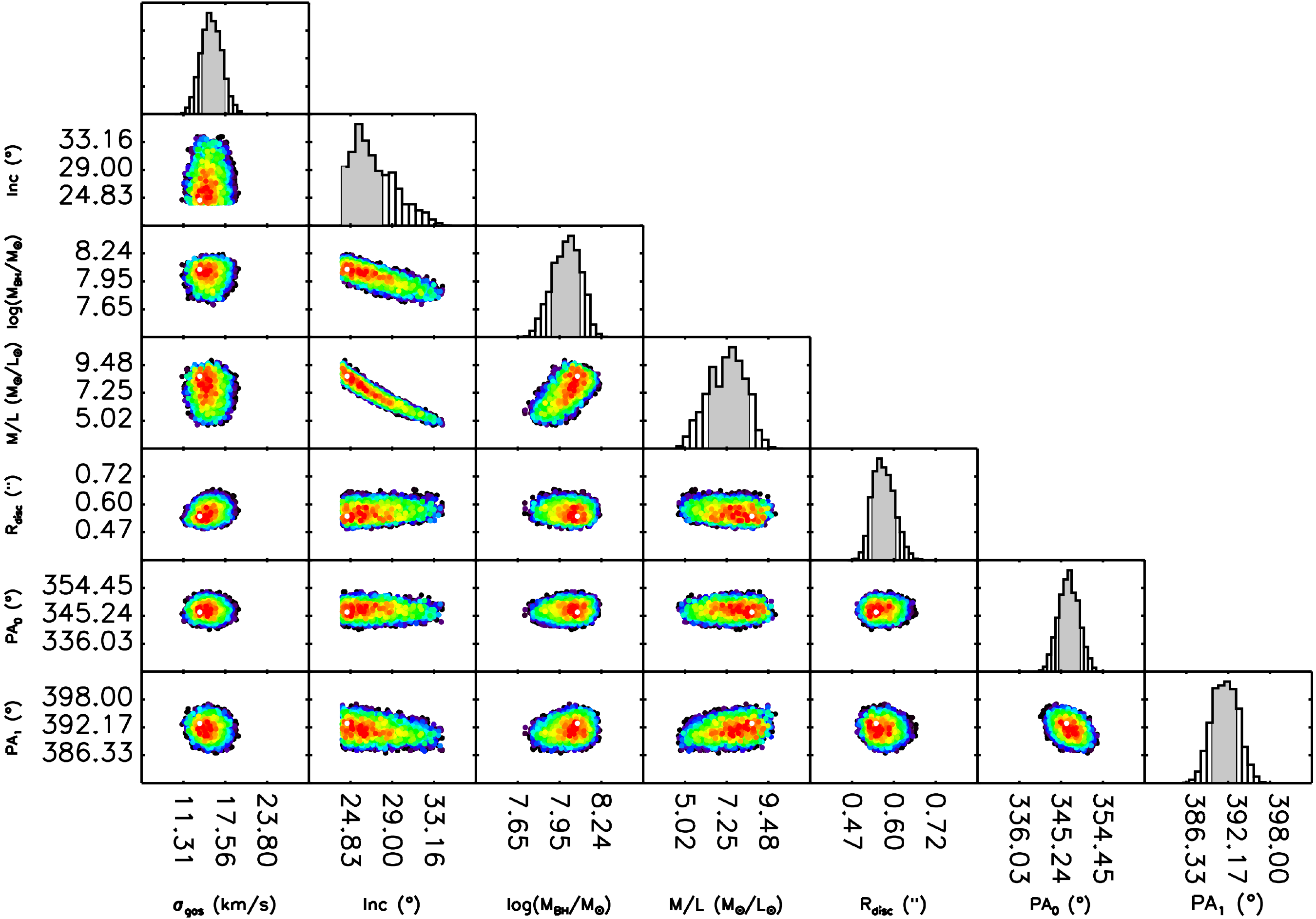}
\caption[]{As in Figure~\ref{fig:NGC612_conts}, but for NGC\,1574. \label{fig:NGC1574_conts}}
\end{figure*}

\begin{figure*}
\centering
\includegraphics[scale=0.6]{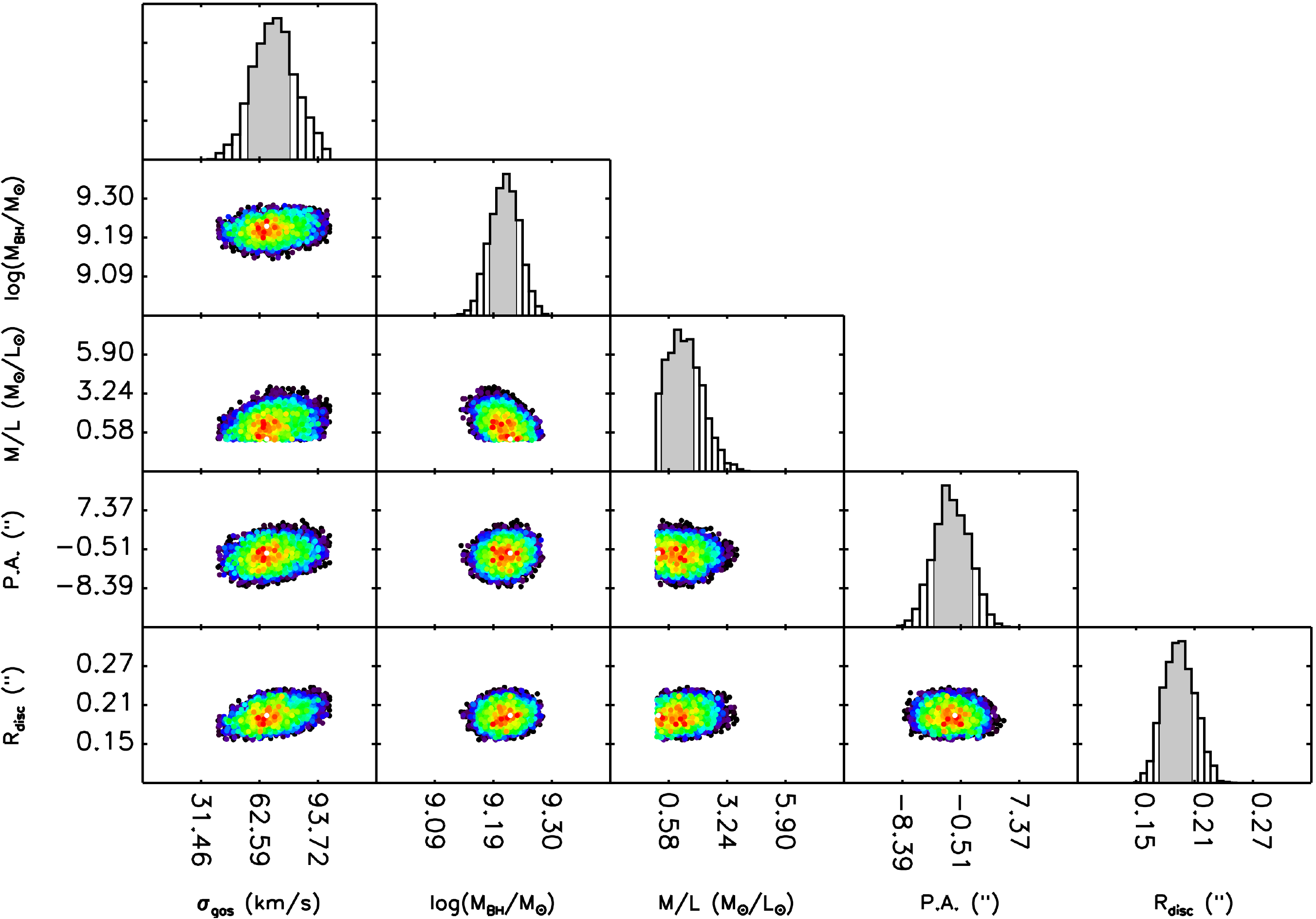}
\caption[]{As in Figure~\ref{fig:NGC612_conts}, but for NGC\,4261. \label{fig:NGC4261_conts}}
\end{figure*}

\section{MGE solutions}\label{sec:MGE_solutions_appendix}
We report here the MGE solutions obtained for the parametrisation of the light profile in NGC\,1574, and reproduce those adopted for NGC\,4261 from \citet{Boizelle21}.

\begin{table}
\centering
\caption{MGE parameterisation of the NGC\,1574 light profile.}
\label{tab:NGC1574_MGE_parameters}
\begin{tabular}{c c c}
\hline
\multicolumn{1}{c}{ $\log_{10}I^{'}_{\rm j}$} &
\multicolumn{1}{c}{ $\sigma_{\rm j}$ } &
\multicolumn{1}{c}{ $q^{'}_{j}$} \\
\multicolumn{1}{c}{ (L$_{\odot, i} pc^{-2}$) } &
\multicolumn{1}{c}{ ('') } &
\multicolumn{1}{c}{  } \\
\multicolumn{1}{c}{ (1) } &
\multicolumn{1}{c}{ (2) } &
\multicolumn{1}{c}{ (3) } \\
\hline
4.983   & 0.035 & 1.000 \\
3.957 & 0.150 & 1.000 \\
4.045  &   0.393   &   1.000 \\
3.797   &   1.279   &   1.000 \\
3.491   &   2.359   &   1.000 \\
3.277  &    4.650   &   1.000 \\
2.718  &    10.289  &   0.375 \\
\hline
\end{tabular}
\parbox[t]{8.5cm}{ \textit{Notes.} $-$ Columns: Intensity (1), width (2), and axis ratio (3) of each Gaussian component.}
\end{table}

\begin{table}
\centering
\caption{MGE parameterisation of the NGC\,4261 light profile, reproduced from \citet[][]{Boizelle21}.}
\label{tab:NGC4261_MGE_param}
\begin{tabular}{c c c}
\hline
\multicolumn{1}{c}{ $\log_{10}I^{'}_{\rm j}$} &
\multicolumn{1}{c}{ $\sigma_{\rm j}$ } &
\multicolumn{1}{c}{ $q^{'}_{j}$} \\
\multicolumn{1}{c}{ (L$_{\odot,H} pc^{-2}$) } &
\multicolumn{1}{c}{ ('') } &
\multicolumn{1}{c}{  } \\
\multicolumn{1}{c}{ (1) } &
\multicolumn{1}{c}{ (2) } &
\multicolumn{1}{c}{ (3) } \\
\hline
4.261  & 1.075 & 0.830 \\
4.143  & 2.131 & 0.717 \\
3.964  & 3.837 & 0.729 \\
3.437 & 8.191 & 0.719 \\
3.194  & 13.54 & 0.834 \\
2.595  & 23.79 & 0.816 \\
2.381  & 49.05 & 0.862 \\
1.392  & 144.4 & 0.820\\
\hline
\end{tabular}
\parbox[t]{8.5cm}{ \textit{Notes.} $-$ Columns: Intensity (1), width (2), and axis ratio (3) of each Gaussian component.}
\end{table}


\bsp	
\label{lastpage}
\end{document}